\documentclass[
10pt,aps,amsmath,amssymb,
reprint
,pra
,longbibliography
,notitlepage
,superscriptaddress
]{revtex4-1}
\usepackage{dcolumn}
\usepackage{bm}
\usepackage{braket}
\usepackage{xcolor}
\usepackage{graphicx}
\usepackage{mathtools}
\usepackage{multirow}
\usepackage[colorlinks=true,linkcolor=blue,urlcolor=black,citecolor=blue,bookmarksopen=true]{hyperref}
\usepackage{tabularx}
\usepackage[caption=false]{subfig}
\usepackage{physics}
\usepackage{mathrsfs}
\usepackage{xifthen}

\newcommand{\mP}{\mathscr{P}}
\newcommand{\mcP}{\mathcal{P}}
\newcommand{\CP}[2][]{%
\ifthenelse{\isempty{#1}}{\textrm{CP}(#2)}{\textrm{CP}^{(#1)}(#2)}
}

\begin{document}

\title{Protocol for nearly deterministic parity projection on two photonic qubits}
\author{Chenxu Liu}
\affiliation{Department of Physics, Virginia Tech, Blacksburg, Virginia, 24061, USA}
\affiliation{Virginia Tech Center for Quantum Information Science and Engineering, Blacksburg, VA 24061, USA}

\author{Rafail Frantzeskakis}
\affiliation{Department of Physics, University of Crete, Heraklion, 71003, Greece}

\author{Sophia E. Economou}
\affiliation{Department of Physics, Virginia Tech, Blacksburg, Virginia, 24061, USA}
\affiliation{Virginia Tech Center for Quantum Information Science and Engineering, Blacksburg, VA 24061, USA}

\author{Edwin Barnes}
\email{Email: efbarnes@vt.edu}
\affiliation{Department of Physics, Virginia Tech, Blacksburg, Virginia, 24061, USA}
\affiliation{Virginia Tech Center for Quantum Information Science and Engineering, Blacksburg, VA 24061, USA}

\date{\today}

\begin{abstract}
Photonic parity projection plays an important role in photonic quantum information processing. Non-destructive parity projections normally require high-fidelity Controlled-Z gates between photonic and matter qubits, which can be experimentally demanding. In this paper, we propose a nearly deterministic parity projection protocol on two photonic qubits which only requires stable matter-photon Controlled-Phase gates. The fact that our protocol does not require perfect Controlled-Z gates makes it more amenable to experimental implementation.
\end{abstract}

\maketitle

\section{Introduction} \label{sec:intro}

Quantum information technologies promise to provide significant speedup on certain classically hard problems~\cite{Grover1996,Shor1994,Shor1997,Simon1997, Arute2019, Bouland2019, Zhong2021, Wu2021}, more secure communications~\cite{Bennett1984, Ekert1991, Shor2000, Sasaki2014, Ursin2007, Liao2017, Bennett1993, Bouwmeester1997, Ma2012, Yin2012, Li2022}, and more accurate measurements~\cite{elzerman2004single,atature2006quantum,ramsay2008fast,robledo2011high,javadi2018spin}. Because photons do not naturally interact with each other, photonic qubits can have long lifetimes and coherence times, which makes them a prominent candidate for long-range quantum communication~\cite{Ursin2007, Ma2012, Yin2012, Liao2017} and large-scale quantum computing~\cite{Zhong2021, Bartolucci2021}. However, photonic quantum information processing requires generating quantum entanglement and performing two-qubit gates between photonic qubits, and these are challenging precisely because of the absence of a natural interaction between photons.

As passive linear-optical operations cannot induce interactions between two photons, applying two-qubit gates between photonic qubits requires either nonlinear media or photonic measurements and feed-forward~\cite{Knill2001}. With the latter approach, photonic Controlled-NOT (CNOT) and other two-qubit gates can be implemented probabilistically~\cite{Knill2001, Ralph2002, OBrien2003, Knill2002, Pittman2003, Gasparoni2004, Zhao2005, Bao2007, Zeuner2018, Li2021}. On the other hand, with the help of a strong nonlinearity provided by, e.g., strongly coupled cavity-QED systems, photon-photon Controlled-Z (CZ) gates can be implemented deterministically~\cite{xia2011efficient,Hacker2016,reuer2022realization}. However, how to implement these operations efficiently and accurately remains a key outstanding question in photonic quantum information processing.

Among all the possible protocols to generate photonic entanglement, projecting onto multiphoton parity subspaces plays a fundamental role in many photonic quantum information processing tasks. It not only facilitates gate operations between photonic qubits and entanglement generatation~\cite{Koashi2001, Pittman2001, Nemoto2004, Kok2007, Ionicioiu2008, Wang2012, Wei2013, Wei2014}, but is also widely used in photonic entanglement purification, distillation, and concentration protocols~\cite{Song2005, Qian2005, Deng2011, Sheng2012, Deng2012, Choudhury2013, Zhu2015, Liu2018, Daiss2019}. One way to perform parity projections on photonic qubits is to use beam splitters combined with photonic measurements~\cite{Pittman2001, Pittman2002, Pittman2003, Browne2005}. In this method, the photonic qubits that undergo the projection are destroyed after the operation. Though this method only relies on linear-optical elements and is thus easily implemented, the fact that photonic qubits are destroyed after the projection limits its efficiency. An alternative way to perform a photonic parity projection relies on a nonlinearity provided by optical nonlinear media to implement a non-destructive parity projection~\cite{Bennett2005,Nemoto2004, Qian2005, Sheng2008, Qing2009, Qi2011, Wang2012}. Because Kerr nonlinearities are typically weak, a strong coherent beam needs to be applied to boost the efficiency, but this is often at the expense of an increase in noise, in addition to the noise stemming from the nonlinearity itself~\cite{Kok2008}. Moreover, because the Kerr nonlinearity is weak, the resulting even-parity projection is usually considered as `incomplete', while the odd-parity projection fails since the two states can be distinguished~\cite{Nemoto2004, Bennett2005}.

Owing to recent progress in fabricating strongly coupled cavity-QED and circuit-QED systems, strong nonlinear coupling between matter qubits and photonic qubits can now be systematically achieved~\cite{Fushman2008, Loo2012, kim2013quantum, Sun2016, Tiarks2016,  Sun2018, Wells2019}. These advances have enabled the realization of matter-photon CZ gates, which can be used to implement non-destructive parity projections~\cite{NielsenBook, Hacker2016, Wei2013, Wei2014, Wang2013, Ren2013}. However, because of limited photon pulse bandwidths~\cite{Shapiro2006, GBanacloche2010} and other experimental imperfections~\cite{Sun2016, Sun2018, Wells2019}, these CZ gates are challenging to achieve with high fidelity. In the presence of coherent phase errors, the CZ gates are actually Controlled-Phase (CPhase) gates, a term we reserve for the general case in which the phase is less than $\pi$ \cite{Campbell_PRA2007}. Thus, an important outstanding question is whether it is still possible to perform high-quality parity projections efficiently if we only have access to matter-photon CPhase gates.

In this paper, we propose a nearly deterministic protocol for implementing non-destructive photonic parity projections. Our protocol utilizes CPhase gates between photonic and matter qubits instead of requiring perfect CZ gates ($\pi$ phase in CPhase gates). Furthermore, our protocol performs a complete parity projection to both even and odd-parity subspaces. Moreover, we demonstrate that our protocol can tolerate moderate Gaussian phase errors in the CPhase gates as well as Pauli errors on the matter qubits. We stress that although our discussion focuses on optical photonic systems, our protocol is generally applicable to other physical systems, including microwave photonic systems and photonic entanglement concentration protocols~\cite{RafailWork}.  

This paper is organized as follows. In Sec.~\ref{sec:perfect_protocol}, we present our nearly-deterministic parity projection protocol with CPhase gates. We show the average channel fidelity and parity measurement error probability of our protocol relative to a perfect parity projection operation. In Sec.~\ref{sec:errors}, we consider the effects of possible imperfections. We consider coherent errors on the CPhase gates in Sec.~\ref{sec:errors:coh_error}, where the different CPhase gates can have different phases. We then extend our discussion to include random CPhase angle fluctuations, which is modeled as random Gaussian noise, in Sec.~\ref{sec:errors:incoh_cphase}. We calculate the channel fidelity and error probability of our protocol in this case. In addition, we discuss the effect of possible Pauli errors that happen in different stages of the protocol in Sec.~\ref{sec:errors:pauli}. Finally we conclude in Sec.~\ref{sec:summary}.

\section{Nearly-deterministic Parity projection protocol} \label{sec:perfect_protocol}

\begin{figure}[h]
\includegraphics[width =  \columnwidth]{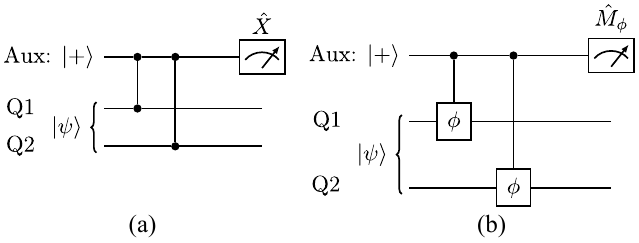}
\caption{Gate sequences for the parity projection on two qubits, Q1 and Q2. (a) The gate sequence for a parity projection implemented with CZ gates between the auxiliary qubit and qubits Q1 and Q2. (b) The gate sequence for a single round of our parity projection protocol, which only requires CPhase gates $\CP{\phi}$. The measurement basis is $\hat{M}_{\phi}$, which is given by Eq.~\eqref{eq:m_basis}.}
\label{fig:parity_proj_circ}
\end{figure}

As demonstrated in Ref.~\cite{NielsenBook}, a non-destructive parity projection on a pair of qubits can be performed using two CZ gates (see Fig.~\ref{fig:parity_proj_circ}a). In this protocol, an auxiliary qubit is initialized to $\ket{+} = \frac{1}{\sqrt{2}}\left( \ket{0}  + \ket{1}\right)$, and then two CZ gates are applied as shown in Fig.~\ref{fig:parity_proj_circ}a. Afterward, the auxiliary qubit is measured in the $X$ basis. Depending on the measurement outcomes, a parity projection onto the even or odd subspace is performed on the two qubits (Q1 and Q2).

When performing the standard non-destructive parity projection shown in Fig.~\ref{fig:parity_proj_circ}a, one of the bottlenecks is implementing the CZ gates with high fidelity, especially for photonic qubits. As demonstrated in Refs.~\cite{Loo2012, kim2013quantum, Sun2016, Tiarks2016,  Sun2018, Wells2019}, precisely implementing CZ gates between a matter qubit and a photonic qubit can be challenging. Instead, the resulting gates are often CPhase gates, which can be represented in the computational basis as
\begin{equation}
	\CP{\phi} = \left[ 
	\begin{array}{cccc}
	1 & & & \\
	& 1 & & \\
	& & 1 & \\
	& & & e^{i\phi}
	\end{array} \right],
	\label{eq:cphase}
\end{equation}
where the phase angle $\phi$ is not exactly $\pi$. If one attempts the parity projection shown in Fig.~\ref{fig:parity_proj_circ}a, but with imperfect CZ gates such that $\phi \neq \pi$, one ends up performing the following imperfect projections:
\begin{align}
	\mcP_{\text{even}}^{\text{err}} & = \mcP_{00} + e^{i\phi/2} \cos\left(\frac{\phi}{2}\right) (\mcP_{01} + \mcP_{10} ) + e^{i\phi} \cos(\phi) \mcP_{11} \label{eq:naive_1}\\
	\mcP_{\text{odd}}^{\text{err}} & = - i e^{i\phi/2} \sin \left(\frac{\phi}{2}\right) (\mcP_{01} + \mcP_{10} ) - i e^{i\phi} \sin(\phi) \mcP_{11}, \label{eq:naive_2}
\end{align}
where $\mcP_{ij}=\ket{i,j}\bra{i,j}$ for $i,j=0,1$ are projectors onto the two-qubit computational states. The quantum operation is not a perfect parity projection operation any more. In what follows, we refer to this as the ``naive" parity projection protocol.

Here we present a protocol to nearly deterministically perform parity projection on two qubits with CPhase gates between the matter qubit and the photonic qubits. Instead of requiring CZ gates between qubits as shown in Fig.~\ref{fig:parity_proj_circ}a, our protocol can use CPhase gates with arbitrary phase angle $\phi$. Our protocol, which only requires $\phi$ to be known and the same for both CPhase gates, is summarized as follows:
\begin{enumerate}
    \item Prepare the auxiliary (matter) qubit in the $\ket{+}$ state and apply two $\text{CP}(\phi)$ gates between the matter qubit and two photonic qubits.
    \item Perform a measurement on the matter qubit in the eigenbasis of the operator
    \begin{equation}
    	\hat{M}_{\phi} = R_z(\phi) \hat X R_z^{\dagger}(\phi),
	\label{eq:m_basis}
    \end{equation}
    where $R_z$ is a Pauli-$Z$ rotation.
    \item If the measurement result is $-1$, the two photonic qubits are projected onto the even subspace.
    \item If the measurement result is $+1$, repeat Step~1 and~2.
\end{enumerate}
By repeating this protocol enough times, provided the measurement outcomes are all $-1$, the state is nearly perfectly projected onto the odd subspace.

To understand how our parity projection works, we first show the Kraus operators for a single round of operation, for which the circuit is shown in Fig.~\ref{fig:parity_proj_circ}b. Notice that a CPhase gate can be decomposed as
\begin{equation}
	\CP[\alpha, \beta]{\phi} = \mcP_{0}^{(\alpha)} I^{(\beta)} + \mcP_{1}^{(\alpha)} S^{(\beta)}(\phi),
\end{equation}
where the superscripts label qubits, $\mcP_{i}= \dyad{i}$ are the projectors onto the computational states $\ket{i}$ with $i=0,1$, $I$ denotes the identity operator, and $S(\phi)$ is the single-qubit phase gate,
\begin{equation}
    S(\phi) = \left( \begin{array}{cc}
        1 &  \\
         & e^{i \phi}
    \end{array} \right) = e^{i \phi/2} R_z(\phi).
\end{equation}
If we take the projective measurement along the basis of $\hat{M}_{\phi}$, i.e., projecting onto the states
\begin{align}
    \ket{\pm_\phi} = \frac{1}{\sqrt{2}} \left( e^{-i \phi/2} \ket{0} \pm e^{i \phi/2} \ket{1} \right),
\end{align}
when the measurement outcome is $-1$, the Kraus operator acting on the two qubits is
\begin{equation}
    \hat{E}_{-1} = i \sin \left( \frac{\phi}{2}\right) \left(\mcP_{00} - e^{i\phi} \mcP_{11} \right),
    \label{eq:first_round_even}
\end{equation}
which perfectly projects the two photonic qubits onto the even subspace. However, if the measurement result is $+1$, the Kraus operator is
\begin{equation}
    \hat{E}_{+1} = \cos \left(\frac{\phi}{2}\right) \left(\mcP_{00} + e^{i\phi} \mcP_{11}\right) + e^{i\phi/2}(\mcP_{01} + \mcP_{10}),
    \label{eq:first_round_odd}
\end{equation}
which does not project the two-qubit state onto the odd subspace perfectly. However, the projection can be viewed as a projection onto the odd subspace with less certainty, where the even subspace is suppressed by $\cos(\phi/2)$. If the measurement outcome is $+1$, the state is more likely to be in the odd-parity subspace. 

Therefore, after we get $+1$ outcome, we can repeat this operation again. In the second round, if the measurement outcome becomes $-1$, the corresponding Kraus operator is
\begin{align}
    \hat{E}_{+1, -1} &= \hat{E}_{-1} \hat{E}_{+1} 
     = i \frac{1}{2} \sin (\phi) \left( \mcP_{00} - e^{i 2 \phi} \mcP_{11}\right),
\end{align}
which shows that the state is projected onto the even subspace again.
If we get $+1$ again, the Kraus operator is
\begin{align}
    \hat{E}_{+1,+1} = \cos^2 \left(\frac{\phi}{2}\right) \left(\mcP_{00} + e^{i 2 \phi} \mcP_{11}\right) + e^{i\phi} (\mcP_{01} + \mcP_{10}).
\end{align}
Although this is still not a perfect odd-parity projection, comparing to the first round operation [Eq.~\eqref{eq:first_round_odd}], the even-parity subspace is further suppressed. Therefore, we can repeat the operation another time to suppress the even-parity subspace even further.

If the operation is performed $n$ rounds in total, and we get $+1$ measurement outcomes for the first $n-1$ rounds but $-1$ for the $n$-th round, the Kraus operator is
\begin{align}
    \hat{E}_{\text{even}}^{(n)} & = i \sin(\phi/2) \left[ \cos(\phi/2) \right]^{n-1}  \left(\mcP_{00} - e^{i n \phi} \mcP_{11}\right).
    \label{eq:even_n}
\end{align}
If we get $+1$ up to round $n$, the Kraus operator is
\begin{align}
    \hat{E}^{(n)}_{\text{odd}} & = e^{i n \phi/2} \left( \mcP_{01} + \mcP_{10} \right) 
    + \cos^{n}(\phi/2) \left(\mcP_{00} + e^{i n \phi} \mcP_{11}\right).
    \label{eq:odd_n}
\end{align}
Although the Kraus operators shown in Eq.~\eqref{eq:even_n} are different from the perfect even-parity projection by a relative phase factor between the projectors $\mcP_{00}$ and $\mcP_{11}$, which depends on the operation cycle $n$ at which a $-1$ measurement outcome is achieved, this phase factor can be corrected by applying a Pauli-$Z$ rotation $R_Z(-n \phi + \pi)$ on either of the two qubits after the measurement. Furthermore, these Pauli-$Z$ rotations can usually be virtually applied by adjusting the lab frame, which does not introduce any error. 

With these Pauli-$Z$ rotations applied to the qubits Q1 or Q2 conditioned on $-1$ measurement outcomes, the Kraus operators in Eq.~\eqref{eq:even_n} become
\begin{equation}
    \hat{E}_{\text{even}}^{(m)} = \sin(\phi/2) \left[ \cos (\phi/2) \right]^{m-1} \left(\mcP_{00} +\mcP_{11}\right),
    \label{eq:correct_even_m}
\end{equation}
where $m = 1, 2, ... , n$. Combining these with the odd-parity projection Kraus operator $\hat{E}_{\text{odd}}^{(n)}$ in Eq.~\eqref{eq:odd_n}, the process can be described as a quantum channel in the operator sum representation with these $n+1$ Kraus operators. The Kraus operators in Eq.~\eqref{eq:correct_even_m} correspond to perfect even-parity projections, while Eq.~\eqref{eq:odd_n} is the imperfect odd-parity projection, where the imperfection is exponentially suppressed as the operation cycle $n$ increases. When $n \rightarrow \infty$, this procedure is exactly a parity projection of the two qubits Q1 and Q2.

In order to evaluate the performance of our protocol and compare it with a perfect parity projection operation, we first treat the process as a quantum channel and show how close it is to the perfect parity projection operation. This is helpful in order to understand how effective the protocol can be as an entanglement generation operation. 
As pointed out in Ref.~\cite{Gilchrist2005}, the average fidelity of the output states of two quantum channels can be used as a measure of their closeness; we use this average channel fidelity of our protocol relative to the ideal parity projection as a figure of merit. The average channel fidelity is defined as
\begin{align}
    \langle F \rangle = \int F \left( \rho_{\text{ideal}}, \rho \right) d\ket{\psi},
    \label{eq:fidelity}
\end{align}
where the integral is over the Haar measure of the two-qubit state space, and the state fidelity is given by
\begin{align}
    F(\rho_{\text{ideal}}, \rho) = \text{Tr}\left( \sqrt{\sqrt{\rho} \rho_{\text{ideal}} \sqrt{\rho}} \right)^2.
\end{align} 
The state $\rho$ is the output from our protocol with $n$ operation cycles when the two qubits are initialized in $\ket{\psi}$: 
\begin{equation}
    \rho = \hat{E}_{\text{odd}}^{(n)} \dyad{\psi} \hat{E}_{\text{odd}}^{(n) \dagger} + \sum_{j=1}^{n} \hat{E}_{\text{even}}^{(j)} \dyad{\psi} \hat{E}_{\text{even}}^{(j) \dagger},
\end{equation}
where $\hat{E}_{\text{odd}}^{(n)}$ and $\hat{E}_{\text{even}}^{(j)}$ are given by Eqs.~\eqref{eq:odd_n} and~~\eqref{eq:correct_even_m}, respectively. The output state from a perfect parity projection operation is
\begin{equation}\label{eq:rho_ideal}
    \rho_{\text{ideal}} = \mcP_{\text{even}} \dyad{\psi} \mcP_{\text{even}} + \mcP_{\text{odd}} \dyad{\psi} \mcP_{\text{odd}},
\end{equation}
where $\mcP_{\text{even}}  = \mcP_{00} + \mcP_{11}$, and $\mcP_{\text{odd}} = \mcP_{01} + \mcP_{10}$.

\begin{figure}[h]
    \centering
    \includegraphics[width =  \columnwidth]{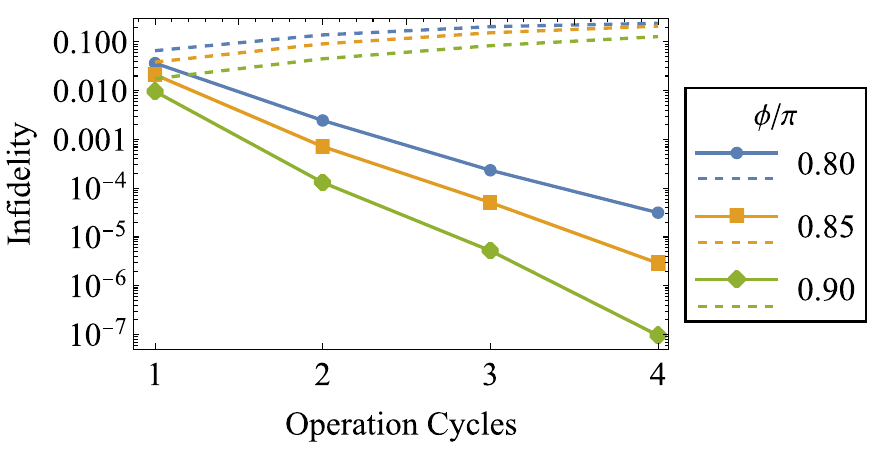}
    \caption{The average channel infidelity of our protocol (solid lines) relative to the perfect parity projection operation. We also plot the average infidelity of the naive parity projection from Eqs.~\eqref{eq:naive_1} and~\eqref{eq:naive_2} as a comparison (dashed lines). The average is taken over $5000$ uniformly randomly sampled two-qubit states. The error bars show the standard deviation of the corresponding samples.}
    \label{fig:perfect_avg_channel_fidelity}
\end{figure}

In Fig.~\ref{fig:perfect_avg_channel_fidelity}, we calculate the average channel fidelity by randomly sampling $5000$ two-qubit initial states, and averaging the output state fidelity relative to the ones obtained from perfect parity projections. Further details about the procedure can be found in Appendix~\ref{appsec:fidelity}. We consider three different CPhase gates: $\phi = 0.8\pi$, $0.85 \pi$, and $0.9\pi$ as examples. As the number of operation cycles $n$ increases, the infidelity is suppressed exponentially. When the phase angle $\phi$ is close to $\pi$, the infidelity is suppressed further for a fixed number of operation cycles, as one would expect based on Eq.~\eqref{eq:odd_n}. We also stress that though our protocol is a perfect parity projection only when $n \rightarrow \infty$, over a range of CPhase gate angles (e.g., $\phi \geq 0.8 \pi$) and with just two or three operation cycles, the channel infidelity can be suppressed below $0.001$, which performs much better compared to directly using the CPhase gates in the parity projection in Fig.~\ref{fig:parity_proj_circ}a (see Fig.~\ref{fig:perfect_avg_channel_fidelity} dashed lines). Note that when the CPhase gates in Fig.~\ref{fig:parity_proj_circ}a are not CZ gates, the naive protocol is not a perfect parity projection anymore. Unlike our protocol, where the infidelity is suppressed with more operation cycles, nesting the naive protocol with more operation cycles increases the channel infidelity.

We can also characterize the quality of our protocol as a two-qubit parity measurement using the `error probability':
\begin{equation}
    \mP_{\text{err}} = \mP_{\text{even}} \mP_{\text{even},\text{err}} + \mP_{\text{odd}} \mP_{\text{odd},\text{err}},
    \label{eq:error_prob}
\end{equation}
where $\mP_{\text{even}}$ ($\mP_{\text{odd}}$) is the probability for our protocol to return an even (odd) parity result, while $\mP_{\text{even},\text{err}}$ ($\mP_{\text{odd},\text{err}}$) is the probability of obtaining an odd (even) parity result after performing $\hat{E}_{\text{even}}$ ($\hat{E}_{\text{odd}}$). Therefore, the error probability is the probability of measuring the opposite parity after a given projection.

We can compute $\mP_{\text{err}}$ by starting from a general pure two-qubit state:
\begin{equation}
    \ket{\psi} = c_{00} \ket{00} + c_{01} \ket{01} + c_{10} \ket{10} + c_{11} \ket{11},
    \label{eq:random_state}
\end{equation}
where $c_{j,k}$ are random complex coefficients that satisfy $\sum_{j,k} \vert c_{j,k} \vert^2  = 1$.
In the first round, the probability of getting an even- or odd-parity projection is
\begin{align}
    \mP_{\text{even}}^{(1)} & = \vert \vert \hat{E}_{\text{even}}^{(1)} \ket{\psi} \vert \vert^2 = \sin^{2}(\phi/2) \left(\vert c_{00} \vert^2 + \vert c_{11} \vert^2 \right), \\ 
    \mP_{\text{odd}}^{(1)} & = \vert \vert \hat{E}_{\text{odd}}^{(1)} \ket{\psi} \vert \vert^2 = \cos^{2}(\phi/2) \left(\vert c_{00} \vert^2 + \vert c_{11} \vert^2 \right) \nonumber \\
    & \qquad \qquad \qquad \quad +\left(\vert c_{01} \vert^2 + \vert c_{10} \vert^2 \right).
\end{align}
The corresponding states are
\begin{align}
    \ket{\psi_{\text{e}}} & = \frac{1}{\sqrt{\mP_{\text{even}}^{(1)}}}\sin \left(\phi/2 \right) \left( c_{00} \ket{00} + c_{11} \ket{11} \right), \label{eq:first_even_perfect}\\
    \ket{\psi_{\text{o}}} & = \frac{1}{\sqrt{\mP_{\text{odd}}^{(1)}}} \left[ \cos \left(\phi/2 \right) \left( e^{-i\phi/2} c_{00} \ket{00} 
+ e^{i\phi/2} c_{11} \ket{11} \right) \right. \nonumber \\
    & \qquad \qquad \qquad \left. + \left(c_{01} \ket{01} + c_{10} \ket{10}\right) \right], \label{eq:first_odd_perfect}
\end{align}
up to global phases. Therefore, if we only apply our protocol with a single cycle, the probability of getting an even- or odd-parity projection result is $\mP_{\text{even}} = \mP_{\text{even}}^{(1)}$ and $\mP_{\text{odd}} = \mP_{\text{odd}}^{(1)}$. As the even-parity projection perfectly projects the state onto the even subspace, there is no chance to get an odd-parity projection out of the state ($\mP_{\text{even,err}} = 0$). However, there is a chance that we can project onto the even-parity subspace if we start from $\ket{\psi_{o}}$. The probability is
\begin{equation}
    \mP_{\text{odd},\text{err}} = \frac{1}{\mP_{\text{odd}}^{(1)}} \cos^2 (\phi/2) \left( \vert c_{00} \vert^2 + \vert c_{11} \vert^2 \right).
\end{equation}
Therefore the error probability is
\begin{equation}
    \mP^{(1)}_{\text{err}} = \mP_{\text{odd}} \mP_{\text{odd},\text{err}} = \cos^2 (\phi/2) \left( \vert c_{00} \vert^2 + \vert c_{11} \vert^2 \right).
\end{equation}
Generalizing this to $n$ operation cycles, the error probability is 
\begin{equation}
    \mP^{(n)}_{\text{err}} = \cos^{(2n)}(\phi/2) \left( \vert c_{00} \vert^2 + \vert c_{11} \vert^2 \right).
    \label{eq:error_p_perfect}
\end{equation}
We see that the error probability depends on the input state $\ket{\psi}$. The maximum error probability is achieved when $\ket{\psi}$ has even parity, in which case it is
\begin{equation}
    \mP^{(n)}_{\text{err,max}} = \cos^{2n}(\phi/2).
    \label{eq:max_error_perfect}
\end{equation}

\begin{figure}[h]
    \centering
    \includegraphics[width =  \columnwidth]{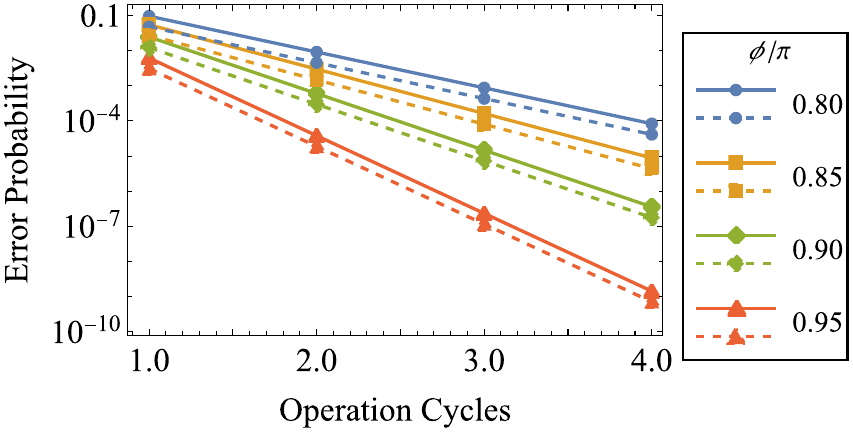}
    \caption{The maximum (solid lines) and average (dashed lines) error probability of our protocol as a function of the number of operation cycles $n$ for four different values of the CPhase angle $\phi$. Results for the average error probability are obtained by averaging Eq.~\eqref{eq:error_p_perfect} over all possible initial pure states.}
    \label{fig:perfect_error_p}
\end{figure}

In Fig.~\ref{fig:perfect_error_p}, we plot the maximum error probability of our protocol as a function of the number of operation cycles $n$ for different CPhase gates (solid lines). Similar to the average channel fidelity, the error probability is also suppressed exponentially with increasing $n$. Across a large range of CPhase gate angles, and with only a few cycles of operation, the error probability can be suppressed by orders of magnitude. For example, for the CPhase gate $\CP{0.9\pi}$ (green line and diamonds), which is close to the gate shown in Ref.~\cite{Wells2019}, the error probability can be reduced to $6.0\times 10^{-4}$ with only two cycles. To further understand the average performance of our protocol, we also calculate the error probability averaged over $4000$ Haar random 2-qubit states, which are shown as the dashed lines in Fig.~\ref{fig:perfect_error_p}. The average error probability using $\CP{0.9\pi}$ gates and $n=2$ cycles is $3.0\times 10^{-4}$. 

\section{Performance under imperfections} \label{sec:errors}

In this section, we consider how robust our parity projection protocol is against possible imperfections. To make our discussion generally applicable to different physical platforms, we survey various sorts of imperfections that can arise and investigate their effects on the performance of the protocol. We consider both coherent and incoherent errors. Specifically, in Sec.~\ref{sec:errors:coh_error}, we consider the effect of systematic errors in the CPhase gates. In Sec.~\ref{sec:errors:incoh_cphase}, we extend the discussion and investigate the impact of random fluctuations in CPhase gate angles. In Sec.~\ref{sec:errors:pauli}, we calculate the effect of incoherent Pauli errors happening in different stages of our protocol. We consider dephasing errors as well as Pauli-$X$ and Pauli-$Y$ errors on the matter qubit between the two CPhase gates. In all cases, we calculate the error probability of our protocol in the presence of such errors.

\begin{figure*}[htbp]
    \centering
    \subfloat[]{\includegraphics[width = 0.23 \textwidth]{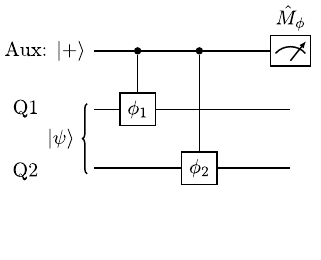}} \quad
    \subfloat[]{\includegraphics[width = 0.35 \textwidth]{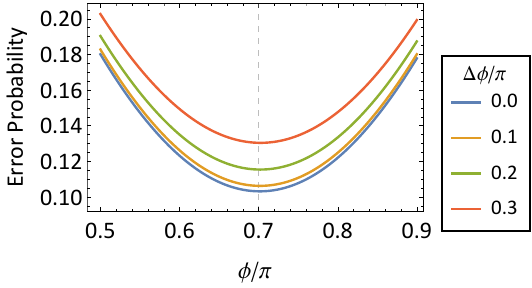}} \ 
    \subfloat[]{\includegraphics[width = 0.35 \textwidth]{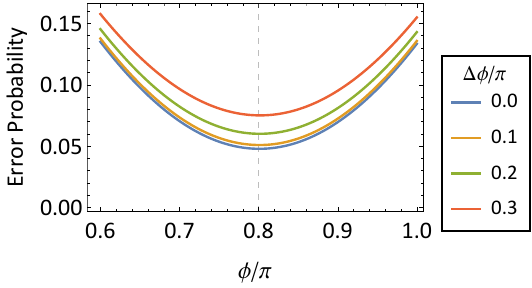}}
    \caption{The parity projection protocol with imbalanced CPhase gates. (a) The gate sequence for a single operation cycle of our protocol when the CPhase gates are imbalanced. (b, c) The average error probability as we change the measurement basis angle $\phi$. Results are obtained by averaging Eq.~\eqref{eq:coh_error_error} over all possible initial pure states. We consider $n=1$ case in both (b) and (c). The two CPhase gate angles are fixed at $\phi_{1,2}=0.7\pi \pm \Delta \phi/2$ for (b) and $\phi_{1,2}=0.8 \pi \pm \Delta \phi/2$ for (c).}
    \label{fig:coh_error}
\end{figure*}

\subsection{Imbalanced CPhase gates} \label{sec:errors:coh_error}

As demonstrated in the previous proposals, the photonic parity projection can be implemented using cavity-QED systems, e.g., a quantum dot qubit strongly coupled to an optical resonator~\cite{Wei2013, Wei2014, Wang2013, Ren2013}. However, the phase angles of the CPhase gates between the matter qubit and the reflected (or transmitted) photonic qubits are not well controlled in experiments~\cite{Loo2012,kim2013quantum,Sun2016,Sun2018,Wells2019,Androvitsaneas2019}. In this section, we consider that the two CPhase gates in each operation cycle are imbalanced, i.e., the two CPhase gates have different phase angles $\phi_1$ and $\phi_2$ (see Fig.~\ref{fig:coh_error}a), and we investigate how our protocol is affected by this.

We take the measurement basis to be $\hat{M}_{\phi}$, where the angle $\phi$ is supposed to be close to $\phi_{1,2}$. We further assume the CPhase gates in different operation cycles are stable, i.e., the same $\phi_1$ and $\phi_2$ are realized every cycle. This assumption will be relaxed in the next section. If we keep the measurement basis $\phi$ the same across cycles, by expressing $\phi_{1,2} = \phi + \delta_{1,2}$, the Kraus operators after $n$ cycles of our protocol are
\begin{widetext}
\begin{align}
    \hat{E}_{\text{even}}^{(m)} & =  i \sin\left(\frac{\phi}{2}\right) \left[ \cos \left(\frac{\phi}{2}\right) \right]^{m-1} \mcP_{00} - i e^{i m \phi + i\frac{m(\delta_1+\delta_2)}{2}} \sin\left(\frac{\phi+\delta_1+\delta_2}{2}\right) \left[\cos \left(\frac{\phi+\delta_1+\delta_2}{2}\right)\right]^{m-1} \mcP_{11}\nonumber \\
    & \quad - i e^{i \frac{m (\phi+\delta_2)}{2}} \sin\left(\frac{\delta_2}{2}\right)\left[ \cos\left(\frac{\delta_2}{2}\right)\right]^{m-1} \mcP_{01} -  i e^{i \frac{m (\phi+\delta_1)}{2}} \sin\left(\frac{\delta_1}{2}\right)\left[\cos\left(\frac{\delta_1}{2}\right)\right]^{m-1} \mcP_{10}, \label{eq:coh_kraus_even}\\
    \hat{E}_{\text{odd}} & = \left[ \cos \left(\frac{\phi}{2}\right) \right]^n \mcP_{00} + e^{i \frac{n (\phi+\delta_2)}{2}} \left[ \cos \left(\frac{\delta_2}{2}\right) \right]^n \mcP_{01} + e^{i \frac{n (\phi+\delta_1)}{2}} \left[ \cos \left(\frac{\delta_1}{2}\right) \right]^n \mcP_{10} + e^{i n \phi + i \frac{n(\delta_1+\delta_2)}{2}}\nonumber \\
    & \quad \times \left[ \cos \left(\frac{\phi+\delta_1+\delta_2}{2}\right) \right]^n \mcP_{11},
    \label{eq:coh_kraus_odd}
\end{align}
\end{widetext}
where in Eq.~\eqref{eq:coh_kraus_even}, $m = 1, ..., n$. The Kraus operator $\hat{E}_{\text{even}}^{(m)}$ corresponds to the case in which we obtain $+1$ up to the $(m-1)$-th cycle, but get $-1$ in the $m$-th cycle. These operators describe imperfect even-parity projections. The Kraus operator $\hat{E}_{\text{odd}}$ in Eq.~\eqref{eq:coh_kraus_odd} is for the case where we get all $+1$ measurement outcomes in the $n$ cycles of operation. It corresponds to the odd-parity projection. Similar to our perfect protocol, the relative phase between the $\mcP_{00}$ and $\mcP_{11}$ terms in the even-parity projection operators, and between the $\mcP_{10}$ and $\mcP_{01}$ terms in the odd-parity projection operation, can be fixed by single-qubit Pauli-$Z$ rotations. However, unlike our perfect protocol, the $-1$ measurement outcome no longer heralds a pure even-parity projection, which introduces more errors.

To determine the optimal measurement basis, we calculate the error probability [Eq.~\eqref{eq:error_prob}] with the CPhase coherent error included. Assuming the initial state of the two photonic qubits is an arbitrary pure state, as Eq.~\eqref{eq:random_state} shows, the error probability after $n$ cycles is
\begin{align}
    & \mP^{(n)}_{\text{err}} =  \cos^{2n}(\phi/2) \vert c_{00} \vert^2 + \cos^{2n} [(\phi+\delta_1 + \delta_2)/2] \vert c_{11} \vert^2 \nonumber \\
    & \; + [1 - \cos^{2n}(\delta_2/2))] \vert  c_{01} \vert^2 + [1 - \cos^{2n}(\delta_1/2)] \vert c_{10} \vert^2,  
    \label{eq:coh_error_error}
\end{align}
the detailed derivation of which can be found in Appendix~\ref{appsec:coh_err}. In Eq.~\eqref{eq:coh_error_error}, the first two terms are from the odd-parity projection [Eq.~\eqref{eq:coh_kraus_odd}], while the last two are from the imperfect even-parity projections [Eq.~\eqref{eq:coh_kraus_even}]. 

Supposing the phase differences are small, i.e., $\delta_1, \delta_2 \ll 1$, the coefficients of the terms from the even-parity projections can be expanded around $0$ as
\begin{align}
    1 - \cos^{2n}(\delta/2) \sim n \delta^2/4 + O(\delta^4),
    \label{eq:coh_err_expand}
\end{align}
which shows that the error from the even-parity projection grows as the number of rounds increases. Note that the optimal measurement basis is specified by the angle $\phi = (\phi_1 + \phi_2)/2$, which corresponds to $\delta_1=-\delta_2$. When this basis is used, the error from the even-parity projections is distributed equally on the odd-parity basis states $\ket{01}$ and $\ket{10}$. In addition, the error from the odd-parity projection is also minimized on average. In Fig.~\ref{fig:coh_error}b and~c, we show the average error probability for a few choices of CPhase and measurement basis angles as examples. It is clear that the minimal single-round error probability is reached for the measurement basis with $\phi = (\phi_1+\phi_2)/2$.

\begin{figure}[h]
    \centering
    \includegraphics[width = \columnwidth]{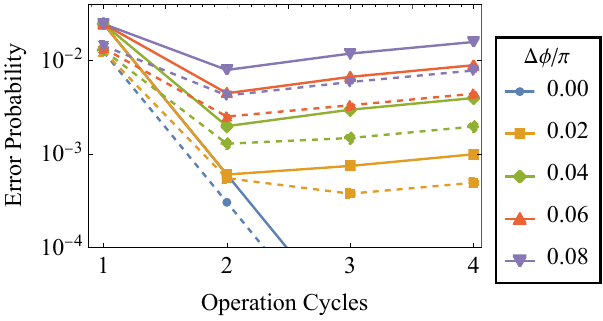}
    \caption{The maximum (solid lines) and average (dashed lines) error probability of our protocol with coherent error on the CPhase gates. We take the measurement basis to be $\phi = (\phi_1 + \phi_2)/2$ and $\Delta\phi/2 = \phi_1 - \phi = \phi - \phi_2$, with $\phi = 0.9 \pi$. Results for the average error probability are obtained by averaging Eq.~\eqref{eq:coh_error_error} over all possible initial pure states.}
    \label{fig:coh_err_err_p}
\end{figure}

In the rest of this subsection, we assume that we always choose the optimal measurement basis $\phi = (\phi_1 + \phi_2)/2$, for which $\delta_1 = - \delta_2 = \Delta \phi/2$, where $\Delta \phi = \vert \phi_1 - \phi_2 \vert$. In addition, we assume that we have applied Pauli-$Z$ rotations after the measurements to compensate for the relative phase factors between $\mcP_{01}$ and $\mcP_{10}$ in the odd-parity projection as well as between $\mcP_{00}$ and $\mcP_{11}$ in the even-parity projections. To investigate how robust our protocol is against the coherent error on the CPhase gates, we calculate the average error probability and quantum channel fidelity [Eq.~\eqref{eq:fidelity}].

Figure~\ref{fig:coh_err_err_p} shows the maximum and average error probability. We take the two CPhase gates to have a mean angle $\phi = 0.9 \pi$~\footnote{The reason we choose this value is that it is close to the CPhase gates demonstrated in Ref.~\cite{Wells2019}.} and sweep the difference $\Delta\phi$ of the two gate angles from $0$ (ideal protocol) to $0.08 \pi$. We observe that the error probability decreases for small $n$, while there is a turning point beyond which the error probability starts to increase for larger $n$. As shown in Eq.~\eqref{eq:coh_err_expand}, this is because the error from the even-parity projection increases as the number of operation cycles increases. After the turning point, the error from the even-parity projection dominates over that of the odd-parity projection. However, even with a large CPhase gate mismatch of $\Delta\phi = \phi_1 - \phi_2 = 0.08 \pi$, with only two operation cycles, the error probability can be suppressed to $1.2\%$. 

\begin{figure*}[htbp]
\subfloat[]{
\includegraphics[width = 0.5 \textwidth]{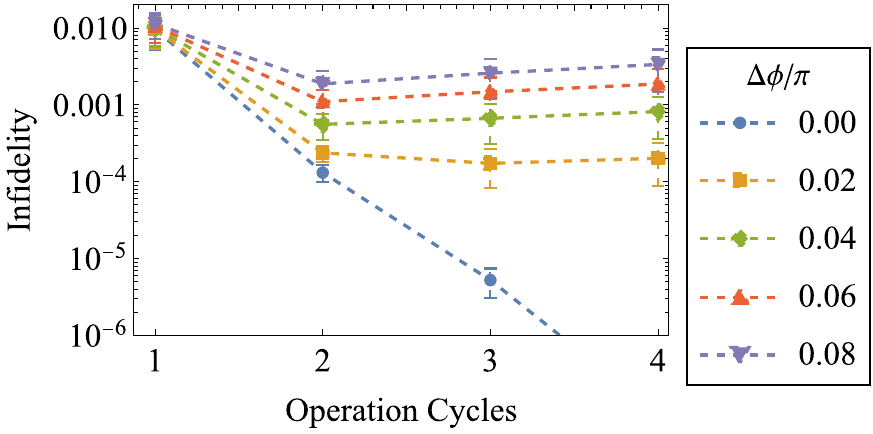}} \quad
\subfloat[]{
\includegraphics[width = 0.4 \textwidth]{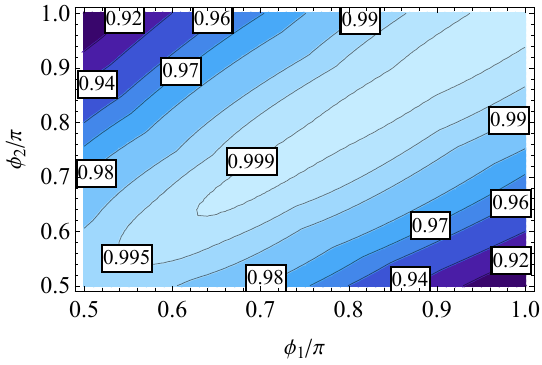}}
\caption{The average channel fidelity of our protocol relative to the perfect parity projection. (a) The average channel fidelity as a function of the number of operation cycles $n$ for several different values of the CPhase angle difference $\Delta \phi =\vert \phi_1-\phi_2 \vert$, with the mean angle set to $\phi=(\phi_1+\phi_2)/2=0.9\pi$. (b) The average channel fidelity after 5 cycles of our protocol as a function of the two CPhase gate angles $\phi_1$ and $\phi_2$.}
\label{fig:coh_error_fid}
\end{figure*}

Next we calculate the average channel fidelity of our protocol relative to a perfect parity projection as a second figure of merit. The average channel fidelity is calculated according to Eq.~\eqref{eq:fidelity}. In Fig.~\ref{fig:coh_error_fid}a, we show the average channel fidelity of our protocol with unbalanced CPhase gates. We again consider the two CPhase gates to have the mean angle $\phi = 0.9 \pi$ and sweep the difference $\Delta \phi$ of the two angles from $0$ (ideal protocol) to $0.08 \pi$. Similar to the error probability above, the average channel infidelity initially decreases and then starts to increase as $n$ becomes larger. With only two or three cycles of our protocol, the infidelity can still be suppressed. Because the average fidelity can increase as $n$ increases with a pair of given CPhase gates, we consider the best performance that can be reached using up to 5 cycles of our protocol. Specifically, we calculate the average channel fidelity with two CPhase gates with angles ($\phi_1$ and $\phi_2$) as we increase $n$ to 5. We then extract the optimal fidelity we can achieve as the figure of merit. In Fig.~\ref{fig:coh_error_fid}b, we sweep the two CPhase angles ($\phi_1$ and $\phi_2$) and plot the optimal average channel fidelity up to 5 cycles of our protocol. There is a wide range of CPhase angles that we can choose from to achieve relatively high fidelity, which shows that our protocol can tolerate a significant amount of coherent error in the CPhase gates.

\subsection{Gaussian random phase errors on the CPhase gates} \label{sec:errors:incoh_cphase}

\begin{figure*}[htbp]
    \centering
    \subfloat[]{\includegraphics[width= \columnwidth]{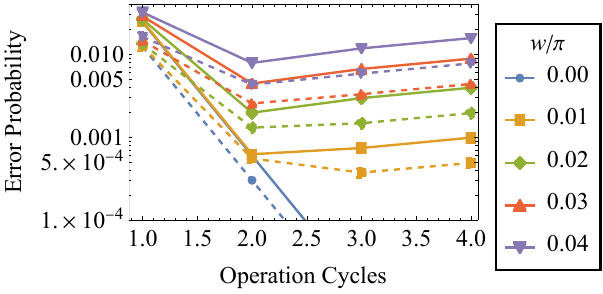}} \quad 
    \subfloat[]{\includegraphics[width = \columnwidth]{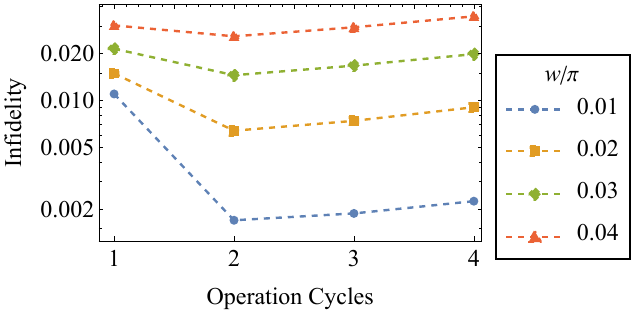}}
    \caption{The performance of our parity-projection protocol with Gaussian random noise on the CPhase gates. (a) The maximum (solid lines) and average (dashed lines) error probability of our protocol as a function of the number of operation cycles $n$. Results for the average error probability are obtained by averaging Eq.~\eqref{eq:incoh_error_cphase_error} over all possible initial pure states. (b) The average channel fidelity of our protocol relative to a perfect parity projection as a function of $n$. In both panels, the CPhase angles are $\phi_{1,2} = \phi + \delta_{1,2}$, where $\phi = 0.9 \pi$, and $\delta_{1,2}$ are sampled from a Gaussian distribution with width $w$. Results for several values of $w$ are shown.}
    \label{fig:gaussian}
\end{figure*}

In this subsection, we consider the case in which the phase angles of the two CPhase gates are not well controlled and exhibit random fluctuations. We extend the discussion in Sec.~\ref{sec:errors:coh_error} and now take the phase errors $\delta_1 = \phi_1 - \phi$ and $\delta_2 = \phi_2 - \phi$ to be two independent Gaussian random variables that are sampled from the probability distribution,
\begin{equation}
    p(\delta) = \frac{1}{\sqrt{2\pi} w} e^{- \frac{\delta^2}{2 w^2}},
    \label{eq:gaussian_distribution}
\end{equation}
where $w$ characterizes the spread in the CPhase angles. Here, we also allow the CPhase angles to vary from one cycle to the next. In this scenario, the error probability after the $n$th cycle is
\begin{align}
    & \langle \mP^{(n)}_{\text{err}} \rangle = \int_{\delta}  \mP^{(n)}_{\text{err}} \prod_{j = 1}^{n}\frac{1}{2\pi w^2} e^{- \frac{(\delta_1^{(j)})^2}{2 w^2}} e^{- \frac{(\delta_2^{(j)})^2}{2 w^2}} d \delta_1^{(j)} d \delta_2^{(j)},
\end{align}    
where the superscript $j$ labels the operation cycle and where now
\begin{align}
\mP^{(n)}_{\text{err}}& =\cos^{2n}(\phi/2)|c_{00}|^2\nonumber\\
& +\prod_{k=1}^n\cos^{2}[(\phi+\delta_1^{(k)}+\delta_2^{(k)})/2]|c_{11}|^2\nonumber\\
& + \left[ 1 - \prod_{k=1}^{n}\cos^{2}(\delta_2^{(k)}/2)\right] |c_{01}|^2\nonumber\\
& + \left[ 1 - \prod_{k=1}^n\cos^{2}(\delta_1^{(k)}/2)\right]|c_{10}|^2, \nonumber
\end{align}
After performing the integration, we find
\begin{align}
    & \langle \mP^{(n)}_{\text{err}} \rangle = \left[ \cos\left(\frac{\phi}{2} \right) \right]^{2n} \vert c_{00} \vert^2 + \left[ \frac{1 + \cos(\phi) e^{-w^2}}{2} \right]^n \vert c_{11}\vert^2 \nonumber \\
    & \quad + \left[ 1 - \frac{1}{2^n} \left(1+e^{-w^2/2} \right)^n \right] \left(\vert c_{01} \vert^2 + \vert c_{10}\vert^2 \right),
    \label{eq:incoh_error_cphase_error}
\end{align}
where the first two terms are from the odd-parity projection, and the last two terms are from the even-parity projection. The coefficients of the first two terms in Eq.~\eqref{eq:incoh_error_cphase_error} are suppressed exponentially as the operation cycle increases. However, when the spread in the CPhase angles is small, $w \ll 1$, the coefficient of the last two terms can be expanded as 
\begin{equation}
    \left[ 1 - \frac{1}{2^n} \left(1+e^{-w^2/2} \right)^n \right] \sim \frac{n w^2}{4} + O(w^4).
\end{equation}
This shows that the error from the even-parity projection grows as $n$ increases. The wider the Gaussian distribution is, the more error is introduced into our protocol. Therefore, there is a turning point where the even-parity projection error will start to be the dominant error. After the turning point, the error probability will grow.

In Fig.~\ref{fig:gaussian}a, we show the maximum error probability as we increase the operation cycles $n$. We observe the same trend as the coherent CPhase error in Fig.~\ref{fig:coh_error_fid}a. Namely, there exists a turning point that depends on the CPhase angle, which also matches our analytical analysis. However, the error suppression of our protocol can tolerate fluctuations in the CPhase gates. With a CPhase angle spread of $w = 0.04 \pi$, the second cycle of our protocol can suppress the error probability from $3.2$\% to $0.8$\%. 

We next consider the channel fidelity. In Fig.~\ref{fig:gaussian}b, we plot the average channel fidelity of the parity projection for various numbers of cycles. The detailed derivation and calculation are given in Appendix~\ref{appsec:gaussian}. The average channel fidelity also shows a turning point as $n$ increases. Even with a Gaussian fluctuation with width $w = 0.04 \pi$, our protocol using CPhase gates with angle $\phi = 0.9 \pi$ can reach $0.96$. We also notice that with the Gaussian noise fluctuation on the CPhase gates, as we do not know exactly the angles of two CPhase gates, the performance of our protocol is worse compared to the coherent error case analyzed above, in which the information of CPhase gates can be used to optimize the protocol performance. We also notice that the suppression of the error probability is greater compared to the average channel fidelity. This means the average output of our protocol with Gaussian noise on the CPhase gates is slightly different from the output state of the perfect parity projection. However, the probability of correctly identifying the state parity using our protocol is less affected.

\subsection{Effects of Pauli errors} \label{sec:errors:pauli}

In this subsection, we consider possible Pauli errors. As our protocol is applicable to a wide variety of physical platforms, we keep the analysis general and survey different Pauli errors and their respective impact on the performance. Understanding which Pauli errors are most detrimental can be helpful in selecting or designing physical platforms in which these errors are less likely to happen. We specifically focus on errors on the matter qubits. In Sec.~\ref{sec:error:pauli:before}, we consider dephasing errors that occur before the two CPhase gates are applied. We also comment on the effect of Pauli-$X$ and Pauli-$Y$ errors occurring at this point in the protocol. In Sec.~\ref{sec:error:pauli:after}, we focus on Pauli errors between the two CPhase gates. As shown in previous sections, both the average channel fidelity and the error probability are useful metrics for the performance of our protocol. The error probability has the additional benefit of being analytically computable. Therefore, in the remainder of this section, we focus on using the maximum and average error probability as the figure of merit of our protocol to keep the discussion concise.

\subsubsection{Dephasing errors on the matter qubit before the two CPhase gates} \label{sec:error:pauli:before}

\begin{figure}[h]
    \centering
    \includegraphics[width = 0.3 \textwidth]{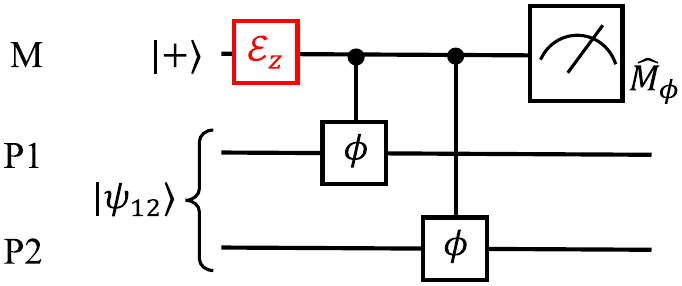}
    \caption{Circuit diagram for a single round of our parity-projection protocol with a possible dephasing error (labeled with a red $\mathcal{E}_z$) on the matter qubit before the two CPhase gates are applied.}
    \label{fig:dephasing_error_before}
\end{figure}

In this subsection, we consider the case in which there is a dephasing error on the matter qubit before the two CPhase gates are applied. The dephasing error is described by the quantum channel
\begin{equation}
    \mathcal{E}_{z} (\rho_\text{m}) = (1 - p_z) \rho_\text{m} + p_z \hat{Z} \rho_\text{m} \hat{Z},
\end{equation}
where $p_z$ is the error probability. A single operation cycle of our protocol is shown in Fig.~\ref{fig:dephasing_error_before}, where the red $\mathcal{E}_z$ represents the dephasing error.

We start by understanding the effect of the dephasing error on a single cycle of operation. We stress that the matter qubit is initialized in the state $\ket{+}$. When there is a Pauli-$Z$ error on the matter qubit before we apply the two CPhase gates, the error flips the matter qubit state to $\ket{-}$. Supposing the initial state of the two photonic qubits is in the state $\rho_0$, the single cycle of operation transforms this state into
\begin{align}
    \rho = \sum_{j = \pm 1} (1-p_z) \hat{E}_{j} \rho_0 \hat{E}_{j}^{\dagger} + p_z \hat{E}_{j, \text{err}} \rho_0 \hat{E}_{j, \text{err}}^{\dagger},
\end{align}
where $j = \pm1$ is the measurement outcome, and $\hat E_{\pm 1, \text{err}}$ are the Kraus operators corresponding to the case when the error occurs. These satisfy
\begin{align}
    \hat{E}_{\pm1, \text{err}} = \hat{E}_{\mp 1},
\end{align}
where $\hat{E}_{\mp 1}$ are the Kraus operators without the Pauli-$Z$ error (see Appendix~\ref{appsec:pauli_z_before} for a detailed derivation).

Assuming we apply our protocol up to the $n$-th cycle and we get all $+1$ measurement results, the un-normalized state of the two photonic qubits is
\begin{align}
    \tilde{\rho}_{n:+1} & =  \sum_{j=0}^{n} {n \choose j} (1-p_z)^{n-j} p_z^j \nonumber \\
    & \quad \times \hat{E}_{+1}^{n-j} \hat{E}_{-1}^{j} \rho_0 (\hat{E}_{-1}^{\dagger})^j (\hat{E}_{+1}^{\dagger})^{n-j},
    \label{eq:n_round_pz_odd_init}
\end{align}
where we use the fact that $[\hat{E}_{+1}, \hat{E}_{-1}] = 0$. This state is considered as an odd-parity projection onto the initial state $\rho_0$ using our protocol. On the other hand, if we get $+1$ up to round $m-1$, and get $-1$ in round $m$, we get the even-parity projection on the initial state, which is
\begin{align}
    \tilde{\rho}_{m,-1} = \mathcal{E}_{-1} (\tilde{\rho}_{m-1:+1}),
\end{align}
where $\tilde{\rho}_{m-1:+1}$ can be derived using Eq.~\eqref{eq:n_round_pz_odd_init} by replacing $n$ with $m-1$, $\mathcal{E}_{-1}$ is the quantum channel for one cycle of our protocol with the $-1$ measurement outcome (see Appendix~\ref{appsec:pauli_z_before}).

\begin{figure}[h]
    \centering
    \subfloat[]{\includegraphics[width = \columnwidth]{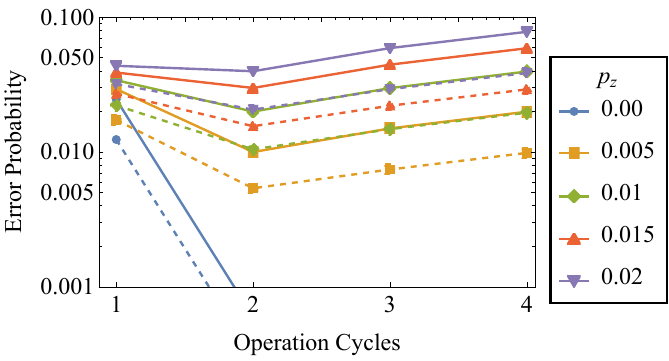}}
    \caption{Maximum (solid lines) and average (dashed lines) error probability as a function of the number of operation cycles $n$ when there are dephasing errors on the matter qubit before the CPhase gates. Results are shown for five different values of the dephasing error probability $p_z$. Results for the average error probability are obtained from Eq.~\eqref{eq:pauli_z_error_prob} by averaging over all initial pure states. The CPhase gate angles are $\phi_1=\phi_2=\phi=0.9 \pi$.}
    \label{fig:pauli_z_before}
\end{figure}

If we assume the initial state of the two photonic qubits is given in Eq.~\eqref{eq:random_state}, the error probability is
\begin{align}
    \mP^{(n)}_{\text{err}} & =  \left[ \frac{1}{2} + \left(\frac{1}{2} - p_z \right) \cos(\phi) \right]^n (\vert c_{00} \vert^2 + \vert c_{11} \vert^2) \nonumber \\
    & \quad + (\vert c_{01} \vert^2 + \vert c_{10} \vert^2) \left[ 1 - (1-p_z)^n \right],
    \label{eq:pauli_z_error_prob}
\end{align}
where the maximum error probability is
\begin{align}
    \mP^{(n)}_{\text{err,max}} = \text{max} &\bigg\lbrace \left[ \frac{1}{2} + \left(\frac{1}{2} - p_z \right) \cos(\phi) \right]^n,\nonumber\\& 1 - (1-p_z)^n  \bigg\rbrace.
    \label{eq:z_before_max_prob}
\end{align}

If we assume the dephasing error is small ($p_z \ll 1$), the two terms in Eq.~\eqref{eq:z_before_max_prob} can be expanded around $p_z = 0$ as
\begin{align}
    & \left[ \frac{1}{2} + \left(\frac{1}{2} - p_z \right) \cos(\phi) \right]^n  \sim \cos^{2n}(\phi/2) \nonumber \\
    & \qquad \qquad \qquad - n \cos^{2n-2}(\phi/2) \cos(\phi) p_z + O\left(p_z^3\right) \label{eq:dephasing_term_1}\\
    & 1 - (1-p_z)^n \sim n p_z + \frac{n}{2}(n-1) p_z^2 + O\left(p_z^3\right). \label{eq:dephasing_term_2}
\end{align}
Both terms will grow as the dephasing error probability $p_z$ increases. From the leading orders of the two terms, the first term is suppressed as the number of operation cycles $n$ increases, while the second increases. Similar to the CPhase error discussion above, with the matter qubit dephasing error, there will be a turning point in the protocol error probability. This is a result of the competition between the suppressed imperfect odd-parity projection [the term in Eq.~\eqref{eq:dephasing_term_1}] and the effect of the dephasing error [the term in Eq.~\eqref{eq:dephasing_term_2}]. 

In Fig.~\ref{fig:pauli_z_before}, we show the maximum and average error probability as we increase the number of operation cycles. As our analytical analysis shows, when the dephasing error is nonzero ($p_z \neq 0$), there are turning points in the error probability. Note that the dephasing error causes the initial state of the matter qubit to flip to $\ket{-}$, which directly causes our protocol to project the two-qubit state into the opposite parity subspace. So the dephasing error on the matter qubit is problematic and should be avoided or suppressed. With $p_z \sim 0.02$, the average error probability can be suppressed from $3.2\%$  to $2.0\%$, but the maximum error probability shows almost no suppression. In the presence of the dephasing error, it is therefore important to combine our protocol with dynamical decoupling on the matter qubit to suppress dephasing~\cite{de2010universal, naydenov2011dynamical}. Dynamical decoupling has been shown experimentally to be quite effective across a variety of quantum emitter systems~\cite{wang2012comparison, shim2012robust, Sukachev2017,wood2022long, varwig2016advanced, kosarev2022extending, zaporski2023ideal}.

At the end of this subsection, we would like to comment on the effects of other types of Pauli errors and of depolarizing error on the matter qubit. We note that the matter qubit is initialized in state $\ket{+}$. The Pauli-$X$ error does not affect the matter qubit state. For Pauli-$Y$ errors, we notice that  
\begin{equation}
    \hat{Y} \dyad{+} \hat{Y} = \dyad{-} = \hat{Z} \dyad{+} \hat{Z},
\end{equation}
which means the Pauli-$Y$ error will affect our protocol in exactly the same way as the dephasing error. Therefore, all our discussions in this subsection about the dephasing error can apply to Pauli-$Y$ errors. The depolarizing errors can be described by the following quantum channel
\begin{equation}
    \mathcal{E}(\rho) = (1-p) \rho + \frac{p}{3} \left(\hat{X} \rho \hat{X} + \hat{Y} \rho \hat{Y} + \hat{Z} \rho \hat{Z} \right),
\end{equation}
where $p$ is the error probability. This error is equivalent to a dephasing error with error probability $p_z = 2p/3$, as can be seen from 
\begin{align}
    \mathcal{E}(\dyad{+}) & = (1-p) \dyad{+} + \frac{p}{3} \left(\dyad{+} + 2 \dyad{-} \right) \nonumber \\
    & = \left( 1- \frac{2p}{3}\right) \dyad{+} + \frac{2p}{3} \hat{Z} \dyad{+} \hat{Z}.
\end{align}

\subsubsection{Incoherent Pauli errors on the matter qubit between the two CPhase gates} \label{sec:error:pauli:after}

\begin{figure}[h]
    \centering
    \includegraphics[width = 0.3 \textwidth]{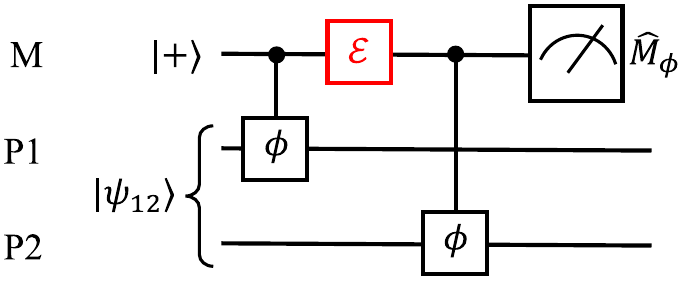}
    \caption{Circuit diagram for a single round of our parity-projection protocol with possible Pauli errors (labeled with a red $\mathcal{E}$) on the matter qubit between the two CPhase gates.}
    \label{fig:pauli_between_circuit}
\end{figure}

In this subsection, we consider the case in which the Pauli errors occur between the two CPhase gates. A single cycle of operation is shown in Fig.~\ref{fig:pauli_between_circuit}, where the red $\mathcal{E}$ represents the Pauli error. Note that since $\hat Z$ commutes with the CPhase gates, the effect of the dephasing error occurring between two CPhase gates is identical to the discussion in Sec.~\ref{sec:error:pauli:before}. Therefore, we only consider Pauli-$X$ and Pauli-$Y$ errors in this subsection.

\paragraph{Pauli-X errors.}

We start by considering Pauli-$X$ errors, which are described by the quantum channel
\begin{equation}
    \mathcal{E}_{x} (\rho) = (1-p_x) \rho + p_x \hat{X} \rho \hat{X},
\end{equation}
where $p_x$ is the error probability. We follow the treatment of Sec.~\ref{sec:error:pauli:before}. The Kraus operators for a single cycle of our protocol with Pauli-$X$ errors included are (see Appendix~\ref{appsec:pauli_x_mid})
\begin{align}
    \hat{E}^{(\text{x-err})}_{+1} & = \cos\left(\frac{\phi}{2}\right) \left(\mcP_{00} + e^{i \phi} \mcP_{11} \right)  \nonumber \\
    & \quad + e^{i \phi/2} \left( \mcP_{01}+ \cos(\phi) \mcP_{10}  \right), \label{eq:x_err_even_error} \\
    \hat{E}^{(\text{x-err})}_{-1} & = i \sin\left(\frac{\phi}{2}\right) \left(\mcP_{00}  + e^{i \phi} P_{11} \right) \nonumber \\
    & \quad + i \sin\left(\phi\right) e^{i\phi/2} \mcP_{10}. \label{eq:x_err_odd_error}
\end{align}

Note that the operators $\hat{E}_{\pm 1}$ and $\hat{E}^{(\text{x-err})}_{\pm 1}$ all commute with each other. The state of the two photonic qubits after $n$ rounds of the protocol with all $+1$ measurement outcomes is
\begin{align}
    \tilde{\rho}_{n:+1} & = \sum_{j=0}^{n} {n \choose j} (1-p_x)^{n-j} p_x^j  \nonumber \\
    & \times \hat{E}_{+1}^{n-j} \left(\hat{E}_{+1}^{(\text{x-err})}\right)^j \rho_0 \left(\hat{E}_{+1}^{(\text{x-err}) \dagger}\right)^j \left(\hat{E}_{+1}^\dagger\right)^{n-j},
    \label{eq:incoh_state_odd}
\end{align}
where $\rho_0$ is the initial state of the two qubits. On the other hand, if $-1$ is obtained in the final $m$th round, the resulting state is $\tilde{\rho}_{m,-1} = \mathcal{E}_{-1}(\tilde{\rho}_{m-1:+1})$, where $\tilde{\rho}_{m-1:+1}$ is the state after we get $+1$ for all previous $m-1$ cycles. 

We further notice that $\mathcal{P}_{\text{odd}} \hat{E}_{-1} = 0$, where $\mcP_{\text{odd}}$ is the perfect odd-parity projection, which greatly simplifies our derivation of the error probability in this case. Assuming an initial pure state for the two photons as in Eq.~\eqref{eq:random_state}, the result is (see Appendix~\ref{appsec:pauli_x_mid})
\begin{align}
    \mP^{(n)}_{\text{err}} & = \cos^{2n}\left(\frac{\phi}{2}\right) \left( \vert c_{00} \vert^2 + \vert c_{11} \vert^2 \right) \nonumber \\
    & \qquad \qquad + \left( 1 - \left[1-p_x \sin^2 (\phi) \right]^n \right) \vert c_{10} \vert^2.
    \label{eq:pauli_x_error}
\end{align}
The maximum error probability is then
\begin{align}
    \mP^{(n)}_{\text{err,max}} = \text{max} \left\lbrace \cos^{2n}\left(\frac{\phi}{2}\right), \left( 1 - \left[1-p_x \sin^2 (\phi) \right]^n \right)  \right\rbrace.\label{eq:pauli_x_error_max}
\end{align}

\begin{figure}
    \centering
    \subfloat[]{\includegraphics[width = \columnwidth]{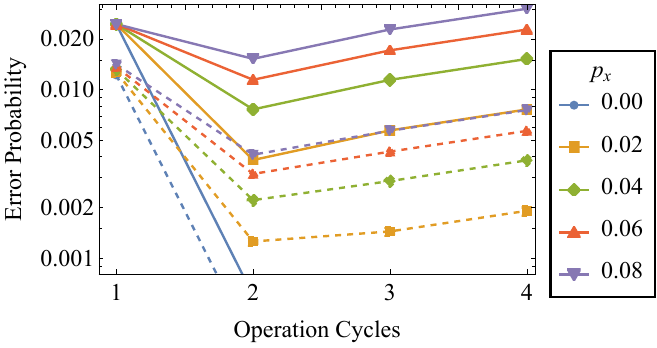}} \\
    \subfloat[]{\includegraphics[width = \columnwidth]{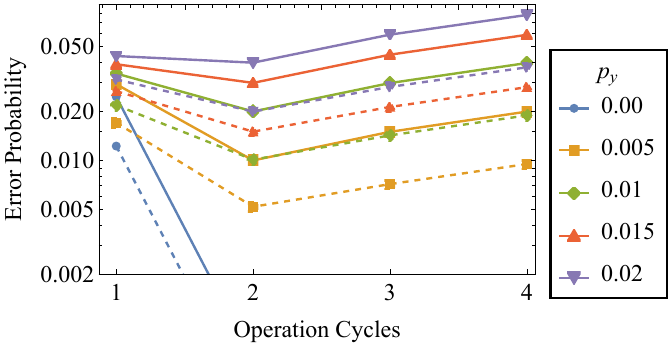}}
    \caption{The maximum (solid lines) and average (dashed lines) error probabilities of our protocol in the presence of (a) Pauli-$X$ and (b) Pauli-$Y$ errors between the CPhase gates for several values of the error probabilities $p_x$ and $p_y$. Results for the average error probabilities are obtained from Eqs.~\eqref{eq:pauli_x_error} and \eqref{eq:y_err_err_prob} by averaging over all initial pure states. Here, the CPhase angles are $\phi_1=\phi_2=\phi=0.9 \pi$.}
    \label{fig:pauli_between}
\end{figure}

In Fig.~\ref{fig:pauli_between}a, we plot the maximum and average error probability of our protocol in the presence of Pauli-$X$ errors for several values of the error probability $p_x$. We notice that when $p_x$ is nonzero, similar to the other types of errors, the error probability exhibits a turning point at $n=2$ cycles of our protocol. The increase in the error probability beyond the turning point can be understood from Eq.~\eqref{eq:pauli_x_error}. As $n$ increases beyond 2, the contribution to $\mP^{(n)}_{\text{err}}$ from the Pauli error increases and dominates over the error from the imperfect even-parity projection. However, it is still the case that over a wide range of $p_x$ values, our protocol with $n=2$ cycles greatly suppresses the error probability. Even with $p_x = 0.08$, the second operation cycle can suppress the average error probability from $1.6\%$ to $0.8\%$ ($2.4\%$ to $1.5\%$ for the maximum error probability). Compared with the dephasing error, our protocol is less sensitive to Pauli-$X$ errors between the two CPhase gates. Comparing Eqs.~\eqref{eq:x_err_even_error} and~\eqref{eq:x_err_odd_error} with the Kraus operators without error, we notice that the faulty Kraus operators have a similar structure, except for the term involving $\mcP_{10}$. This is because the Pauli-$X$ error essentially flips the first CPhase gate from $\CP{\phi}$ to $\CP{-\phi}$. This will only affect our protocol when the first qubit is in the state $\ket{1}$. 

\paragraph{Pauli-Y errors.}

Pauli-$Y$ errors on the matter qubit can be analyzed using the same approach. Note that, unlike the Pauli-$X$ error, the Pauli-$Y$ error not only flips the sign of the first CPhase gate, it also flips the matter qubit state from $\ket{+}$ to $\ket{-}$. The faulty Kraus operators when the Pauli-$Y$ error occurs are
\begin{align}
    \hat{E}^{(\text{y-err})}_{+1} & =  \sin\left(\frac{\phi}{2}\right) \left(P_{00}  + e^{i \phi} P_{11} \right) + \sin(\phi) e^{i \phi/2} P_{10} ,\label{eq:yerr_p1} \\
    \hat{E}^{(\text{y-err})}_{-1} & = -i \cos\left(\frac{\phi}{2}\right) \left(P_{00}  + e^{i \phi} P_{11} \right) \nonumber \\
    & \quad - i e^{i\phi/2} \left( P_{01}+ \cos(\phi) P_{10}  \right). \label{eq:yerr_m1}
\end{align}

Denoting the probability for a Pauli-$Y$ error to occur by $p_y$, the error probability for our protocol after $n$ operation cycles is
\begin{align}
    & \mP^{(n)}_{\text{err}} =  \left[ \frac{1}{2} + \left( \frac{1}{2} - p_y \right) \cos(\phi)\right]^n \left( \vert c_{00} \vert^2 + \vert c_{11} \vert^2\right) \nonumber \\ & + \left[1 - (1-p_y)^n\right] \vert c_{01} \vert^2 + \left\lbrace 1 - \left[1-p_y \cos^{2} (\phi) \right]^n\right\rbrace \vert c_{10} \vert^2.
    \label{eq:y_err_err_prob}
\end{align}
The maximum error probability is
\begin{align}
    \mP^{(n)}_{\text{err,max}} & = \text{max}\left\lbrace \left[ \frac{1}{2} + \left( \frac{1}{2} - p_y \right) \cos(\phi)\right]^n, \right. \nonumber \\
    & \qquad \qquad \left. \left[1 - (1-p_y)^n\right] \right\rbrace,
    \label{eq:y_error}
\end{align}
where we used the fact that $1-p_y \leq 1-p_y \cos^{2n} (\phi)$ implies that the second term in Eq.~\eqref{eq:y_err_err_prob} always has a larger maximal value compared to that of the third term.

In Fig.~\ref{fig:pauli_between}b, we plot the maximum and average error probability of our protocol in the presence of Pauli-$Y$ errors. Compared with Fig.~\ref{fig:pauli_between}a and Fig.~\ref{fig:pauli_z_before}, the performance of our protocol with Pauli-$Y$ errors is similar to the case with Pauli-$Z$ errors. This can also be seen from Eqs.~\eqref{eq:y_err_err_prob} and~\eqref{eq:y_error}. The error probability expression is similar to that of the Pauli-$Z$ error case. It can also be understood from the fact that a Pauli-$Y$ error can be decomposed into a combination of Pauli-$X$ and Pauli-$Z$ errors. As our protocol is more robust against Pauli-$X$ errors, the effect of Pauli-$Z$ errors dominates. 

\section{Summary and outlook} \label{sec:summary}

In summary, we presented a protocol for performing parity projection on two photonic qubits using a matter qubit and CPhase gates. In particular, our protocol does not require perfect matter-photon CZ gates; CPhase gates with angles $\phi<\pi$ suffice. We analyze the performance of our protocol in the presence of various possible imperfections, including coherent errors on the CPhase gates, random CPhase angle fluctuations, and Pauli errors that occur on the matter qubit either before or in between the two CPhase gates. We demonstrated that our protocol is robust against most types of error, exhibiting high fidelities and low error probabilities in the resulting parity measurement. Dephasing errors on the matter qubit can be mitigated with dynamical decoupling. These results show that our protocol can be used not only as a tool to induce quantum entanglement between photonic qubits, but also as a tool to perform reliable photonic parity measurements.

\section{Acknowledgments}

E.B. acknowledges support from the Virginia Commonwealth Cyber Initiative (CCI) and the National Science Foundation (grant no. 2137953). S.E.E. acknowledges the Department of Energy Office of Science, National Quantum Information Science Research Centers, Co-design Center for Quantum Advantage (C2QA), contract number DE-SC0012704.  R. F. acknowledges the EU Horizon 2020 programme (GA 862035
QLUSTER).

\appendix

\section{Average channel fidelity to an ideal parity projection channel} \label{appsec:fidelity}

The fidelity between two states $\rho$ and $\sigma$ is defined as 
\begin{equation}
    F(\rho, \sigma) = \text{tr}(\sqrt{\sqrt{\rho} \sigma \sqrt{\rho}})^2.
\end{equation}
For a general pure two-qubit state $\ket{\psi}$, after the perfect parity projection, the state becomes
\begin{align}
    \rho_{\text{ideal}} = \mP_{\text{e}}\dyad{\psi_{\text{e}}} + \mP_{\text{o}} \dyad{\psi_{\text{o}}},
\end{align}
where $\ket{\psi_{\text{e}}}$ and $\ket{\psi_{\text{o}}}$ are the normalized states corresponding to the even- and odd-parity parts of the state $\ket{\psi}$, and $\mP_{\text{e}}$ and $\mP_{\text{o}}$ are the probabilities for even and odd parity projections to occur. Viewing our protocol as a quantum channel $\mathcal{E}$, we can write
\begin{align}
    & \sqrt{\rho_{\text{ideal}}} \mathcal{E}(\dyad{\psi}) \sqrt{\rho_{\text{ideal}}}  \nonumber \\
    & \quad = \sum_{i,j = \text{o, e}} \sqrt{\mP_{i} \mP_{j}} \mel{\psi_{i}}{\mathcal{E}(\dyad{\psi})}{\psi_{j}} \dyad{\psi_{i}}{\psi_{j}},
\end{align}
which can be treated in a $2 \times 2$ matrix form in the basis of $\ket{\psi_\text{o}}$ and $\ket{\psi_{\text{e}}}$, as these two states are orthogonal. We can then diagonalize this matrix to calculate the fidelity:
\begin{align}
    F &= \langle \mathcal{E}(\dyad{\psi}) \rangle_{\text{ee}} + \langle \mathcal{E}(\dyad{\psi}) \rangle_{\text{oo}} \nonumber \\
    & + 2\sqrt{\langle \mathcal{E}(\dyad{\psi}) \rangle_{\text{oo}}\langle \mathcal{E}(\dyad{\psi}) \rangle_{\text{ee}} - \left\vert \langle \mathcal{E}(\dyad{\psi}) \rangle_{\text{eo}} \right\vert^2}, 
\end{align}
where $\langle \mathcal{E}(\dyad{\psi}) \rangle_{ij} = \sqrt{\mP_{i} \mP_{j}}\mel{\psi_{i}}{\mathcal{E}(\dyad{\psi})}{\psi_{j}}$. This is the equation of the state fidelity that is used in calculating average channel fidelity relative to a perfect parity projection in our manuscript. 

\section{Calculating the error probability of our protocol with unbalanced CPhase gates}  \label{appsec:coh_err}

In the case of unbalanced CPhase gates, the Kraus operators of our protocol are given in Eqs.~\eqref{eq:coh_kraus_even} and~\eqref{eq:coh_kraus_odd}. In this section, we show how we calculate the error probability of our protocol using these Kraus operators.

The error probability is defined in Eq.~\eqref{eq:error_prob}. The error probability of the odd-parity projection is
\begin{align}
    \mP_{\text{odd}} \mP_{\text{odd}, \text{err}} & = \cos^{2n} \left( \phi/2 \right) \vert c_{00}\vert^2 \nonumber \\
    & \; \quad + \cos^{2n}\left[ (\phi+\delta_1+\delta_2)/2\right] \vert c_{11}\vert^2.
\end{align}
As there are multiple Kraus operators corresponding to even-parity projections, the error probability in this case is
\begin{align}
    \sum_{m=1}^{n} \mP_{\text{even}}^{(m)} \mP_{\text{even, error}}^{(m)},
    \label{eq:coh_even_error}
\end{align}
where the summation is over all the operation cycles. Each term in Eq.~\eqref{eq:coh_even_error} evaluates to
\begin{align}
    & \mP_{\text{even}}^{(m)} \mP_{\text{even, error}}^{(m)} = \sin^{2}(\delta_2/2) \left( \cos(\delta_2/2)\right)^{2m-2} \vert c_{01} \vert^2 \nonumber\\
    & \qquad \qquad + \sin^{2}(\delta_1/2) \left( \cos(\delta_1/2)\right)^{2m-2} \vert c_{10} \vert^2.
    \label{eq:coh_error_even}
\end{align}
Therefore, the error probability of our protocol after the $n$-th cycle is
\begin{align}
    & \mP^{(n)}_{\text{err}} =  \cos^{2n}(\phi/2) \vert c_{00} \vert^2 + \cos^{2n} [(\phi+\delta_1 + \delta_2)/2] \vert c_{11} \vert^2 \nonumber \\
    & \; + [1 - \cos^{2n}(\delta_2/2))] \vert  c_{01} \vert^2 + [1 - \cos^{2n}(\delta_1/2)] \vert c_{10} \vert^2,  
\end{align}
where the last two terms are from the summation over $m$ in Eq.~\eqref{eq:coh_error_even} from $1$ to $n$. This is  Eq.~\eqref{eq:coh_error_error} in the main text.

\section{Calculation of expected channel fidelity with Gaussian CPhase angle fluctuations} \label{appsec:gaussian}

In Sec.~\ref{sec:errors:incoh_cphase}, we show the average channel fidelity of our protocol relative to the perfect parity projection when our protocol suffers from Gaussian random CPhase angle fluctuations. The angle fluctuations are sampled from the probability distribution defined in Eq.~\eqref{eq:gaussian_distribution}. The average channel fidelity is calculated from
\begin{widetext}
\begin{align}
    F = \int \prod_{j=1}^n d \delta_1^{(j)} d \delta_2^{(j)} \int d\ket{\psi} F \left[ \mcP(\dyad{\psi}, \mathcal{E}_{\{\delta_1^{(j)}, \delta_2^{(j)} \}}(\dyad{\psi}) \right]  \mP(\delta_1^{(j)}) \mP(\delta_2^{(j)}),
\end{align}
where $\delta_{1,2} = \phi_{1,2} - \phi$ are treated as two independent random variables, and the integration $\int d\ket{\psi}$ is over the uniform space of two-qubit states, and the superscript $j$ labels the operation cycle. When numerically evaluating the average channel fidelity, for a given number of operation cycles, we randomly sample all the CPhase gate angles in the protocol according to the Gaussian distribution 1000 times independently. We then calculate the average channel fidelity for each sample of CPhase angles. Finally, we average over the 1000 samples to estimate the average channel fidelity in the presence of Gaussian CPhase fluctuations.

\section{Error Kraus operators with matter qubit Pauli-$Z$ errors before CPhase gates} \label{appsec:pauli_z_before}
In this appendix, we derive the Kraus operators for one cycle of our parity-projection protocol in the case where Pauli-$Z$ errors occur before the two CPhase gates. We also compute the error probability in the presence of such errors.

We denote the state of the two photonic qubits by $\rho_{0}^{(1,2)}$. The CPhase gate between a photonic qubit (Q1 or Q2) and the matter qubit (`m') can be expressed as
\begin{align}
    \text{CP}_{\alpha,\text{m}} (\phi) = P_{0}^{(\alpha)} I^{(\text{m})} + P_{1}^{(\alpha)} S^{(\text{m})} (\phi),
\end{align}
where $\alpha = 1, 2$ labels the two photonic qubits. According to our protocol, we can express the state of the two photonic qubits after a single  measurement on the matter qubit with outcome $\pm 1$ as
\begin{align}
    \tilde{\rho}^{(1,2)}_{\pm 1} & = \bra{\pm}_{\text{m}} R_{z}^{(\text{m}) \, \dagger}(\phi) \text{CP}_{2,\text{m}}(\phi) \text{CP}_{1,\text{m}}(\phi) \mathcal{E}_z(\dyad{+}) \otimes \rho_{0}^{(1,2)} \text{CP}_{1,\text{m}}^{\dagger}(\phi) \text{CP}_{2,\text{m}}^{\dagger}(\phi) R_{z}^{(\text{m})}(\phi) \ket{\pm}_{\text{m}} \nonumber \\
    & = (1-p_z) \bra{\pm}_{\text{m}} R_{z}^{(\text{m}) \, \dagger}(\phi) \text{CP}_{2,\text{m}}(\phi) \text{CP}_{1,\text{m}}(\phi) \ket{+}_\text{m} \rho_{0}^{(1,2)} \bra{+}_\text{m} \text{CP}_{1,\text{m}}^{\dagger}(\phi) \text{CP}_{2,\text{m}}^{\dagger}(\phi) R_{z}^{(\text{m})}(\phi) \ket{\pm}_{\text{m}} \nonumber \\ 
    & \qquad \ +  p_z \bra{\pm}_{\text{m}} R_{z}^{(\text{m}) \, \dagger}(\phi) \text{CP}_{2,\text{m}}(\phi) \text{CP}_{1,\text{m}}(\phi) \hat{Z}_\text{m} \ket{+}_\text{m} \rho_{0}^{(1,2)} \bra{+}_\text{m} \hat{Z}_\text{m} \text{CP}_{1,\text{m}}^{\dagger}(\phi) \text{CP}_{2,\text{m}}^{\dagger}(\phi) R_{z}^{(\text{m})}(\phi) \ket{\pm}_{\text{m}},
\end{align}
where the tilde means the density operator is not normalized. The trace of these density matrices is the probability to get the corresponding measurement results. Therefore, we can define the single-operator operators as
\begin{align}
    \hat{E}_{\pm 1} & \equiv \bra{\pm}_{\text{m}} R_{z}^{(\text{m}) \, \dagger}(\phi) \text{CP}_{2,\text{m}}(\phi) \text{CP}_{1,\text{m}}(\phi) \ket{+}_\text{m}, \\
    \hat{E}_{\pm 1}^{(\text{err})} & \equiv \bra{\pm}_{\text{m}} R_{z}^{(\text{m}) \, \dagger}(\phi) \text{CP}_{2,\text{m}}(\phi) \text{CP}_{1,\text{m}}(\phi) \ket{-}_\text{m},
\end{align}
where the operators $\hat{E}_{\pm 1}$ are given in Eqs.~\eqref{eq:first_round_even} and~\eqref{eq:first_round_odd}. We further notice that
\begin{align}
    \hat{E}_{\pm 1}^{(\text{err})} & = P_{00} \bra{\pm}_{\text{m}} R_{z}^{(\text{m}) \, \dagger} (\phi) \ket{-}_\text{m} + P_{11} e^{i\phi} \bra{\pm}_{\text{m}} R_{z}^{(\text{m})} (\phi) \ket{-}_\text{m} + P_{01} e^{i\phi/2} \braket{\pm}{-} + P_{10} e^{i\phi/2} \braket{\pm}{-} = \hat{E}_{\mp 1},
\end{align}
and so the photonic state after measuring the matter qubit is
\begin{align}
    \tilde{\rho}^{(1,2)}_{\pm 1} \equiv \mathcal{E}_{\pm 1} (\rho^{(1,2)}) = (1-p_z) \hat{E}_{\pm 1} \rho^{(1,2)} \hat{E}_{\pm 1}^{\dagger} + p_z \hat{E}_{\mp 1} \rho^{(1,2)} \hat{E}_{\mp 1}^\dagger.
\end{align}

Therefore, the state after a successful even-parity projection in round $k$ is
\begin{align}
    \tilde{\rho}^{(1,2)}_{k,-1} & = \sum_{j=0}^{k-1} {k-1 \choose j} (1-p_z)^{k-j} p_z^j \hat{E}_{+1}^{k-1-j} \hat{E}_{-1}^{j+1} \rho^{(1,2)} (\hat{E}_{-1}^{\dagger})^{j+1} (\hat{E}_{+1}^{\dagger})^{k-1-j} \nonumber \\
    & \quad + \sum_{j=0}^{k-1} {k-1 \choose j} (1-p_z)^{k-1-j} p_z^{j+1} \hat{E}_{+1}^{k-j} \hat{E}_{-1}^{j} \rho^{(1,2)} (\hat{E}_{-1}^{\dagger})^{j} (\hat{E}_{+1}^{\dagger})^{k-j}, \quad \text{for } k \geq 1.
    \label{eq:n_round_pz_even_init}
\end{align}

We denote the perfect even and odd-parity projection operators as $\mathcal{P}_{\text{even}}$ and $\mathcal{P}_{\text{odd}}$ (see the main text below Eq.~\eqref{eq:rho_ideal}). 
The error probability after we apply our protocol $n$ times is
\begin{align}
    \mP^{(n)}_{\text{err}} & = \mP_{\text{odd}} \mP_{\text{odd}, \text{err}} + \sum_{k=1}^{n} \mP_{\text{even}}^{(k)} \mP_{\text{even}, \text{err}}^{(k)} \nonumber \\
    & = \tr \left( \mathcal{P}_{\text{even}} \tilde{\rho}^{(1,2)}_{n:+1} \mathcal{P}_{\text{even}}\right) + \sum_{k=1}^{n} \tr \left( \mathcal{P}_{\text{odd}} \tilde{\rho}^{(1,2)}_{k,-1} \mathcal{P}_{\text{odd}}\right),
    \label{eq:general_incoh_errP}
\end{align}
where the states $\tilde{\rho}^{(1,2)}_{n:+1}$ and $\tilde{\rho}^{(1,2)}_{k,-1}$ are given in Eqs.~\eqref{eq:n_round_pz_odd_init} and~\eqref{eq:n_round_pz_even_init}.

Further, we consider the case when the initial state of the two photonic qubits is a pure state, which can be expressed in Eq.~\eqref{eq:random_state}, which gives the density matrix $\rho^{(1,2)} = \dyad{\psi}$. We then notice that
\begin{align}
    \tr \left( \mathcal{P}_{\text{even}} \tilde{\rho}^{(1,2)}_{n:+1} \mathcal{P}_{\text{even}}\right) & = \sum_{j=0}^{n} {n \choose j} (1-p_z)^{n-j} p_z^j \tr \left( \mathcal{P}_\text{even} \hat{E}_{+1}^{n-j} \hat{E}_{-1}^{j} \dyad{\psi} (\hat{E}_{-1}^{\dagger})^j (\hat{E}_{+1}^{\dagger})^{n-j} \mathcal{P}_\text{even} \right) \nonumber \\
    & = \sum_{j=0}^{n} {n \choose j} (1-p_z)^{n-j} p_z^j \left\Vert \mathcal{P}_\text{even} \hat{E}_{+1}^{n-j} \hat{E}_{-1}^{j} \ket{\psi} \right\Vert^2
    \label{eq:even_error_pauli_z}
\end{align}
where $\Vert \dots \Vert$ is the norm of the state vector. Similarly, the second term in Eq.~\eqref{eq:general_incoh_errP}
\begin{align}
    \tr \left( \mathcal{P}_{\text{odd}} \tilde{\rho}^{(1,2)}_{k,-1} \mathcal{P}_{\text{odd}}\right) & = \sum_{j=0}^{k-1} {k-1 \choose j} (1-p_z)^{k-1-j} p_z^j \left\lbrace (1-p_z) \left\Vert \mathcal{P}_\text{odd} \hat{E}_{+1}^{k-1-j} \hat{E}_{-1}^{j+1} \ket{\psi} \right\Vert^2 + p_z \left\Vert \mathcal{P}_\text{odd} \hat{E}_{+1}^{k-j} \hat{E}_{-1}^{j} \ket{\psi} \right\Vert^2 \right\rbrace.
    \label{eq:odd_error_pauli_z}
\end{align}
We want to point out that $\mathcal{P}_{\text{odd}} \hat{E}_{-1} = 0$, which shows that the even-parity projection has probability $(1-p_z)$ to be a perfect parity projection. Combining both terms in Eqs.~\eqref{eq:even_error_pauli_z} and~\eqref{eq:odd_error_pauli_z}, we can derive the error probability Eq.~\eqref{eq:pauli_z_error_prob} in the main text.

\section{Pauli-X error between two CPhase gates} \label{appsec:pauli_x_mid}

In this appendix, we derive the Kraus operators for one cycle of our parity-projection protocol in the case where Pauli-$X$ errors occur between the two CPhase gates. We also compute the error probability in the presence of such errors.

The matter qubit is initialized in the $\ket{+}$ state, and we denote the initial state of the photonic qubits by $\rho^{(1,2)}$. The state of these two photonic qubits after the matter qubit is measured with outcome $\pm1$ is
\begin{align}
    \tilde{\rho}^{(1,2)}_{\pm 1} & = \bra{\pm}_{\text{m}} R_{z}^{(\text{m}) \, \dagger}(\phi) \text{CP}_{2,\text{m}}(\phi) \ \mathcal{I}^{(1,2)} \circ \mathcal{E}_{x}^{(\text{m})} \left[ \text{CP}_{1,\text{m}}(\phi) \dyad{+} \otimes \rho_{0}^{(1,2)} \text{CP}_{1,\text{m}}^{\dagger}(\phi) \right] \text{CP}_{2,\text{m}}^{\dagger}(\phi) R_{z}^{(\text{m})}(\phi) \ket{\pm}_{\text{m}}, 
\end{align}
where $\mathcal{I}^{(1,2)}$ is an identity quantum channel on the photonic qubits, $\mathcal{E}_{x}^{(\text{m})}$ is the Pauli-$X$ error channel on the matter qubit, and $\circ$ denotes a composition of the quantum channels on the combined system. The tilde means the density matrix is not normalized, with the trace equal to the probability to get the corresponding measurement result. The state can be explicitly written as
\begin{align}
    \tilde{\rho}^{(1,2)}_{\pm 1} & = (1-p_x) \bra{\pm}_{\text{m}} R_{z}^{(\text{m}) \, \dagger}(\phi) \text{CP}_{2,\text{m}}(\phi) \text{CP}_{1,\text{m}}(\phi) \ket{+}_\text{m} \rho_{0}^{(1,2)} \bra{+}_\text{m} \text{CP}_{1,\text{m}}^{\dagger}(\phi) \text{CP}_{2,\text{m}}^{\dagger}(\phi) R_{z}^{(\text{m})}(\phi) \ket{\pm}_{\text{m}} \nonumber \\ 
    & \qquad \ +  p_x \bra{\pm}_{\text{m}} R_{z}^{(\text{m}) \, \dagger}(\phi) \text{CP}_{2,\text{m}}(\phi) \hat{X}_\text{m} \text{CP}_{1,\text{m}}(\phi)  \ket{+}_\text{m} \rho_{0}^{(1,2)} \bra{+}_\text{m} \text{CP}_{1,\text{m}}^{\dagger}(\phi)  \hat{X}_\text{m} \text{CP}_{2,\text{m}}^{\dagger}(\phi) R_{z}^{(\text{m})}(\phi) \ket{\pm}_{\text{m}}.
\end{align}
Note that the first term is identical to the no-Pauli-error case, where we can still define the Kraus operators $\hat{E}_{\pm 1}$ as in Eqs.~\eqref{eq:first_round_even} and~\eqref{eq:first_round_odd}. The second term shows the effect of the Pauli-$X$ error, where the faulty Kraus operators are
\begin{align}
    \hat{E}^{(\text{x-err})}_{\pm 1} & = \bra{\pm}_{\text{m}} R_{z}^{(\text{m}) \, \dagger}(\phi)  \text{CP}_{2,\text{m}}(\phi) \hat{X}_{\text{m}}\text{CP}_{1,\text{m}}(\phi) \ket{+}_\text{m} \nonumber \\
    & = P_{00} \bra{\pm}_{\text{m}} R_{z}^{(\text{m}) \, \dagger} (\phi) \ket{+}_\text{m} + P_{11} \bra{\pm}_{\text{m}} R_{z}^{(\text{m})} (-\phi) e^{i \phi} \ket{+}_\text{m} + P_{01}e^{i\phi/2} \braket{\pm}{+} + P_{10} \bra{\pm}e^{i\phi/2} R_{z}^{(\text{m})}(-2\phi)\ket{+}, \label{eq:intermediate}
\end{align}
where we use the fact that $\hat{X} \ket{+} = \ket{+}$ and $\hat{X} R_{z}(\phi) \hat{X} = R_{z}(-\phi)$. After working out the overlapping terms in Eq.~\eqref{eq:intermediate}, the faulty Kraus operators for Pauli-X error are
\begin{align}
    \hat{E}^{(\text{x-err})}_{+1} & = \cos\left(\frac{\phi}{2}\right) \left( P_{00}  + e^{i \phi} P_{11} \right) + e^{i\phi/2}\left[P_{01}+ \cos(\phi) P_{10}  \right] \\
    \hat{E}^{(\text{x-err})}_{-1} & = i \sin\left(\frac{\phi}{2}\right) \left( P_{00}  + e^{i \phi} P_{11} + 2 e^{i\phi/2}\cos(\phi/2) P_{10} \right).
\end{align}
The state after we get a $-1$ at round $k$ while all the previous $k-1$ rounds yielded $+1$ is
\begin{align}
    \tilde{\rho}^{(1,2)}_{k,-1} & = \sum_{j=0}^{k-1} {k-1 \choose j} (1-p_x)^{k-1-j} p_x^j \left\lbrace (1-p_x) \hat{E}_{-1} \hat{E}_{+1}^{k-1-j} \left(\hat{E}_{+1}^{(\text{x-err})}\right)^{j} \rho^{(1,2)} \left(\hat{E}_{+1}^{(\text{x-err}) \dagger}\right)^{j} \left(\hat{E}_{+1}^{\dagger}\right)^{k-1-j} \hat{E}_{-1}^\dagger \right. \nonumber \\
    & \qquad \qquad \qquad + \left.  p_x \hat{E}_{-1}^{(\text{x-err})} \hat{E}_{+1}^{k-1-j} \left(\hat{E}_{+1}^{(\text{x-err})}\right)^{j} \rho^{(1,2)} \left(\hat{E}_{+1}^{(\text{x-err}) \dagger}\right)^{j} \left(\hat{E}_{+1}^{\dagger}\right)^{k-1-j} (\hat{E}_{-1}^{(\text{x-err})})^\dagger \right\rbrace.
    \label{eq:incoh_state_even}
\end{align}

Similar to the analysis in Sec.~\ref{sec:error:pauli:before}, the error probability $\mP^{(n)}_{\text{err}}$ is given by Eq.~\eqref{eq:general_incoh_errP}, where the first and second terms are calculated as
\begin{align}
    \mP_{\text{odd}} \mP_{\text{odd}, \text{err}} & = \tr \left( \mathcal{P}_{\text{even}} \tilde{\rho}^{(1,2)}_{n:+1} \mathcal{P}_{\text{even}}\right) = \sum_{j=0}^{n} {n \choose j} (1-p_x)^{n-j} p_x^j \left\Vert \mathcal{P}_{\text{even}} \hat{E}_{+1}^{n-j} \left(\hat{E}_{+1}^{(\text{x-err})}\right)^j \ket{\psi} \right\Vert^2, \\
    \mP_{\text{even}} \mP_{\text{even}, \text{err}} & = \sum_{k=1}^{n} \sum_{j=0}^{k-1} {k-1 \choose j} (1-p_x)^{k-1-j} p_x^j \left\lbrace (1-p_x) \left\Vert \mathcal{P}_{\text{odd}} \hat{E}_{-1} \hat{E}_{+1}^{k-1-j} \left(\hat{E}_{+1}^{(\text{x-err})}\right)^{j} \ket{\psi} \right\Vert^2 \right. \nonumber \\
    & \qquad \qquad \qquad \qquad \qquad \qquad + \left. p_x \left\Vert \mathcal{P}_{\text{odd}} \hat{E}_{-1}^{(\text{x-err})} \hat{E}_{+1}^{k-1-j} \left(\hat{E}_{+1}^{(\text{x-err})}\right)^{j} \ket{\psi}\right\Vert^2 \right\rbrace,
\end{align}
where we assume the initial two-photon state is pure and can be expressed as in Eq.~\eqref{eq:random_state}.

\end{widetext}

\bibliography{ref}

\begin{thebibliography}{88}%
\makeatletter
\providecommand \@ifxundefined [1]{%
 \@ifx{#1\undefined}
}%
\providecommand \@ifnum [1]{%
 \ifnum #1\expandafter \@firstoftwo
 \else \expandafter \@secondoftwo
 \fi
}%
\providecommand \@ifx [1]{%
 \ifx #1\expandafter \@firstoftwo
 \else \expandafter \@secondoftwo
 \fi
}%
\providecommand \natexlab [1]{#1}%
\providecommand \enquote  [1]{``#1''}%
\providecommand \bibnamefont  [1]{#1}%
\providecommand \bibfnamefont [1]{#1}%
\providecommand \citenamefont [1]{#1}%
\providecommand \href@noop [0]{\@secondoftwo}%
\providecommand \href [0]{\begingroup \@sanitize@url \@href}%
\providecommand \@href[1]{\@@startlink{#1}\@@href}%
\providecommand \@@href[1]{\endgroup#1\@@endlink}%
\providecommand \@sanitize@url [0]{\catcode `\\12\catcode `\$12\catcode
  `\&12\catcode `\#12\catcode `\^12\catcode `\_12\catcode `\%12\relax}%
\providecommand \@@startlink[1]{}%
\providecommand \@@endlink[0]{}%
\providecommand \url  [0]{\begingroup\@sanitize@url \@url }%
\providecommand \@url [1]{\endgroup\@href {#1}{\urlprefix }}%
\providecommand \urlprefix  [0]{URL }%
\providecommand \Eprint [0]{\href }%
\providecommand \doibase [0]{http://dx.doi.org/}%
\providecommand \selectlanguage [0]{\@gobble}%
\providecommand \bibinfo  [0]{\@secondoftwo}%
\providecommand \bibfield  [0]{\@secondoftwo}%
\providecommand \translation [1]{[#1]}%
\providecommand \BibitemOpen [0]{}%
\providecommand \bibitemStop [0]{}%
\providecommand \bibitemNoStop [0]{.\EOS\space}%
\providecommand \EOS [0]{\spacefactor3000\relax}%
\providecommand \BibitemShut  [1]{\csname bibitem#1\endcsname}%
\let\auto@bib@innerbib\@empty
%</preamble>
\bibitem [{\citenamefont {Grover}(1996)}]{Grover1996}%
  \BibitemOpen
  \bibfield  {author} {\bibinfo {author} {\bibfnamefont {Lov~K.}\ \bibnamefont
  {Grover}},\ }\bibfield  {title} {\enquote {\bibinfo {title} {A fast quantum
  mechanical algorithm for database search},}\ }in\ \href {\doibase
  10.1145/237814.237866} {\emph {\bibinfo {booktitle} {Proceedings of the
  Twenty-eighth Annual ACM Symposium on Theory of Computing}}},\ \bibinfo
  {series and number} {STOC '96}\ (\bibinfo  {publisher} {ACM},\ \bibinfo
  {address} {New York, NY, USA},\ \bibinfo {year} {1996})\ pp.\ \bibinfo
  {pages} {212--219}\BibitemShut {NoStop}%
\bibitem [{\citenamefont {Shor}(1994)}]{Shor1994}%
  \BibitemOpen
  \bibfield  {author} {\bibinfo {author} {\bibfnamefont {P.~W.}\ \bibnamefont
  {Shor}},\ }\bibfield  {title} {\enquote {\bibinfo {title} {Algorithms for
  quantum computation: Discrete logarithms and factoring},}\ }in\ \href
  {\doibase 10.1109/SFCS.1994.365700} {\emph {\bibinfo {booktitle} {Proceedings
  of the 35th Annual Symposium on Foundations of Computer Science}}},\ \bibinfo
  {series and number} {SFCS '94}\ (\bibinfo  {publisher} {IEEE Computer
  Society},\ \bibinfo {address} {Washington, DC, USA},\ \bibinfo {year}
  {1994})\ pp.\ \bibinfo {pages} {124--134}\BibitemShut {NoStop}%
\bibitem [{\citenamefont {Shor}(1997)}]{Shor1997}%
  \BibitemOpen
  \bibfield  {author} {\bibinfo {author} {\bibfnamefont {Peter~W.}\
  \bibnamefont {Shor}},\ }\bibfield  {title} {\enquote {\bibinfo {title}
  {Polynomial-time algorithms for prime factorization and discrete logarithms
  on a quantum computer},}\ }\href {\doibase 10.1137/S0097539795293172}
  {\bibfield  {journal} {\bibinfo  {journal} {SIAM Journal on Computing}\
  }\textbf {\bibinfo {volume} {26}},\ \bibinfo {pages} {1484--1509} (\bibinfo
  {year} {1997})}\BibitemShut {NoStop}%
\bibitem [{\citenamefont {Simon}(1997)}]{Simon1997}%
  \BibitemOpen
  \bibfield  {author} {\bibinfo {author} {\bibfnamefont {Daniel~R.}\
  \bibnamefont {Simon}},\ }\bibfield  {title} {\enquote {\bibinfo {title} {On
  the power of quantum computation},}\ }\href {\doibase
  10.1137/S0097539796298637} {\bibfield  {journal} {\bibinfo  {journal} {SIAM
  Journal on Computing}\ }\textbf {\bibinfo {volume} {26}},\ \bibinfo {pages}
  {1474--1483} (\bibinfo {year} {1997})}\BibitemShut {NoStop}%
\bibitem [{\citenamefont {Arute}\ \emph {et~al.}(2019)\citenamefont {Arute},
  \citenamefont {Arya}, \citenamefont {Babbush}, \citenamefont {Bacon},
  \citenamefont {Bardin}, \citenamefont {Barends}, \citenamefont {Biswas},
  \citenamefont {Boixo}, \citenamefont {Brandao}, \citenamefont {Buell},
  \citenamefont {Burkett}, \citenamefont {Chen}, \citenamefont {Chen},
  \citenamefont {Chiaro}, \citenamefont {Collins}, \citenamefont {Courtney},
  \citenamefont {Dunsworth}, \citenamefont {Farhi}, \citenamefont {Foxen},
  \citenamefont {Fowler}, \citenamefont {Gidney}, \citenamefont {Giustina},
  \citenamefont {Graff}, \citenamefont {Guerin}, \citenamefont {Habegger},
  \citenamefont {Harrigan}, \citenamefont {Hartmann}, \citenamefont {Ho},
  \citenamefont {Hoffmann}, \citenamefont {Huang}, \citenamefont {Humble},
  \citenamefont {Isakov}, \citenamefont {Jeffrey}, \citenamefont {Jiang},
  \citenamefont {Kafri}, \citenamefont {Kechedzhi}, \citenamefont {Kelly},
  \citenamefont {Klimov}, \citenamefont {Knysh}, \citenamefont {Korotkov},
  \citenamefont {Kostritsa}, \citenamefont {Landhuis}, \citenamefont
  {Lindmark}, \citenamefont {Lucero}, \citenamefont {Lyakh}, \citenamefont
  {Mandr{\`a}}, \citenamefont {McClean}, \citenamefont {McEwen}, \citenamefont
  {Megrant}, \citenamefont {Mi}, \citenamefont {Michielsen}, \citenamefont
  {Mohseni}, \citenamefont {Mutus}, \citenamefont {Naaman}, \citenamefont
  {Neeley}, \citenamefont {Neill}, \citenamefont {Niu}, \citenamefont {Ostby},
  \citenamefont {Petukhov}, \citenamefont {Platt}, \citenamefont {Quintana},
  \citenamefont {Rieffel}, \citenamefont {Roushan}, \citenamefont {Rubin},
  \citenamefont {Sank}, \citenamefont {Satzinger}, \citenamefont {Smelyanskiy},
  \citenamefont {Sung}, \citenamefont {Trevithick}, \citenamefont
  {Vainsencher}, \citenamefont {Villalonga}, \citenamefont {White},
  \citenamefont {Yao}, \citenamefont {Yeh}, \citenamefont {Zalcman},
  \citenamefont {Neven},\ and\ \citenamefont {Martinis}}]{Arute2019}%
  \BibitemOpen
  \bibfield  {author} {\bibinfo {author} {\bibfnamefont {Frank}\ \bibnamefont
  {Arute}}, \bibinfo {author} {\bibfnamefont {Kunal}\ \bibnamefont {Arya}},
  \bibinfo {author} {\bibfnamefont {Ryan}\ \bibnamefont {Babbush}}, \bibinfo
  {author} {\bibfnamefont {Dave}\ \bibnamefont {Bacon}}, \bibinfo {author}
  {\bibfnamefont {Joseph~C.}\ \bibnamefont {Bardin}}, \bibinfo {author}
  {\bibfnamefont {Rami}\ \bibnamefont {Barends}}, \bibinfo {author}
  {\bibfnamefont {Rupak}\ \bibnamefont {Biswas}}, \bibinfo {author}
  {\bibfnamefont {Sergio}\ \bibnamefont {Boixo}}, \bibinfo {author}
  {\bibfnamefont {Fernando G. S.~L.}\ \bibnamefont {Brandao}}, \bibinfo
  {author} {\bibfnamefont {David~A.}\ \bibnamefont {Buell}}, \bibinfo {author}
  {\bibfnamefont {Brian}\ \bibnamefont {Burkett}}, \bibinfo {author}
  {\bibfnamefont {Yu}~\bibnamefont {Chen}}, \bibinfo {author} {\bibfnamefont
  {Zijun}\ \bibnamefont {Chen}}, \bibinfo {author} {\bibfnamefont {Ben}\
  \bibnamefont {Chiaro}}, \bibinfo {author} {\bibfnamefont {Roberto}\
  \bibnamefont {Collins}}, \bibinfo {author} {\bibfnamefont {William}\
  \bibnamefont {Courtney}}, \bibinfo {author} {\bibfnamefont {Andrew}\
  \bibnamefont {Dunsworth}}, \bibinfo {author} {\bibfnamefont {Edward}\
  \bibnamefont {Farhi}}, \bibinfo {author} {\bibfnamefont {Brooks}\
  \bibnamefont {Foxen}}, \bibinfo {author} {\bibfnamefont {Austin}\
  \bibnamefont {Fowler}}, \bibinfo {author} {\bibfnamefont {Craig}\
  \bibnamefont {Gidney}}, \bibinfo {author} {\bibfnamefont {Marissa}\
  \bibnamefont {Giustina}}, \bibinfo {author} {\bibfnamefont {Rob}\
  \bibnamefont {Graff}}, \bibinfo {author} {\bibfnamefont {Keith}\ \bibnamefont
  {Guerin}}, \bibinfo {author} {\bibfnamefont {Steve}\ \bibnamefont
  {Habegger}}, \bibinfo {author} {\bibfnamefont {Matthew~P.}\ \bibnamefont
  {Harrigan}}, \bibinfo {author} {\bibfnamefont {Michael~J.}\ \bibnamefont
  {Hartmann}}, \bibinfo {author} {\bibfnamefont {Alan}\ \bibnamefont {Ho}},
  \bibinfo {author} {\bibfnamefont {Markus}\ \bibnamefont {Hoffmann}}, \bibinfo
  {author} {\bibfnamefont {Trent}\ \bibnamefont {Huang}}, \bibinfo {author}
  {\bibfnamefont {Travis~S.}\ \bibnamefont {Humble}}, \bibinfo {author}
  {\bibfnamefont {Sergei~V.}\ \bibnamefont {Isakov}}, \bibinfo {author}
  {\bibfnamefont {Evan}\ \bibnamefont {Jeffrey}}, \bibinfo {author}
  {\bibfnamefont {Zhang}\ \bibnamefont {Jiang}}, \bibinfo {author}
  {\bibfnamefont {Dvir}\ \bibnamefont {Kafri}}, \bibinfo {author}
  {\bibfnamefont {Kostyantyn}\ \bibnamefont {Kechedzhi}}, \bibinfo {author}
  {\bibfnamefont {Julian}\ \bibnamefont {Kelly}}, \bibinfo {author}
  {\bibfnamefont {Paul~V.}\ \bibnamefont {Klimov}}, \bibinfo {author}
  {\bibfnamefont {Sergey}\ \bibnamefont {Knysh}}, \bibinfo {author}
  {\bibfnamefont {Alexander}\ \bibnamefont {Korotkov}}, \bibinfo {author}
  {\bibfnamefont {Fedor}\ \bibnamefont {Kostritsa}}, \bibinfo {author}
  {\bibfnamefont {David}\ \bibnamefont {Landhuis}}, \bibinfo {author}
  {\bibfnamefont {Mike}\ \bibnamefont {Lindmark}}, \bibinfo {author}
  {\bibfnamefont {Erik}\ \bibnamefont {Lucero}}, \bibinfo {author}
  {\bibfnamefont {Dmitry}\ \bibnamefont {Lyakh}}, \bibinfo {author}
  {\bibfnamefont {Salvatore}\ \bibnamefont {Mandr{\`a}}}, \bibinfo {author}
  {\bibfnamefont {Jarrod~R.}\ \bibnamefont {McClean}}, \bibinfo {author}
  {\bibfnamefont {Matthew}\ \bibnamefont {McEwen}}, \bibinfo {author}
  {\bibfnamefont {Anthony}\ \bibnamefont {Megrant}}, \bibinfo {author}
  {\bibfnamefont {Xiao}\ \bibnamefont {Mi}}, \bibinfo {author} {\bibfnamefont
  {Kristel}\ \bibnamefont {Michielsen}}, \bibinfo {author} {\bibfnamefont
  {Masoud}\ \bibnamefont {Mohseni}}, \bibinfo {author} {\bibfnamefont {Josh}\
  \bibnamefont {Mutus}}, \bibinfo {author} {\bibfnamefont {Ofer}\ \bibnamefont
  {Naaman}}, \bibinfo {author} {\bibfnamefont {Matthew}\ \bibnamefont
  {Neeley}}, \bibinfo {author} {\bibfnamefont {Charles}\ \bibnamefont {Neill}},
  \bibinfo {author} {\bibfnamefont {Murphy~Yuezhen}\ \bibnamefont {Niu}},
  \bibinfo {author} {\bibfnamefont {Eric}\ \bibnamefont {Ostby}}, \bibinfo
  {author} {\bibfnamefont {Andre}\ \bibnamefont {Petukhov}}, \bibinfo {author}
  {\bibfnamefont {John~C.}\ \bibnamefont {Platt}}, \bibinfo {author}
  {\bibfnamefont {Chris}\ \bibnamefont {Quintana}}, \bibinfo {author}
  {\bibfnamefont {Eleanor~G.}\ \bibnamefont {Rieffel}}, \bibinfo {author}
  {\bibfnamefont {Pedram}\ \bibnamefont {Roushan}}, \bibinfo {author}
  {\bibfnamefont {Nicholas~C.}\ \bibnamefont {Rubin}}, \bibinfo {author}
  {\bibfnamefont {Daniel}\ \bibnamefont {Sank}}, \bibinfo {author}
  {\bibfnamefont {Kevin~J.}\ \bibnamefont {Satzinger}}, \bibinfo {author}
  {\bibfnamefont {Vadim}\ \bibnamefont {Smelyanskiy}}, \bibinfo {author}
  {\bibfnamefont {Kevin~J.}\ \bibnamefont {Sung}}, \bibinfo {author}
  {\bibfnamefont {Matthew~D.}\ \bibnamefont {Trevithick}}, \bibinfo {author}
  {\bibfnamefont {Amit}\ \bibnamefont {Vainsencher}}, \bibinfo {author}
  {\bibfnamefont {Benjamin}\ \bibnamefont {Villalonga}}, \bibinfo {author}
  {\bibfnamefont {Theodore}\ \bibnamefont {White}}, \bibinfo {author}
  {\bibfnamefont {Z.~Jamie}\ \bibnamefont {Yao}}, \bibinfo {author}
  {\bibfnamefont {Ping}\ \bibnamefont {Yeh}}, \bibinfo {author} {\bibfnamefont
  {Adam}\ \bibnamefont {Zalcman}}, \bibinfo {author} {\bibfnamefont {Hartmut}\
  \bibnamefont {Neven}}, \ and\ \bibinfo {author} {\bibfnamefont {John~M.}\
  \bibnamefont {Martinis}},\ }\bibfield  {title} {\enquote {\bibinfo {title}
  {Quantum supremacy using a programmable superconducting processor},}\ }\href
  {\doibase 10.1038/s41586-019-1666-5} {\bibfield  {journal} {\bibinfo
  {journal} {Nature}\ }\textbf {\bibinfo {volume} {574}},\ \bibinfo {pages}
  {505--510} (\bibinfo {year} {2019})}\BibitemShut {NoStop}%
\bibitem [{\citenamefont {Bouland}\ \emph {et~al.}(2019)\citenamefont
  {Bouland}, \citenamefont {Fefferman}, \citenamefont {Nirkhe},\ and\
  \citenamefont {Vazirani}}]{Bouland2019}%
  \BibitemOpen
  \bibfield  {author} {\bibinfo {author} {\bibfnamefont {Adam}\ \bibnamefont
  {Bouland}}, \bibinfo {author} {\bibfnamefont {Bill}\ \bibnamefont
  {Fefferman}}, \bibinfo {author} {\bibfnamefont {Chinmay}\ \bibnamefont
  {Nirkhe}}, \ and\ \bibinfo {author} {\bibfnamefont {Umesh}\ \bibnamefont
  {Vazirani}},\ }\bibfield  {title} {\enquote {\bibinfo {title} {On the
  complexity and verification of quantum random circuit sampling},}\ }\href
  {\doibase 10.1038/s41567-018-0318-2} {\bibfield  {journal} {\bibinfo
  {journal} {Nature Physics}\ }\textbf {\bibinfo {volume} {15}},\ \bibinfo
  {pages} {159--163} (\bibinfo {year} {2019})}\BibitemShut {NoStop}%
\bibitem [{\citenamefont {Zhong}\ \emph {et~al.}(2021)\citenamefont {Zhong},
  \citenamefont {Deng}, \citenamefont {Qin}, \citenamefont {Wang},
  \citenamefont {Chen}, \citenamefont {Peng}, \citenamefont {Luo},
  \citenamefont {Wu}, \citenamefont {Gong}, \citenamefont {Su}, \citenamefont
  {Hu}, \citenamefont {Hu}, \citenamefont {Yang}, \citenamefont {Zhang},
  \citenamefont {Li}, \citenamefont {Li}, \citenamefont {Jiang}, \citenamefont
  {Gan}, \citenamefont {Yang}, \citenamefont {You}, \citenamefont {Wang},
  \citenamefont {Li}, \citenamefont {Liu}, \citenamefont {Renema},
  \citenamefont {Lu},\ and\ \citenamefont {Pan}}]{Zhong2021}%
  \BibitemOpen
  \bibfield  {author} {\bibinfo {author} {\bibfnamefont {Han-Sen}\ \bibnamefont
  {Zhong}}, \bibinfo {author} {\bibfnamefont {Yu-Hao}\ \bibnamefont {Deng}},
  \bibinfo {author} {\bibfnamefont {Jian}\ \bibnamefont {Qin}}, \bibinfo
  {author} {\bibfnamefont {Hui}\ \bibnamefont {Wang}}, \bibinfo {author}
  {\bibfnamefont {Ming-Cheng}\ \bibnamefont {Chen}}, \bibinfo {author}
  {\bibfnamefont {Li-Chao}\ \bibnamefont {Peng}}, \bibinfo {author}
  {\bibfnamefont {Yi-Han}\ \bibnamefont {Luo}}, \bibinfo {author}
  {\bibfnamefont {Dian}\ \bibnamefont {Wu}}, \bibinfo {author} {\bibfnamefont
  {Si-Qiu}\ \bibnamefont {Gong}}, \bibinfo {author} {\bibfnamefont {Hao}\
  \bibnamefont {Su}}, \bibinfo {author} {\bibfnamefont {Yi}~\bibnamefont {Hu}},
  \bibinfo {author} {\bibfnamefont {Peng}\ \bibnamefont {Hu}}, \bibinfo
  {author} {\bibfnamefont {Xiao-Yan}\ \bibnamefont {Yang}}, \bibinfo {author}
  {\bibfnamefont {Wei-Jun}\ \bibnamefont {Zhang}}, \bibinfo {author}
  {\bibfnamefont {Hao}\ \bibnamefont {Li}}, \bibinfo {author} {\bibfnamefont
  {Yuxuan}\ \bibnamefont {Li}}, \bibinfo {author} {\bibfnamefont {Xiao}\
  \bibnamefont {Jiang}}, \bibinfo {author} {\bibfnamefont {Lin}\ \bibnamefont
  {Gan}}, \bibinfo {author} {\bibfnamefont {Guangwen}\ \bibnamefont {Yang}},
  \bibinfo {author} {\bibfnamefont {Lixing}\ \bibnamefont {You}}, \bibinfo
  {author} {\bibfnamefont {Zhen}\ \bibnamefont {Wang}}, \bibinfo {author}
  {\bibfnamefont {Li}~\bibnamefont {Li}}, \bibinfo {author} {\bibfnamefont
  {Nai-Le}\ \bibnamefont {Liu}}, \bibinfo {author} {\bibfnamefont {Jelmer~J.}\
  \bibnamefont {Renema}}, \bibinfo {author} {\bibfnamefont {Chao-Yang}\
  \bibnamefont {Lu}}, \ and\ \bibinfo {author} {\bibfnamefont {Jian-Wei}\
  \bibnamefont {Pan}},\ }\bibfield  {title} {\enquote {\bibinfo {title}
  {Phase-programmable gaussian boson sampling using stimulated squeezed
  light},}\ }\href {\doibase 10.1103/PhysRevLett.127.180502} {\bibfield
  {journal} {\bibinfo  {journal} {Phys. Rev. Lett.}\ }\textbf {\bibinfo
  {volume} {127}},\ \bibinfo {pages} {180502} (\bibinfo {year}
  {2021})}\BibitemShut {NoStop}%
\bibitem [{\citenamefont {Wu}\ \emph {et~al.}(2021)\citenamefont {Wu},
  \citenamefont {Bao}, \citenamefont {Cao}, \citenamefont {Chen}, \citenamefont
  {Chen}, \citenamefont {Chen}, \citenamefont {Chung}, \citenamefont {Deng},
  \citenamefont {Du}, \citenamefont {Fan}, \citenamefont {Gong}, \citenamefont
  {Guo}, \citenamefont {Guo}, \citenamefont {Guo}, \citenamefont {Han},
  \citenamefont {Hong}, \citenamefont {Huang}, \citenamefont {Huo},
  \citenamefont {Li}, \citenamefont {Li}, \citenamefont {Li}, \citenamefont
  {Li}, \citenamefont {Liang}, \citenamefont {Lin}, \citenamefont {Lin},
  \citenamefont {Qian}, \citenamefont {Qiao}, \citenamefont {Rong},
  \citenamefont {Su}, \citenamefont {Sun}, \citenamefont {Wang}, \citenamefont
  {Wang}, \citenamefont {Wu}, \citenamefont {Xu}, \citenamefont {Yan},
  \citenamefont {Yang}, \citenamefont {Yang}, \citenamefont {Ye}, \citenamefont
  {Yin}, \citenamefont {Ying}, \citenamefont {Yu}, \citenamefont {Zha},
  \citenamefont {Zhang}, \citenamefont {Zhang}, \citenamefont {Zhang},
  \citenamefont {Zhang}, \citenamefont {Zhao}, \citenamefont {Zhao},
  \citenamefont {Zhou}, \citenamefont {Zhu}, \citenamefont {Lu}, \citenamefont
  {Peng}, \citenamefont {Zhu},\ and\ \citenamefont {Pan}}]{Wu2021}%
  \BibitemOpen
  \bibfield  {author} {\bibinfo {author} {\bibfnamefont {Yulin}\ \bibnamefont
  {Wu}}, \bibinfo {author} {\bibfnamefont {Wan-Su}\ \bibnamefont {Bao}},
  \bibinfo {author} {\bibfnamefont {Sirui}\ \bibnamefont {Cao}}, \bibinfo
  {author} {\bibfnamefont {Fusheng}\ \bibnamefont {Chen}}, \bibinfo {author}
  {\bibfnamefont {Ming-Cheng}\ \bibnamefont {Chen}}, \bibinfo {author}
  {\bibfnamefont {Xiawei}\ \bibnamefont {Chen}}, \bibinfo {author}
  {\bibfnamefont {Tung-Hsun}\ \bibnamefont {Chung}}, \bibinfo {author}
  {\bibfnamefont {Hui}\ \bibnamefont {Deng}}, \bibinfo {author} {\bibfnamefont
  {Yajie}\ \bibnamefont {Du}}, \bibinfo {author} {\bibfnamefont {Daojin}\
  \bibnamefont {Fan}}, \bibinfo {author} {\bibfnamefont {Ming}\ \bibnamefont
  {Gong}}, \bibinfo {author} {\bibfnamefont {Cheng}\ \bibnamefont {Guo}},
  \bibinfo {author} {\bibfnamefont {Chu}\ \bibnamefont {Guo}}, \bibinfo
  {author} {\bibfnamefont {Shaojun}\ \bibnamefont {Guo}}, \bibinfo {author}
  {\bibfnamefont {Lianchen}\ \bibnamefont {Han}}, \bibinfo {author}
  {\bibfnamefont {Linyin}\ \bibnamefont {Hong}}, \bibinfo {author}
  {\bibfnamefont {He-Liang}\ \bibnamefont {Huang}}, \bibinfo {author}
  {\bibfnamefont {Yong-Heng}\ \bibnamefont {Huo}}, \bibinfo {author}
  {\bibfnamefont {Liping}\ \bibnamefont {Li}}, \bibinfo {author} {\bibfnamefont
  {Na}~\bibnamefont {Li}}, \bibinfo {author} {\bibfnamefont {Shaowei}\
  \bibnamefont {Li}}, \bibinfo {author} {\bibfnamefont {Yuan}\ \bibnamefont
  {Li}}, \bibinfo {author} {\bibfnamefont {Futian}\ \bibnamefont {Liang}},
  \bibinfo {author} {\bibfnamefont {Chun}\ \bibnamefont {Lin}}, \bibinfo
  {author} {\bibfnamefont {Jin}\ \bibnamefont {Lin}}, \bibinfo {author}
  {\bibfnamefont {Haoran}\ \bibnamefont {Qian}}, \bibinfo {author}
  {\bibfnamefont {Dan}\ \bibnamefont {Qiao}}, \bibinfo {author} {\bibfnamefont
  {Hao}\ \bibnamefont {Rong}}, \bibinfo {author} {\bibfnamefont {Hong}\
  \bibnamefont {Su}}, \bibinfo {author} {\bibfnamefont {Lihua}\ \bibnamefont
  {Sun}}, \bibinfo {author} {\bibfnamefont {Liangyuan}\ \bibnamefont {Wang}},
  \bibinfo {author} {\bibfnamefont {Shiyu}\ \bibnamefont {Wang}}, \bibinfo
  {author} {\bibfnamefont {Dachao}\ \bibnamefont {Wu}}, \bibinfo {author}
  {\bibfnamefont {Yu}~\bibnamefont {Xu}}, \bibinfo {author} {\bibfnamefont
  {Kai}\ \bibnamefont {Yan}}, \bibinfo {author} {\bibfnamefont {Weifeng}\
  \bibnamefont {Yang}}, \bibinfo {author} {\bibfnamefont {Yang}\ \bibnamefont
  {Yang}}, \bibinfo {author} {\bibfnamefont {Yangsen}\ \bibnamefont {Ye}},
  \bibinfo {author} {\bibfnamefont {Jianghan}\ \bibnamefont {Yin}}, \bibinfo
  {author} {\bibfnamefont {Chong}\ \bibnamefont {Ying}}, \bibinfo {author}
  {\bibfnamefont {Jiale}\ \bibnamefont {Yu}}, \bibinfo {author} {\bibfnamefont
  {Chen}\ \bibnamefont {Zha}}, \bibinfo {author} {\bibfnamefont {Cha}\
  \bibnamefont {Zhang}}, \bibinfo {author} {\bibfnamefont {Haibin}\
  \bibnamefont {Zhang}}, \bibinfo {author} {\bibfnamefont {Kaili}\ \bibnamefont
  {Zhang}}, \bibinfo {author} {\bibfnamefont {Yiming}\ \bibnamefont {Zhang}},
  \bibinfo {author} {\bibfnamefont {Han}\ \bibnamefont {Zhao}}, \bibinfo
  {author} {\bibfnamefont {Youwei}\ \bibnamefont {Zhao}}, \bibinfo {author}
  {\bibfnamefont {Liang}\ \bibnamefont {Zhou}}, \bibinfo {author}
  {\bibfnamefont {Qingling}\ \bibnamefont {Zhu}}, \bibinfo {author}
  {\bibfnamefont {Chao-Yang}\ \bibnamefont {Lu}}, \bibinfo {author}
  {\bibfnamefont {Cheng-Zhi}\ \bibnamefont {Peng}}, \bibinfo {author}
  {\bibfnamefont {Xiaobo}\ \bibnamefont {Zhu}}, \ and\ \bibinfo {author}
  {\bibfnamefont {Jian-Wei}\ \bibnamefont {Pan}},\ }\bibfield  {title}
  {\enquote {\bibinfo {title} {Strong quantum computational advantage using a
  superconducting quantum processor},}\ }\href {\doibase
  10.1103/PhysRevLett.127.180501} {\bibfield  {journal} {\bibinfo  {journal}
  {Phys. Rev. Lett.}\ }\textbf {\bibinfo {volume} {127}},\ \bibinfo {pages}
  {180501} (\bibinfo {year} {2021})}\BibitemShut {NoStop}%
\bibitem [{\citenamefont {Bennett}\ and\ \citenamefont
  {Brassard}(1984)}]{Bennett1984}%
  \BibitemOpen
  \bibfield  {author} {\bibinfo {author} {\bibfnamefont {C.~H.}\ \bibnamefont
  {Bennett}}\ and\ \bibinfo {author} {\bibfnamefont {G.}~\bibnamefont
  {Brassard}},\ }\bibfield  {title} {\enquote {\bibinfo {title} {{Quantum
  cryptography: Public key distribution and coin tossing}},}\ }in\ \href@noop
  {} {\emph {\bibinfo {booktitle} {Proceedings of IEEE International Conference
  on Computers, Systems, and Signal Processing}}}\ (\bibinfo {address}
  {India},\ \bibinfo {year} {1984})\ p.\ \bibinfo {pages} {175}\BibitemShut
  {NoStop}%
\bibitem [{\citenamefont {Ekert}(1991)}]{Ekert1991}%
  \BibitemOpen
  \bibfield  {author} {\bibinfo {author} {\bibfnamefont {Artur~K.}\
  \bibnamefont {Ekert}},\ }\bibfield  {title} {\enquote {\bibinfo {title}
  {Quantum cryptography based on bell's theorem},}\ }\href {\doibase
  10.1103/PhysRevLett.67.661} {\bibfield  {journal} {\bibinfo  {journal} {Phys.
  Rev. Lett.}\ }\textbf {\bibinfo {volume} {67}},\ \bibinfo {pages} {661--663}
  (\bibinfo {year} {1991})}\BibitemShut {NoStop}%
\bibitem [{\citenamefont {Shor}\ and\ \citenamefont
  {Preskill}(2000)}]{Shor2000}%
  \BibitemOpen
  \bibfield  {author} {\bibinfo {author} {\bibfnamefont {Peter~W.}\
  \bibnamefont {Shor}}\ and\ \bibinfo {author} {\bibfnamefont {John}\
  \bibnamefont {Preskill}},\ }\bibfield  {title} {\enquote {\bibinfo {title}
  {Simple proof of security of the bb84 quantum key distribution protocol},}\
  }\href {\doibase 10.1103/PhysRevLett.85.441} {\bibfield  {journal} {\bibinfo
  {journal} {Phys. Rev. Lett.}\ }\textbf {\bibinfo {volume} {85}},\ \bibinfo
  {pages} {441--444} (\bibinfo {year} {2000})}\BibitemShut {NoStop}%
\bibitem [{\citenamefont {Sasaki}\ \emph {et~al.}(2014)\citenamefont {Sasaki},
  \citenamefont {Yamamoto},\ and\ \citenamefont {Koashi}}]{Sasaki2014}%
  \BibitemOpen
  \bibfield  {author} {\bibinfo {author} {\bibfnamefont {Toshihiko}\
  \bibnamefont {Sasaki}}, \bibinfo {author} {\bibfnamefont {Yoshihisa}\
  \bibnamefont {Yamamoto}}, \ and\ \bibinfo {author} {\bibfnamefont {Masato}\
  \bibnamefont {Koashi}},\ }\bibfield  {title} {\enquote {\bibinfo {title}
  {Practical quantum key distribution protocol without monitoring signal
  disturbance},}\ }\href {\doibase 10.1038/nature13303} {\bibfield  {journal}
  {\bibinfo  {journal} {Nature}\ }\textbf {\bibinfo {volume} {509}},\ \bibinfo
  {pages} {475--478} (\bibinfo {year} {2014})}\BibitemShut {NoStop}%
\bibitem [{\citenamefont {Ursin}\ \emph {et~al.}(2007)\citenamefont {Ursin},
  \citenamefont {Tiefenbacher}, \citenamefont {Schmitt-Manderbach},
  \citenamefont {Weier}, \citenamefont {Scheidl}, \citenamefont {Lindenthal},
  \citenamefont {Blauensteiner}, \citenamefont {Jennewein}, \citenamefont
  {Perdigues}, \citenamefont {Trojek}, \citenamefont {{\"O}mer}, \citenamefont
  {F{\"u}rst}, \citenamefont {Meyenburg}, \citenamefont {Rarity}, \citenamefont
  {Sodnik}, \citenamefont {Barbieri}, \citenamefont {Weinfurter},\ and\
  \citenamefont {Zeilinger}}]{Ursin2007}%
  \BibitemOpen
  \bibfield  {author} {\bibinfo {author} {\bibfnamefont {R.}~\bibnamefont
  {Ursin}}, \bibinfo {author} {\bibfnamefont {F.}~\bibnamefont {Tiefenbacher}},
  \bibinfo {author} {\bibfnamefont {T.}~\bibnamefont {Schmitt-Manderbach}},
  \bibinfo {author} {\bibfnamefont {H.}~\bibnamefont {Weier}}, \bibinfo
  {author} {\bibfnamefont {T.}~\bibnamefont {Scheidl}}, \bibinfo {author}
  {\bibfnamefont {M.}~\bibnamefont {Lindenthal}}, \bibinfo {author}
  {\bibfnamefont {B.}~\bibnamefont {Blauensteiner}}, \bibinfo {author}
  {\bibfnamefont {T.}~\bibnamefont {Jennewein}}, \bibinfo {author}
  {\bibfnamefont {J.}~\bibnamefont {Perdigues}}, \bibinfo {author}
  {\bibfnamefont {P.}~\bibnamefont {Trojek}}, \bibinfo {author} {\bibfnamefont
  {B.}~\bibnamefont {{\"O}mer}}, \bibinfo {author} {\bibfnamefont
  {M.}~\bibnamefont {F{\"u}rst}}, \bibinfo {author} {\bibfnamefont
  {M.}~\bibnamefont {Meyenburg}}, \bibinfo {author} {\bibfnamefont
  {J.}~\bibnamefont {Rarity}}, \bibinfo {author} {\bibfnamefont
  {Z.}~\bibnamefont {Sodnik}}, \bibinfo {author} {\bibfnamefont
  {C.}~\bibnamefont {Barbieri}}, \bibinfo {author} {\bibfnamefont
  {H.}~\bibnamefont {Weinfurter}}, \ and\ \bibinfo {author} {\bibfnamefont
  {A.}~\bibnamefont {Zeilinger}},\ }\bibfield  {title} {\enquote {\bibinfo
  {title} {Entanglement-based quantum communication over 144{\thinspace}km},}\
  }\href {\doibase 10.1038/nphys629} {\bibfield  {journal} {\bibinfo  {journal}
  {Nature Physics}\ }\textbf {\bibinfo {volume} {3}},\ \bibinfo {pages}
  {481--486} (\bibinfo {year} {2007})}\BibitemShut {NoStop}%
\bibitem [{\citenamefont {Liao}\ \emph {et~al.}(2017)\citenamefont {Liao},
  \citenamefont {Cai}, \citenamefont {Liu}, \citenamefont {Zhang},
  \citenamefont {Li}, \citenamefont {Ren}, \citenamefont {Yin}, \citenamefont
  {Shen}, \citenamefont {Cao}, \citenamefont {Li}, \citenamefont {Li},
  \citenamefont {Chen}, \citenamefont {Sun}, \citenamefont {Jia}, \citenamefont
  {Wu}, \citenamefont {Jiang}, \citenamefont {Wang}, \citenamefont {Huang},
  \citenamefont {Wang}, \citenamefont {Zhou}, \citenamefont {Deng},
  \citenamefont {Xi}, \citenamefont {Ma}, \citenamefont {Hu}, \citenamefont
  {Zhang}, \citenamefont {Chen}, \citenamefont {Liu}, \citenamefont {Wang},
  \citenamefont {Zhu}, \citenamefont {Lu}, \citenamefont {Shu}, \citenamefont
  {Peng}, \citenamefont {Wang},\ and\ \citenamefont {Pan}}]{Liao2017}%
  \BibitemOpen
  \bibfield  {author} {\bibinfo {author} {\bibfnamefont {Sheng-Kai}\
  \bibnamefont {Liao}}, \bibinfo {author} {\bibfnamefont {Wen-Qi}\ \bibnamefont
  {Cai}}, \bibinfo {author} {\bibfnamefont {Wei-Yue}\ \bibnamefont {Liu}},
  \bibinfo {author} {\bibfnamefont {Liang}\ \bibnamefont {Zhang}}, \bibinfo
  {author} {\bibfnamefont {Yang}\ \bibnamefont {Li}}, \bibinfo {author}
  {\bibfnamefont {Ji-Gang}\ \bibnamefont {Ren}}, \bibinfo {author}
  {\bibfnamefont {Juan}\ \bibnamefont {Yin}}, \bibinfo {author} {\bibfnamefont
  {Qi}~\bibnamefont {Shen}}, \bibinfo {author} {\bibfnamefont {Yuan}\
  \bibnamefont {Cao}}, \bibinfo {author} {\bibfnamefont {Zheng-Ping}\
  \bibnamefont {Li}}, \bibinfo {author} {\bibfnamefont {Feng-Zhi}\ \bibnamefont
  {Li}}, \bibinfo {author} {\bibfnamefont {Xia-Wei}\ \bibnamefont {Chen}},
  \bibinfo {author} {\bibfnamefont {Li-Hua}\ \bibnamefont {Sun}}, \bibinfo
  {author} {\bibfnamefont {Jian-Jun}\ \bibnamefont {Jia}}, \bibinfo {author}
  {\bibfnamefont {Jin-Cai}\ \bibnamefont {Wu}}, \bibinfo {author}
  {\bibfnamefont {Xiao-Jun}\ \bibnamefont {Jiang}}, \bibinfo {author}
  {\bibfnamefont {Jian-Feng}\ \bibnamefont {Wang}}, \bibinfo {author}
  {\bibfnamefont {Yong-Mei}\ \bibnamefont {Huang}}, \bibinfo {author}
  {\bibfnamefont {Qiang}\ \bibnamefont {Wang}}, \bibinfo {author}
  {\bibfnamefont {Yi-Lin}\ \bibnamefont {Zhou}}, \bibinfo {author}
  {\bibfnamefont {Lei}\ \bibnamefont {Deng}}, \bibinfo {author} {\bibfnamefont
  {Tao}\ \bibnamefont {Xi}}, \bibinfo {author} {\bibfnamefont {Lu}~\bibnamefont
  {Ma}}, \bibinfo {author} {\bibfnamefont {Tai}\ \bibnamefont {Hu}}, \bibinfo
  {author} {\bibfnamefont {Qiang}\ \bibnamefont {Zhang}}, \bibinfo {author}
  {\bibfnamefont {Yu-Ao}\ \bibnamefont {Chen}}, \bibinfo {author}
  {\bibfnamefont {Nai-Le}\ \bibnamefont {Liu}}, \bibinfo {author}
  {\bibfnamefont {Xiang-Bin}\ \bibnamefont {Wang}}, \bibinfo {author}
  {\bibfnamefont {Zhen-Cai}\ \bibnamefont {Zhu}}, \bibinfo {author}
  {\bibfnamefont {Chao-Yang}\ \bibnamefont {Lu}}, \bibinfo {author}
  {\bibfnamefont {Rong}\ \bibnamefont {Shu}}, \bibinfo {author} {\bibfnamefont
  {Cheng-Zhi}\ \bibnamefont {Peng}}, \bibinfo {author} {\bibfnamefont
  {Jian-Yu}\ \bibnamefont {Wang}}, \ and\ \bibinfo {author} {\bibfnamefont
  {Jian-Wei}\ \bibnamefont {Pan}},\ }\bibfield  {title} {\enquote {\bibinfo
  {title} {Satellite-to-ground quantum key distribution},}\ }\href {\doibase
  10.1038/nature23655} {\bibfield  {journal} {\bibinfo  {journal} {Nature}\
  }\textbf {\bibinfo {volume} {549}},\ \bibinfo {pages} {43--47} (\bibinfo
  {year} {2017})}\BibitemShut {NoStop}%
\bibitem [{\citenamefont {Bennett}\ \emph {et~al.}(1993)\citenamefont
  {Bennett}, \citenamefont {Brassard}, \citenamefont {Cr\'epeau}, \citenamefont
  {Jozsa}, \citenamefont {Peres},\ and\ \citenamefont
  {Wootters}}]{Bennett1993}%
  \BibitemOpen
  \bibfield  {author} {\bibinfo {author} {\bibfnamefont {Charles~H.}\
  \bibnamefont {Bennett}}, \bibinfo {author} {\bibfnamefont {Gilles}\
  \bibnamefont {Brassard}}, \bibinfo {author} {\bibfnamefont {Claude}\
  \bibnamefont {Cr\'epeau}}, \bibinfo {author} {\bibfnamefont {Richard}\
  \bibnamefont {Jozsa}}, \bibinfo {author} {\bibfnamefont {Asher}\ \bibnamefont
  {Peres}}, \ and\ \bibinfo {author} {\bibfnamefont {William~K.}\ \bibnamefont
  {Wootters}},\ }\bibfield  {title} {\enquote {\bibinfo {title} {Teleporting an
  unknown quantum state via dual classical and einstein-podolsky-rosen
  channels},}\ }\href {\doibase 10.1103/PhysRevLett.70.1895} {\bibfield
  {journal} {\bibinfo  {journal} {Phys. Rev. Lett.}\ }\textbf {\bibinfo
  {volume} {70}},\ \bibinfo {pages} {1895--1899} (\bibinfo {year}
  {1993})}\BibitemShut {NoStop}%
\bibitem [{\citenamefont {Bouwmeester}\ \emph {et~al.}(1997)\citenamefont
  {Bouwmeester}, \citenamefont {Pan}, \citenamefont {Mattle}, \citenamefont
  {Eibl}, \citenamefont {Weinfurter},\ and\ \citenamefont
  {Zeilinger}}]{Bouwmeester1997}%
  \BibitemOpen
  \bibfield  {author} {\bibinfo {author} {\bibfnamefont {Dik}\ \bibnamefont
  {Bouwmeester}}, \bibinfo {author} {\bibfnamefont {Jian-Wei}\ \bibnamefont
  {Pan}}, \bibinfo {author} {\bibfnamefont {Klaus}\ \bibnamefont {Mattle}},
  \bibinfo {author} {\bibfnamefont {Manfred}\ \bibnamefont {Eibl}}, \bibinfo
  {author} {\bibfnamefont {Harald}\ \bibnamefont {Weinfurter}}, \ and\ \bibinfo
  {author} {\bibfnamefont {Anton}\ \bibnamefont {Zeilinger}},\ }\bibfield
  {title} {\enquote {\bibinfo {title} {Experimental quantum teleportation},}\
  }\href {\doibase 10.1038/37539} {\bibfield  {journal} {\bibinfo  {journal}
  {Nature}\ }\textbf {\bibinfo {volume} {390}},\ \bibinfo {pages} {575--579}
  (\bibinfo {year} {1997})}\BibitemShut {NoStop}%
\bibitem [{\citenamefont {Ma}\ \emph {et~al.}(2012)\citenamefont {Ma},
  \citenamefont {Herbst}, \citenamefont {Scheidl}, \citenamefont {Wang},
  \citenamefont {Kropatschek}, \citenamefont {Naylor}, \citenamefont
  {Wittmann}, \citenamefont {Mech}, \citenamefont {Kofler}, \citenamefont
  {Anisimova}, \citenamefont {Makarov}, \citenamefont {Jennewein},
  \citenamefont {Ursin},\ and\ \citenamefont {Zeilinger}}]{Ma2012}%
  \BibitemOpen
  \bibfield  {author} {\bibinfo {author} {\bibfnamefont {Xiao-Song}\
  \bibnamefont {Ma}}, \bibinfo {author} {\bibfnamefont {Thomas}\ \bibnamefont
  {Herbst}}, \bibinfo {author} {\bibfnamefont {Thomas}\ \bibnamefont
  {Scheidl}}, \bibinfo {author} {\bibfnamefont {Daqing}\ \bibnamefont {Wang}},
  \bibinfo {author} {\bibfnamefont {Sebastian}\ \bibnamefont {Kropatschek}},
  \bibinfo {author} {\bibfnamefont {William}\ \bibnamefont {Naylor}}, \bibinfo
  {author} {\bibfnamefont {Bernhard}\ \bibnamefont {Wittmann}}, \bibinfo
  {author} {\bibfnamefont {Alexandra}\ \bibnamefont {Mech}}, \bibinfo {author}
  {\bibfnamefont {Johannes}\ \bibnamefont {Kofler}}, \bibinfo {author}
  {\bibfnamefont {Elena}\ \bibnamefont {Anisimova}}, \bibinfo {author}
  {\bibfnamefont {Vadim}\ \bibnamefont {Makarov}}, \bibinfo {author}
  {\bibfnamefont {Thomas}\ \bibnamefont {Jennewein}}, \bibinfo {author}
  {\bibfnamefont {Rupert}\ \bibnamefont {Ursin}}, \ and\ \bibinfo {author}
  {\bibfnamefont {Anton}\ \bibnamefont {Zeilinger}},\ }\bibfield  {title}
  {\enquote {\bibinfo {title} {Quantum teleportation over 143 kilometres using
  active feed-forward},}\ }\href {\doibase 10.1038/nature11472} {\bibfield
  {journal} {\bibinfo  {journal} {Nature}\ }\textbf {\bibinfo {volume} {489}},\
  \bibinfo {pages} {269--273} (\bibinfo {year} {2012})}\BibitemShut {NoStop}%
\bibitem [{\citenamefont {Yin}\ \emph {et~al.}(2012)\citenamefont {Yin},
  \citenamefont {Ren}, \citenamefont {Lu}, \citenamefont {Cao}, \citenamefont
  {Yong}, \citenamefont {Wu}, \citenamefont {Liu}, \citenamefont {Liao},
  \citenamefont {Zhou}, \citenamefont {Jiang}, \citenamefont {Cai},
  \citenamefont {Xu}, \citenamefont {Pan}, \citenamefont {Jia}, \citenamefont
  {Huang}, \citenamefont {Yin}, \citenamefont {Wang}, \citenamefont {Chen},
  \citenamefont {Peng},\ and\ \citenamefont {Pan}}]{Yin2012}%
  \BibitemOpen
  \bibfield  {author} {\bibinfo {author} {\bibfnamefont {Juan}\ \bibnamefont
  {Yin}}, \bibinfo {author} {\bibfnamefont {Ji-Gang}\ \bibnamefont {Ren}},
  \bibinfo {author} {\bibfnamefont {He}~\bibnamefont {Lu}}, \bibinfo {author}
  {\bibfnamefont {Yuan}\ \bibnamefont {Cao}}, \bibinfo {author} {\bibfnamefont
  {Hai-Lin}\ \bibnamefont {Yong}}, \bibinfo {author} {\bibfnamefont {Yu-Ping}\
  \bibnamefont {Wu}}, \bibinfo {author} {\bibfnamefont {Chang}\ \bibnamefont
  {Liu}}, \bibinfo {author} {\bibfnamefont {Sheng-Kai}\ \bibnamefont {Liao}},
  \bibinfo {author} {\bibfnamefont {Fei}\ \bibnamefont {Zhou}}, \bibinfo
  {author} {\bibfnamefont {Yan}\ \bibnamefont {Jiang}}, \bibinfo {author}
  {\bibfnamefont {Xin-Dong}\ \bibnamefont {Cai}}, \bibinfo {author}
  {\bibfnamefont {Ping}\ \bibnamefont {Xu}}, \bibinfo {author} {\bibfnamefont
  {Ge-Sheng}\ \bibnamefont {Pan}}, \bibinfo {author} {\bibfnamefont {Jian-Jun}\
  \bibnamefont {Jia}}, \bibinfo {author} {\bibfnamefont {Yong-Mei}\
  \bibnamefont {Huang}}, \bibinfo {author} {\bibfnamefont {Hao}\ \bibnamefont
  {Yin}}, \bibinfo {author} {\bibfnamefont {Jian-Yu}\ \bibnamefont {Wang}},
  \bibinfo {author} {\bibfnamefont {Yu-Ao}\ \bibnamefont {Chen}}, \bibinfo
  {author} {\bibfnamefont {Cheng-Zhi}\ \bibnamefont {Peng}}, \ and\ \bibinfo
  {author} {\bibfnamefont {Jian-Wei}\ \bibnamefont {Pan}},\ }\bibfield  {title}
  {\enquote {\bibinfo {title} {Quantum teleportation and entanglement
  distribution over 100-kilometre free-space channels},}\ }\href {\doibase
  10.1038/nature11332} {\bibfield  {journal} {\bibinfo  {journal} {Nature}\
  }\textbf {\bibinfo {volume} {488}},\ \bibinfo {pages} {185--188} (\bibinfo
  {year} {2012})}\BibitemShut {NoStop}%
\bibitem [{\citenamefont {Li}\ \emph {et~al.}(2022)\citenamefont {Li},
  \citenamefont {Cao}, \citenamefont {Li}, \citenamefont {Cai}, \citenamefont
  {Liu}, \citenamefont {Ren}, \citenamefont {Liao}, \citenamefont {Wu},
  \citenamefont {Li}, \citenamefont {Li}, \citenamefont {Liu}, \citenamefont
  {Lu}, \citenamefont {Yin}, \citenamefont {Chen}, \citenamefont {Peng},\ and\
  \citenamefont {Pan}}]{Li2022}%
  \BibitemOpen
  \bibfield  {author} {\bibinfo {author} {\bibfnamefont {Bo}~\bibnamefont
  {Li}}, \bibinfo {author} {\bibfnamefont {Yuan}\ \bibnamefont {Cao}}, \bibinfo
  {author} {\bibfnamefont {Yu-Huai}\ \bibnamefont {Li}}, \bibinfo {author}
  {\bibfnamefont {Wen-Qi}\ \bibnamefont {Cai}}, \bibinfo {author}
  {\bibfnamefont {Wei-Yue}\ \bibnamefont {Liu}}, \bibinfo {author}
  {\bibfnamefont {Ji-Gang}\ \bibnamefont {Ren}}, \bibinfo {author}
  {\bibfnamefont {Sheng-Kai}\ \bibnamefont {Liao}}, \bibinfo {author}
  {\bibfnamefont {Hui-Nan}\ \bibnamefont {Wu}}, \bibinfo {author}
  {\bibfnamefont {Shuang-Lin}\ \bibnamefont {Li}}, \bibinfo {author}
  {\bibfnamefont {Li}~\bibnamefont {Li}}, \bibinfo {author} {\bibfnamefont
  {Nai-Le}\ \bibnamefont {Liu}}, \bibinfo {author} {\bibfnamefont {Chao-Yang}\
  \bibnamefont {Lu}}, \bibinfo {author} {\bibfnamefont {Juan}\ \bibnamefont
  {Yin}}, \bibinfo {author} {\bibfnamefont {Yu-Ao}\ \bibnamefont {Chen}},
  \bibinfo {author} {\bibfnamefont {Cheng-Zhi}\ \bibnamefont {Peng}}, \ and\
  \bibinfo {author} {\bibfnamefont {Jian-Wei}\ \bibnamefont {Pan}},\ }\bibfield
   {title} {\enquote {\bibinfo {title} {Quantum state transfer over 1200 km
  assisted by prior distributed entanglement},}\ }\href {\doibase
  10.1103/PhysRevLett.128.170501} {\bibfield  {journal} {\bibinfo  {journal}
  {Phys. Rev. Lett.}\ }\textbf {\bibinfo {volume} {128}},\ \bibinfo {pages}
  {170501} (\bibinfo {year} {2022})}\BibitemShut {NoStop}%
\bibitem [{\citenamefont {Elzerman}\ \emph {et~al.}(2004)\citenamefont
  {Elzerman}, \citenamefont {Hanson}, \citenamefont {Willems~van Beveren},
  \citenamefont {Witkamp}, \citenamefont {Vandersypen},\ and\ \citenamefont
  {Kouwenhoven}}]{elzerman2004single}%
  \BibitemOpen
  \bibfield  {author} {\bibinfo {author} {\bibfnamefont {JM}~\bibnamefont
  {Elzerman}}, \bibinfo {author} {\bibfnamefont {R}~\bibnamefont {Hanson}},
  \bibinfo {author} {\bibfnamefont {LH}~\bibnamefont {Willems~van Beveren}},
  \bibinfo {author} {\bibfnamefont {B}~\bibnamefont {Witkamp}}, \bibinfo
  {author} {\bibfnamefont {LMK}\ \bibnamefont {Vandersypen}}, \ and\ \bibinfo
  {author} {\bibfnamefont {Leo~P}\ \bibnamefont {Kouwenhoven}},\ }\bibfield
  {title} {\enquote {\bibinfo {title} {Single-shot read-out of an individual
  electron spin in a quantum dot},}\ }\href@noop {} {\bibfield  {journal}
  {\bibinfo  {journal} {nature}\ }\textbf {\bibinfo {volume} {430}},\ \bibinfo
  {pages} {431--435} (\bibinfo {year} {2004})}\BibitemShut {NoStop}%
\bibitem [{\citenamefont {Atature}\ \emph {et~al.}(2006)\citenamefont
  {Atature}, \citenamefont {Dreiser}, \citenamefont {Badolato}, \citenamefont
  {Hogele}, \citenamefont {Karrai},\ and\ \citenamefont
  {Imamoglu}}]{atature2006quantum}%
  \BibitemOpen
  \bibfield  {author} {\bibinfo {author} {\bibfnamefont {Mete}\ \bibnamefont
  {Atature}}, \bibinfo {author} {\bibfnamefont {Jan}\ \bibnamefont {Dreiser}},
  \bibinfo {author} {\bibfnamefont {Antonio}\ \bibnamefont {Badolato}},
  \bibinfo {author} {\bibfnamefont {Alexander}\ \bibnamefont {Hogele}},
  \bibinfo {author} {\bibfnamefont {Khaled}\ \bibnamefont {Karrai}}, \ and\
  \bibinfo {author} {\bibfnamefont {Atac}\ \bibnamefont {Imamoglu}},\
  }\bibfield  {title} {\enquote {\bibinfo {title} {Quantum-dot spin-state
  preparation with near-unity fidelity},}\ }\href@noop {} {\bibfield  {journal}
  {\bibinfo  {journal} {Science}\ }\textbf {\bibinfo {volume} {312}},\ \bibinfo
  {pages} {551--553} (\bibinfo {year} {2006})}\BibitemShut {NoStop}%
\bibitem [{\citenamefont {Ramsay}\ \emph {et~al.}(2008)\citenamefont {Ramsay},
  \citenamefont {Boyle}, \citenamefont {Kolodka}, \citenamefont {Oliveira},
  \citenamefont {Skiba-Szymanska}, \citenamefont {Liu}, \citenamefont
  {Hopkinson}, \citenamefont {Fox},\ and\ \citenamefont
  {Skolnick}}]{ramsay2008fast}%
  \BibitemOpen
  \bibfield  {author} {\bibinfo {author} {\bibfnamefont {AJ}~\bibnamefont
  {Ramsay}}, \bibinfo {author} {\bibfnamefont {SJ}~\bibnamefont {Boyle}},
  \bibinfo {author} {\bibfnamefont {RS}~\bibnamefont {Kolodka}}, \bibinfo
  {author} {\bibfnamefont {Jos{\'e} Br{\'a}s Barreto~de}\ \bibnamefont
  {Oliveira}}, \bibinfo {author} {\bibfnamefont {J}~\bibnamefont
  {Skiba-Szymanska}}, \bibinfo {author} {\bibfnamefont {HY}~\bibnamefont
  {Liu}}, \bibinfo {author} {\bibfnamefont {M}~\bibnamefont {Hopkinson}},
  \bibinfo {author} {\bibfnamefont {AM}~\bibnamefont {Fox}}, \ and\ \bibinfo
  {author} {\bibfnamefont {MS}~\bibnamefont {Skolnick}},\ }\bibfield  {title}
  {\enquote {\bibinfo {title} {Fast optical preparation, control, and readout
  of a single quantum dot spin},}\ }\href@noop {} {\bibfield  {journal}
  {\bibinfo  {journal} {Physical review letters}\ }\textbf {\bibinfo {volume}
  {100}},\ \bibinfo {pages} {197401} (\bibinfo {year} {2008})}\BibitemShut
  {NoStop}%
\bibitem [{\citenamefont {Robledo}\ \emph {et~al.}(2011)\citenamefont
  {Robledo}, \citenamefont {Childress}, \citenamefont {Bernien}, \citenamefont
  {Hensen}, \citenamefont {Alkemade},\ and\ \citenamefont
  {Hanson}}]{robledo2011high}%
  \BibitemOpen
  \bibfield  {author} {\bibinfo {author} {\bibfnamefont {Lucio}\ \bibnamefont
  {Robledo}}, \bibinfo {author} {\bibfnamefont {Lilian}\ \bibnamefont
  {Childress}}, \bibinfo {author} {\bibfnamefont {Hannes}\ \bibnamefont
  {Bernien}}, \bibinfo {author} {\bibfnamefont {Bas}\ \bibnamefont {Hensen}},
  \bibinfo {author} {\bibfnamefont {Paul~FA}\ \bibnamefont {Alkemade}}, \ and\
  \bibinfo {author} {\bibfnamefont {Ronald}\ \bibnamefont {Hanson}},\
  }\bibfield  {title} {\enquote {\bibinfo {title} {High-fidelity projective
  read-out of a solid-state spin quantum register},}\ }\href@noop {} {\bibfield
   {journal} {\bibinfo  {journal} {Nature}\ }\textbf {\bibinfo {volume}
  {477}},\ \bibinfo {pages} {574--578} (\bibinfo {year} {2011})}\BibitemShut
  {NoStop}%
\bibitem [{\citenamefont {Javadi}\ \emph {et~al.}(2018)\citenamefont {Javadi},
  \citenamefont {Ding}, \citenamefont {Appel}, \citenamefont {Mahmoodian},
  \citenamefont {L{\"o}bl}, \citenamefont {S{\"o}llner}, \citenamefont
  {Schott}, \citenamefont {Papon}, \citenamefont {Pregnolato}, \citenamefont
  {Stobbe} \emph {et~al.}}]{javadi2018spin}%
  \BibitemOpen
  \bibfield  {author} {\bibinfo {author} {\bibfnamefont {Alisa}\ \bibnamefont
  {Javadi}}, \bibinfo {author} {\bibfnamefont {Dapeng}\ \bibnamefont {Ding}},
  \bibinfo {author} {\bibfnamefont {Martin~Hayhurst}\ \bibnamefont {Appel}},
  \bibinfo {author} {\bibfnamefont {Sahand}\ \bibnamefont {Mahmoodian}},
  \bibinfo {author} {\bibfnamefont {Matthias~Christian}\ \bibnamefont
  {L{\"o}bl}}, \bibinfo {author} {\bibfnamefont {Immo}\ \bibnamefont
  {S{\"o}llner}}, \bibinfo {author} {\bibfnamefont {R{\"u}diger}\ \bibnamefont
  {Schott}}, \bibinfo {author} {\bibfnamefont {Camille}\ \bibnamefont {Papon}},
  \bibinfo {author} {\bibfnamefont {Tommaso}\ \bibnamefont {Pregnolato}},
  \bibinfo {author} {\bibfnamefont {S{\o}ren}\ \bibnamefont {Stobbe}},  \emph
  {et~al.},\ }\bibfield  {title} {\enquote {\bibinfo {title} {Spin--photon
  interface and spin-controlled photon switching in a nanobeam waveguide},}\
  }\href@noop {} {\bibfield  {journal} {\bibinfo  {journal} {Nature
  nanotechnology}\ }\textbf {\bibinfo {volume} {13}},\ \bibinfo {pages}
  {398--403} (\bibinfo {year} {2018})}\BibitemShut {NoStop}%
\bibitem [{\citenamefont {Bartolucci}\ \emph {et~al.}(2021)\citenamefont
  {Bartolucci}, \citenamefont {Birchall}, \citenamefont {Bombin}, \citenamefont
  {Cable}, \citenamefont {Dawson}, \citenamefont {Gimeno-Segovia},
  \citenamefont {Johnston}, \citenamefont {Kieling}, \citenamefont {Nickerson},
  \citenamefont {Pant}, \citenamefont {Pastawski}, \citenamefont {Rudolph},\
  and\ \citenamefont {Sparrow}}]{Bartolucci2021}%
  \BibitemOpen
  \bibfield  {author} {\bibinfo {author} {\bibfnamefont {Sara}\ \bibnamefont
  {Bartolucci}}, \bibinfo {author} {\bibfnamefont {Patrick}\ \bibnamefont
  {Birchall}}, \bibinfo {author} {\bibfnamefont {Hector}\ \bibnamefont
  {Bombin}}, \bibinfo {author} {\bibfnamefont {Hugo}\ \bibnamefont {Cable}},
  \bibinfo {author} {\bibfnamefont {Chris}\ \bibnamefont {Dawson}}, \bibinfo
  {author} {\bibfnamefont {Mercedes}\ \bibnamefont {Gimeno-Segovia}}, \bibinfo
  {author} {\bibfnamefont {Eric}\ \bibnamefont {Johnston}}, \bibinfo {author}
  {\bibfnamefont {Konrad}\ \bibnamefont {Kieling}}, \bibinfo {author}
  {\bibfnamefont {Naomi}\ \bibnamefont {Nickerson}}, \bibinfo {author}
  {\bibfnamefont {Mihir}\ \bibnamefont {Pant}}, \bibinfo {author}
  {\bibfnamefont {Fernando}\ \bibnamefont {Pastawski}}, \bibinfo {author}
  {\bibfnamefont {Terry}\ \bibnamefont {Rudolph}}, \ and\ \bibinfo {author}
  {\bibfnamefont {Chris}\ \bibnamefont {Sparrow}},\ }\href {\doibase
  10.48550/arxiv.2101.09310} {\enquote {\bibinfo {title} {Fusion-based quantum
  computation},}\ } (\bibinfo {year} {2021})\BibitemShut {NoStop}%
\bibitem [{\citenamefont {Knill}\ \emph {et~al.}(2001)\citenamefont {Knill},
  \citenamefont {Laflamme},\ and\ \citenamefont {Milburn}}]{Knill2001}%
  \BibitemOpen
  \bibfield  {author} {\bibinfo {author} {\bibfnamefont {E.}~\bibnamefont
  {Knill}}, \bibinfo {author} {\bibfnamefont {R.}~\bibnamefont {Laflamme}}, \
  and\ \bibinfo {author} {\bibfnamefont {G.~J.}\ \bibnamefont {Milburn}},\
  }\bibfield  {title} {\enquote {\bibinfo {title} {A scheme for efficient
  quantum computation with linear optics},}\ }\href {\doibase 10.1038/35051009}
  {\bibfield  {journal} {\bibinfo  {journal} {Nature}\ }\textbf {\bibinfo
  {volume} {409}},\ \bibinfo {pages} {46--52} (\bibinfo {year}
  {2001})}\BibitemShut {NoStop}%
\bibitem [{\citenamefont {Ralph}\ \emph {et~al.}(2001)\citenamefont {Ralph},
  \citenamefont {White}, \citenamefont {Munro},\ and\ \citenamefont
  {Milburn}}]{Ralph2002}%
  \BibitemOpen
  \bibfield  {author} {\bibinfo {author} {\bibfnamefont {T.~C.}\ \bibnamefont
  {Ralph}}, \bibinfo {author} {\bibfnamefont {A.~G.}\ \bibnamefont {White}},
  \bibinfo {author} {\bibfnamefont {W.~J.}\ \bibnamefont {Munro}}, \ and\
  \bibinfo {author} {\bibfnamefont {G.~J.}\ \bibnamefont {Milburn}},\
  }\bibfield  {title} {\enquote {\bibinfo {title} {Simple scheme for efficient
  linear optics quantum gates},}\ }\href {\doibase 10.1103/PhysRevA.65.012314}
  {\bibfield  {journal} {\bibinfo  {journal} {Phys. Rev. A}\ }\textbf {\bibinfo
  {volume} {65}},\ \bibinfo {pages} {012314} (\bibinfo {year}
  {2001})}\BibitemShut {NoStop}%
\bibitem [{\citenamefont {O'Brien}\ \emph {et~al.}(2003)\citenamefont
  {O'Brien}, \citenamefont {Pryde}, \citenamefont {White}, \citenamefont
  {Ralph},\ and\ \citenamefont {Branning}}]{OBrien2003}%
  \BibitemOpen
  \bibfield  {author} {\bibinfo {author} {\bibfnamefont {J.~L.}\ \bibnamefont
  {O'Brien}}, \bibinfo {author} {\bibfnamefont {G.~J.}\ \bibnamefont {Pryde}},
  \bibinfo {author} {\bibfnamefont {A.~G.}\ \bibnamefont {White}}, \bibinfo
  {author} {\bibfnamefont {T.~C.}\ \bibnamefont {Ralph}}, \ and\ \bibinfo
  {author} {\bibfnamefont {D.}~\bibnamefont {Branning}},\ }\bibfield  {title}
  {\enquote {\bibinfo {title} {Demonstration of an all-optical quantum
  controlled-not gate},}\ }\href {\doibase 10.1038/nature02054} {\bibfield
  {journal} {\bibinfo  {journal} {Nature}\ }\textbf {\bibinfo {volume} {426}},\
  \bibinfo {pages} {264--267} (\bibinfo {year} {2003})}\BibitemShut {NoStop}%
\bibitem [{\citenamefont {Knill}(2002)}]{Knill2002}%
  \BibitemOpen
  \bibfield  {author} {\bibinfo {author} {\bibfnamefont {E.}~\bibnamefont
  {Knill}},\ }\bibfield  {title} {\enquote {\bibinfo {title} {Quantum gates
  using linear optics and postselection},}\ }\href {\doibase
  10.1103/PhysRevA.66.052306} {\bibfield  {journal} {\bibinfo  {journal} {Phys.
  Rev. A}\ }\textbf {\bibinfo {volume} {66}},\ \bibinfo {pages} {052306}
  (\bibinfo {year} {2002})}\BibitemShut {NoStop}%
\bibitem [{\citenamefont {Pittman}\ \emph {et~al.}(2003)\citenamefont
  {Pittman}, \citenamefont {Fitch}, \citenamefont {Jacobs},\ and\ \citenamefont
  {Franson}}]{Pittman2003}%
  \BibitemOpen
  \bibfield  {author} {\bibinfo {author} {\bibfnamefont {T.~B.}\ \bibnamefont
  {Pittman}}, \bibinfo {author} {\bibfnamefont {M.~J.}\ \bibnamefont {Fitch}},
  \bibinfo {author} {\bibfnamefont {B.~C}\ \bibnamefont {Jacobs}}, \ and\
  \bibinfo {author} {\bibfnamefont {J.~D.}\ \bibnamefont {Franson}},\
  }\bibfield  {title} {\enquote {\bibinfo {title} {Experimental controlled-not
  logic gate for single photons in the coincidence basis},}\ }\href {\doibase
  10.1103/PhysRevA.68.032316} {\bibfield  {journal} {\bibinfo  {journal} {Phys.
  Rev. A}\ }\textbf {\bibinfo {volume} {68}},\ \bibinfo {pages} {032316}
  (\bibinfo {year} {2003})}\BibitemShut {NoStop}%
\bibitem [{\citenamefont {Gasparoni}\ \emph {et~al.}(2004)\citenamefont
  {Gasparoni}, \citenamefont {Pan}, \citenamefont {Walther}, \citenamefont
  {Rudolph},\ and\ \citenamefont {Zeilinger}}]{Gasparoni2004}%
  \BibitemOpen
  \bibfield  {author} {\bibinfo {author} {\bibfnamefont {Sara}\ \bibnamefont
  {Gasparoni}}, \bibinfo {author} {\bibfnamefont {Jian-Wei}\ \bibnamefont
  {Pan}}, \bibinfo {author} {\bibfnamefont {Philip}\ \bibnamefont {Walther}},
  \bibinfo {author} {\bibfnamefont {Terry}\ \bibnamefont {Rudolph}}, \ and\
  \bibinfo {author} {\bibfnamefont {Anton}\ \bibnamefont {Zeilinger}},\
  }\bibfield  {title} {\enquote {\bibinfo {title} {Realization of a photonic
  controlled-not gate sufficient for quantum computation},}\ }\href {\doibase
  10.1103/PhysRevLett.93.020504} {\bibfield  {journal} {\bibinfo  {journal}
  {Phys. Rev. Lett.}\ }\textbf {\bibinfo {volume} {93}},\ \bibinfo {pages}
  {020504} (\bibinfo {year} {2004})}\BibitemShut {NoStop}%
\bibitem [{\citenamefont {Zhao}\ \emph {et~al.}(2005)\citenamefont {Zhao},
  \citenamefont {Zhang}, \citenamefont {Chen}, \citenamefont {Zhang},
  \citenamefont {Du}, \citenamefont {Yang},\ and\ \citenamefont
  {Pan}}]{Zhao2005}%
  \BibitemOpen
  \bibfield  {author} {\bibinfo {author} {\bibfnamefont {Zhi}\ \bibnamefont
  {Zhao}}, \bibinfo {author} {\bibfnamefont {An-Ning}\ \bibnamefont {Zhang}},
  \bibinfo {author} {\bibfnamefont {Yu-Ao}\ \bibnamefont {Chen}}, \bibinfo
  {author} {\bibfnamefont {Han}\ \bibnamefont {Zhang}}, \bibinfo {author}
  {\bibfnamefont {Jiang-Feng}\ \bibnamefont {Du}}, \bibinfo {author}
  {\bibfnamefont {Tao}\ \bibnamefont {Yang}}, \ and\ \bibinfo {author}
  {\bibfnamefont {Jian-Wei}\ \bibnamefont {Pan}},\ }\bibfield  {title}
  {\enquote {\bibinfo {title} {Experimental demonstration of a nondestructive
  controlled-not quantum gate for two independent photon qubits},}\ }\href
  {\doibase 10.1103/PhysRevLett.94.030501} {\bibfield  {journal} {\bibinfo
  {journal} {Phys. Rev. Lett.}\ }\textbf {\bibinfo {volume} {94}},\ \bibinfo
  {pages} {030501} (\bibinfo {year} {2005})}\BibitemShut {NoStop}%
\bibitem [{\citenamefont {Bao}\ \emph {et~al.}(2007)\citenamefont {Bao},
  \citenamefont {Chen}, \citenamefont {Zhang}, \citenamefont {Yang},
  \citenamefont {Zhang}, \citenamefont {Yang},\ and\ \citenamefont
  {Pan}}]{Bao2007}%
  \BibitemOpen
  \bibfield  {author} {\bibinfo {author} {\bibfnamefont {Xiao-Hui}\
  \bibnamefont {Bao}}, \bibinfo {author} {\bibfnamefont {Teng-Yun}\
  \bibnamefont {Chen}}, \bibinfo {author} {\bibfnamefont {Qiang}\ \bibnamefont
  {Zhang}}, \bibinfo {author} {\bibfnamefont {Jian}\ \bibnamefont {Yang}},
  \bibinfo {author} {\bibfnamefont {Han}\ \bibnamefont {Zhang}}, \bibinfo
  {author} {\bibfnamefont {Tao}\ \bibnamefont {Yang}}, \ and\ \bibinfo {author}
  {\bibfnamefont {Jian-Wei}\ \bibnamefont {Pan}},\ }\bibfield  {title}
  {\enquote {\bibinfo {title} {Optical nondestructive controlled-not gate
  without using entangled photons},}\ }\href {\doibase
  10.1103/PhysRevLett.98.170502} {\bibfield  {journal} {\bibinfo  {journal}
  {Phys. Rev. Lett.}\ }\textbf {\bibinfo {volume} {98}},\ \bibinfo {pages}
  {170502} (\bibinfo {year} {2007})}\BibitemShut {NoStop}%
\bibitem [{\citenamefont {Zeuner}\ \emph {et~al.}(2018)\citenamefont {Zeuner},
  \citenamefont {Sharma}, \citenamefont {Tillmann}, \citenamefont {Heilmann},
  \citenamefont {Gr{\"a}fe}, \citenamefont {Moqanaki}, \citenamefont
  {Szameit},\ and\ \citenamefont {Walther}}]{Zeuner2018}%
  \BibitemOpen
  \bibfield  {author} {\bibinfo {author} {\bibfnamefont {Jonas}\ \bibnamefont
  {Zeuner}}, \bibinfo {author} {\bibfnamefont {Aditya~N.}\ \bibnamefont
  {Sharma}}, \bibinfo {author} {\bibfnamefont {Max}\ \bibnamefont {Tillmann}},
  \bibinfo {author} {\bibfnamefont {Ren{\'e}}\ \bibnamefont {Heilmann}},
  \bibinfo {author} {\bibfnamefont {Markus}\ \bibnamefont {Gr{\"a}fe}},
  \bibinfo {author} {\bibfnamefont {Amir}\ \bibnamefont {Moqanaki}}, \bibinfo
  {author} {\bibfnamefont {Alexander}\ \bibnamefont {Szameit}}, \ and\ \bibinfo
  {author} {\bibfnamefont {Philip}\ \bibnamefont {Walther}},\ }\bibfield
  {title} {\enquote {\bibinfo {title} {Integrated-optics heralded
  controlled-not gate for polarization-encoded qubits},}\ }\href {\doibase
  10.1038/s41534-018-0068-0} {\bibfield  {journal} {\bibinfo  {journal} {npj
  Quantum Information}\ }\textbf {\bibinfo {volume} {4}},\ \bibinfo {pages}
  {13} (\bibinfo {year} {2018})}\BibitemShut {NoStop}%
\bibitem [{\citenamefont {Li}\ \emph {et~al.}(2021)\citenamefont {Li},
  \citenamefont {Gu}, \citenamefont {Qin}, \citenamefont {Wu}, \citenamefont
  {You}, \citenamefont {Wang}, \citenamefont {Schneider}, \citenamefont
  {H\"ofling}, \citenamefont {Huo}, \citenamefont {Lu}, \citenamefont {Liu},
  \citenamefont {Li},\ and\ \citenamefont {Pan}}]{Li2021}%
  \BibitemOpen
  \bibfield  {author} {\bibinfo {author} {\bibfnamefont {Jin-Peng}\
  \bibnamefont {Li}}, \bibinfo {author} {\bibfnamefont {Xuemei}\ \bibnamefont
  {Gu}}, \bibinfo {author} {\bibfnamefont {Jian}\ \bibnamefont {Qin}}, \bibinfo
  {author} {\bibfnamefont {Dian}\ \bibnamefont {Wu}}, \bibinfo {author}
  {\bibfnamefont {Xiang}\ \bibnamefont {You}}, \bibinfo {author} {\bibfnamefont
  {Hui}\ \bibnamefont {Wang}}, \bibinfo {author} {\bibfnamefont {Christian}\
  \bibnamefont {Schneider}}, \bibinfo {author} {\bibfnamefont {Sven}\
  \bibnamefont {H\"ofling}}, \bibinfo {author} {\bibfnamefont {Yong-Heng}\
  \bibnamefont {Huo}}, \bibinfo {author} {\bibfnamefont {Chao-Yang}\
  \bibnamefont {Lu}}, \bibinfo {author} {\bibfnamefont {Nai-Le}\ \bibnamefont
  {Liu}}, \bibinfo {author} {\bibfnamefont {Li}~\bibnamefont {Li}}, \ and\
  \bibinfo {author} {\bibfnamefont {Jian-Wei}\ \bibnamefont {Pan}},\ }\bibfield
   {title} {\enquote {\bibinfo {title} {Heralded nondestructive quantum
  entangling gate with single-photon sources},}\ }\href {\doibase
  10.1103/PhysRevLett.126.140501} {\bibfield  {journal} {\bibinfo  {journal}
  {Phys. Rev. Lett.}\ }\textbf {\bibinfo {volume} {126}},\ \bibinfo {pages}
  {140501} (\bibinfo {year} {2021})}\BibitemShut {NoStop}%
\bibitem [{\citenamefont {Xia}\ \emph {et~al.}(2011)\citenamefont {Xia},
  \citenamefont {Song}, \citenamefont {Lu},\ and\ \citenamefont
  {Song}}]{xia2011efficient}%
  \BibitemOpen
  \bibfield  {author} {\bibinfo {author} {\bibfnamefont {Yan}\ \bibnamefont
  {Xia}}, \bibinfo {author} {\bibfnamefont {Jie}\ \bibnamefont {Song}},
  \bibinfo {author} {\bibfnamefont {Pei-Min}\ \bibnamefont {Lu}}, \ and\
  \bibinfo {author} {\bibfnamefont {He-Shan}\ \bibnamefont {Song}},\ }\bibfield
   {title} {\enquote {\bibinfo {title} {Efficient implementation of the
  two-qubit controlled phase gate with cross-kerr nonlinearity},}\ }\href@noop
  {} {\bibfield  {journal} {\bibinfo  {journal} {Journal of Physics B: Atomic,
  Molecular and Optical Physics}\ }\textbf {\bibinfo {volume} {44}},\ \bibinfo
  {pages} {025503} (\bibinfo {year} {2011})}\BibitemShut {NoStop}%
\bibitem [{\citenamefont {Hacker}\ \emph {et~al.}(2016)\citenamefont {Hacker},
  \citenamefont {Welte}, \citenamefont {Rempe},\ and\ \citenamefont
  {Ritter}}]{Hacker2016}%
  \BibitemOpen
  \bibfield  {author} {\bibinfo {author} {\bibfnamefont {Bastian}\ \bibnamefont
  {Hacker}}, \bibinfo {author} {\bibfnamefont {Stephan}\ \bibnamefont {Welte}},
  \bibinfo {author} {\bibfnamefont {Gerhard}\ \bibnamefont {Rempe}}, \ and\
  \bibinfo {author} {\bibfnamefont {Stephan}\ \bibnamefont {Ritter}},\
  }\bibfield  {title} {\enquote {\bibinfo {title} {A photon--photon quantum
  gate based on a single atom in an optical resonator},}\ }\href {\doibase
  10.1038/nature18592} {\bibfield  {journal} {\bibinfo  {journal} {Nature}\
  }\textbf {\bibinfo {volume} {536}},\ \bibinfo {pages} {193--196} (\bibinfo
  {year} {2016})}\BibitemShut {NoStop}%
\bibitem [{\citenamefont {Reuer}\ \emph {et~al.}(2022)\citenamefont {Reuer},
  \citenamefont {Besse}, \citenamefont {Wernli}, \citenamefont {Magnard},
  \citenamefont {Kurpiers}, \citenamefont {Norris}, \citenamefont {Wallraff},\
  and\ \citenamefont {Eichler}}]{reuer2022realization}%
  \BibitemOpen
  \bibfield  {author} {\bibinfo {author} {\bibfnamefont {Kevin}\ \bibnamefont
  {Reuer}}, \bibinfo {author} {\bibfnamefont {Jean-Claude}\ \bibnamefont
  {Besse}}, \bibinfo {author} {\bibfnamefont {Lucien}\ \bibnamefont {Wernli}},
  \bibinfo {author} {\bibfnamefont {Paul}\ \bibnamefont {Magnard}}, \bibinfo
  {author} {\bibfnamefont {Philipp}\ \bibnamefont {Kurpiers}}, \bibinfo
  {author} {\bibfnamefont {Graham~J}\ \bibnamefont {Norris}}, \bibinfo {author}
  {\bibfnamefont {Andreas}\ \bibnamefont {Wallraff}}, \ and\ \bibinfo {author}
  {\bibfnamefont {Christopher}\ \bibnamefont {Eichler}},\ }\bibfield  {title}
  {\enquote {\bibinfo {title} {Realization of a universal quantum gate set for
  itinerant microwave photons},}\ }\href@noop {} {\bibfield  {journal}
  {\bibinfo  {journal} {Physical Review X}\ }\textbf {\bibinfo {volume} {12}},\
  \bibinfo {pages} {011008} (\bibinfo {year} {2022})}\BibitemShut {NoStop}%
\bibitem [{\citenamefont {Koashi}\ \emph {et~al.}(2001)\citenamefont {Koashi},
  \citenamefont {Yamamoto},\ and\ \citenamefont {Imoto}}]{Koashi2001}%
  \BibitemOpen
  \bibfield  {author} {\bibinfo {author} {\bibfnamefont {Masato}\ \bibnamefont
  {Koashi}}, \bibinfo {author} {\bibfnamefont {Takashi}\ \bibnamefont
  {Yamamoto}}, \ and\ \bibinfo {author} {\bibfnamefont {Nobuyuki}\ \bibnamefont
  {Imoto}},\ }\bibfield  {title} {\enquote {\bibinfo {title} {Probabilistic
  manipulation of entangled photons},}\ }\href {\doibase
  10.1103/PhysRevA.63.030301} {\bibfield  {journal} {\bibinfo  {journal} {Phys.
  Rev. A}\ }\textbf {\bibinfo {volume} {63}},\ \bibinfo {pages} {030301}
  (\bibinfo {year} {2001})}\BibitemShut {NoStop}%
\bibitem [{\citenamefont {Pittman}\ \emph {et~al.}(2001)\citenamefont
  {Pittman}, \citenamefont {Jacobs},\ and\ \citenamefont
  {Franson}}]{Pittman2001}%
  \BibitemOpen
  \bibfield  {author} {\bibinfo {author} {\bibfnamefont {T.~B.}\ \bibnamefont
  {Pittman}}, \bibinfo {author} {\bibfnamefont {B.~C.}\ \bibnamefont {Jacobs}},
  \ and\ \bibinfo {author} {\bibfnamefont {J.~D.}\ \bibnamefont {Franson}},\
  }\bibfield  {title} {\enquote {\bibinfo {title} {Probabilistic quantum logic
  operations using polarizing beam splitters},}\ }\href {\doibase
  10.1103/PhysRevA.64.062311} {\bibfield  {journal} {\bibinfo  {journal} {Phys.
  Rev. A}\ }\textbf {\bibinfo {volume} {64}},\ \bibinfo {pages} {062311}
  (\bibinfo {year} {2001})}\BibitemShut {NoStop}%
\bibitem [{\citenamefont {Nemoto}\ and\ \citenamefont
  {Munro}(2004)}]{Nemoto2004}%
  \BibitemOpen
  \bibfield  {author} {\bibinfo {author} {\bibfnamefont {Kae}\ \bibnamefont
  {Nemoto}}\ and\ \bibinfo {author} {\bibfnamefont {W.~J.}\ \bibnamefont
  {Munro}},\ }\bibfield  {title} {\enquote {\bibinfo {title} {Nearly
  deterministic linear optical controlled-not gate},}\ }\href {\doibase
  10.1103/PhysRevLett.93.250502} {\bibfield  {journal} {\bibinfo  {journal}
  {Phys. Rev. Lett.}\ }\textbf {\bibinfo {volume} {93}},\ \bibinfo {pages}
  {250502} (\bibinfo {year} {2004})}\BibitemShut {NoStop}%
\bibitem [{\citenamefont {Kok}\ \emph {et~al.}(2007)\citenamefont {Kok},
  \citenamefont {Munro}, \citenamefont {Nemoto}, \citenamefont {Ralph},
  \citenamefont {Dowling},\ and\ \citenamefont {Milburn}}]{Kok2007}%
  \BibitemOpen
  \bibfield  {author} {\bibinfo {author} {\bibfnamefont {Pieter}\ \bibnamefont
  {Kok}}, \bibinfo {author} {\bibfnamefont {W.~J.}\ \bibnamefont {Munro}},
  \bibinfo {author} {\bibfnamefont {Kae}\ \bibnamefont {Nemoto}}, \bibinfo
  {author} {\bibfnamefont {T.~C.}\ \bibnamefont {Ralph}}, \bibinfo {author}
  {\bibfnamefont {Jonathan~P.}\ \bibnamefont {Dowling}}, \ and\ \bibinfo
  {author} {\bibfnamefont {G.~J.}\ \bibnamefont {Milburn}},\ }\bibfield
  {title} {\enquote {\bibinfo {title} {Linear optical quantum computing with
  photonic qubits},}\ }\href {\doibase 10.1103/RevModPhys.79.135} {\bibfield
  {journal} {\bibinfo  {journal} {Rev. Mod. Phys.}\ }\textbf {\bibinfo {volume}
  {79}},\ \bibinfo {pages} {135--174} (\bibinfo {year} {2007})}\BibitemShut
  {NoStop}%
\bibitem [{\citenamefont {Ionicioiu}\ \emph {et~al.}(2008)\citenamefont
  {Ionicioiu}, \citenamefont {Popescu}, \citenamefont {Munro},\ and\
  \citenamefont {Spiller}}]{Ionicioiu2008}%
  \BibitemOpen
  \bibfield  {author} {\bibinfo {author} {\bibfnamefont {Radu}\ \bibnamefont
  {Ionicioiu}}, \bibinfo {author} {\bibfnamefont {Anca~E}\ \bibnamefont
  {Popescu}}, \bibinfo {author} {\bibfnamefont {William~J}\ \bibnamefont
  {Munro}}, \ and\ \bibinfo {author} {\bibfnamefont {Timothy~P}\ \bibnamefont
  {Spiller}},\ }\bibfield  {title} {\enquote {\bibinfo {title} {Generalized
  parity measurements},}\ }\href@noop {} {\bibfield  {journal} {\bibinfo
  {journal} {Physical Review A}\ }\textbf {\bibinfo {volume} {78}},\ \bibinfo
  {pages} {052326} (\bibinfo {year} {2008})}\BibitemShut {NoStop}%
\bibitem [{\citenamefont {Wang}\ \emph
  {et~al.}(2012{\natexlab{a}})\citenamefont {Wang}, \citenamefont {Zhang},
  \citenamefont {Tang}, \citenamefont {Xie}, \citenamefont {Wang},\ and\
  \citenamefont {Kuang}}]{Wang2012}%
  \BibitemOpen
  \bibfield  {author} {\bibinfo {author} {\bibfnamefont {Xin-Wen}\ \bibnamefont
  {Wang}}, \bibinfo {author} {\bibfnamefont {Deng-Yu}\ \bibnamefont {Zhang}},
  \bibinfo {author} {\bibfnamefont {Shi-Qing}\ \bibnamefont {Tang}}, \bibinfo
  {author} {\bibfnamefont {Li-Jun}\ \bibnamefont {Xie}}, \bibinfo {author}
  {\bibfnamefont {Zhi-Yong}\ \bibnamefont {Wang}}, \ and\ \bibinfo {author}
  {\bibfnamefont {Le-Man}\ \bibnamefont {Kuang}},\ }\bibfield  {title}
  {\enquote {\bibinfo {title} {Photonic two-qubit parity gate with tiny
  cross--kerr nonlinearity},}\ }\href@noop {} {\bibfield  {journal} {\bibinfo
  {journal} {Physical Review A}\ }\textbf {\bibinfo {volume} {85}},\ \bibinfo
  {pages} {052326} (\bibinfo {year} {2012}{\natexlab{a}})}\BibitemShut
  {NoStop}%
\bibitem [{\citenamefont {Wei}\ and\ \citenamefont {Deng}(2013)}]{Wei2013}%
  \BibitemOpen
  \bibfield  {author} {\bibinfo {author} {\bibfnamefont {Hai-Rui}\ \bibnamefont
  {Wei}}\ and\ \bibinfo {author} {\bibfnamefont {Fu-Guo}\ \bibnamefont
  {Deng}},\ }\bibfield  {title} {\enquote {\bibinfo {title} {Scalable photonic
  quantum computing assisted by quantum-dot spin in double-sided optical
  microcavity},}\ }\href {\doibase 10.1364/OE.21.017671} {\bibfield  {journal}
  {\bibinfo  {journal} {Opt. Express}\ }\textbf {\bibinfo {volume} {21}},\
  \bibinfo {pages} {17671--17685} (\bibinfo {year} {2013})}\BibitemShut
  {NoStop}%
\bibitem [{\citenamefont {Wei}\ and\ \citenamefont {Deng}(2014)}]{Wei2014}%
  \BibitemOpen
  \bibfield  {author} {\bibinfo {author} {\bibfnamefont {Hai-Rui}\ \bibnamefont
  {Wei}}\ and\ \bibinfo {author} {\bibfnamefont {Fu-Guo}\ \bibnamefont
  {Deng}},\ }\bibfield  {title} {\enquote {\bibinfo {title} {Universal quantum
  gates on electron-spin qubits with quantum dots inside single-side optical
  microcavities},}\ }\href {\doibase 10.1364/OE.22.000593} {\bibfield
  {journal} {\bibinfo  {journal} {Opt. Express}\ }\textbf {\bibinfo {volume}
  {22}},\ \bibinfo {pages} {593--607} (\bibinfo {year} {2014})}\BibitemShut
  {NoStop}%
\bibitem [{\citenamefont {Song}\ \emph {et~al.}(2005)\citenamefont {Song},
  \citenamefont {Feng}, \citenamefont {Kwek},\ and\ \citenamefont
  {Oh}}]{Song2005}%
  \BibitemOpen
  \bibfield  {author} {\bibinfo {author} {\bibfnamefont {Xin-Guo}\ \bibnamefont
  {Song}}, \bibinfo {author} {\bibfnamefont {Xun-Li}\ \bibnamefont {Feng}},
  \bibinfo {author} {\bibfnamefont {LC}~\bibnamefont {Kwek}}, \ and\ \bibinfo
  {author} {\bibfnamefont {CH}~\bibnamefont {Oh}},\ }\bibfield  {title}
  {\enquote {\bibinfo {title} {Entanglement purification based on photonic
  polarization parity measurements},}\ }\href@noop {} {\bibfield  {journal}
  {\bibinfo  {journal} {Journal of Physics B: Atomic, Molecular and Optical
  Physics}\ }\textbf {\bibinfo {volume} {38}},\ \bibinfo {pages} {2827}
  (\bibinfo {year} {2005})}\BibitemShut {NoStop}%
\bibitem [{\citenamefont {Qian}\ \emph {et~al.}(2005)\citenamefont {Qian},
  \citenamefont {Feng},\ and\ \citenamefont {Gong}}]{Qian2005}%
  \BibitemOpen
  \bibfield  {author} {\bibinfo {author} {\bibfnamefont {Jun}\ \bibnamefont
  {Qian}}, \bibinfo {author} {\bibfnamefont {Xun-Li}\ \bibnamefont {Feng}}, \
  and\ \bibinfo {author} {\bibfnamefont {Shang-Qing}\ \bibnamefont {Gong}},\
  }\bibfield  {title} {\enquote {\bibinfo {title} {Universal
  greenberger-horne-zeilinger-state analyzer based on two-photon polarization
  parity detection},}\ }\href {\doibase 10.1103/PhysRevA.72.052308} {\bibfield
  {journal} {\bibinfo  {journal} {Phys. Rev. A}\ }\textbf {\bibinfo {volume}
  {72}},\ \bibinfo {pages} {052308} (\bibinfo {year} {2005})}\BibitemShut
  {NoStop}%
\bibitem [{\citenamefont {Deng}(2011)}]{Deng2011}%
  \BibitemOpen
  \bibfield  {author} {\bibinfo {author} {\bibfnamefont {Fu-Guo}\ \bibnamefont
  {Deng}},\ }\bibfield  {title} {\enquote {\bibinfo {title} {Efficient
  multipartite entanglement purification with the entanglement link from a
  subspace},}\ }\href@noop {} {\bibfield  {journal} {\bibinfo  {journal}
  {Physical Review A}\ }\textbf {\bibinfo {volume} {84}},\ \bibinfo {pages}
  {052312} (\bibinfo {year} {2011})}\BibitemShut {NoStop}%
\bibitem [{\citenamefont {Sheng}\ \emph {et~al.}(2012)\citenamefont {Sheng},
  \citenamefont {Zhou}, \citenamefont {Zhao},\ and\ \citenamefont
  {Zheng}}]{Sheng2012}%
  \BibitemOpen
  \bibfield  {author} {\bibinfo {author} {\bibfnamefont {Yu-Bo}\ \bibnamefont
  {Sheng}}, \bibinfo {author} {\bibfnamefont {Lan}\ \bibnamefont {Zhou}},
  \bibinfo {author} {\bibfnamefont {Sheng-Mei}\ \bibnamefont {Zhao}}, \ and\
  \bibinfo {author} {\bibfnamefont {Bao-Yu}\ \bibnamefont {Zheng}},\ }\bibfield
   {title} {\enquote {\bibinfo {title} {Efficient single-photon-assisted
  entanglement concentration for partially entangled photon pairs},}\ }\href
  {\doibase 10.1103/PhysRevA.85.012307} {\bibfield  {journal} {\bibinfo
  {journal} {Phys. Rev. A}\ }\textbf {\bibinfo {volume} {85}},\ \bibinfo
  {pages} {012307} (\bibinfo {year} {2012})}\BibitemShut {NoStop}%
\bibitem [{\citenamefont {Deng}(2012)}]{Deng2012}%
  \BibitemOpen
  \bibfield  {author} {\bibinfo {author} {\bibfnamefont {Fu-Guo}\ \bibnamefont
  {Deng}},\ }\bibfield  {title} {\enquote {\bibinfo {title} {Optimal nonlocal
  multipartite entanglement concentration based on projection measurements},}\
  }\href {\doibase 10.1103/PhysRevA.85.022311} {\bibfield  {journal} {\bibinfo
  {journal} {Phys. Rev. A}\ }\textbf {\bibinfo {volume} {85}},\ \bibinfo
  {pages} {022311} (\bibinfo {year} {2012})}\BibitemShut {NoStop}%
\bibitem [{\citenamefont {Choudhury}\ and\ \citenamefont
  {Dhara}(2013)}]{Choudhury2013}%
  \BibitemOpen
  \bibfield  {author} {\bibinfo {author} {\bibfnamefont {Binayak~S.}\
  \bibnamefont {Choudhury}}\ and\ \bibinfo {author} {\bibfnamefont {Arpan}\
  \bibnamefont {Dhara}},\ }\bibfield  {title} {\enquote {\bibinfo {title} {An
  entanglement concentration protocol for cluster states},}\ }\href {\doibase
  10.1007/s11128-013-0549-1} {\bibfield  {journal} {\bibinfo  {journal}
  {Quantum Information Processing}\ }\textbf {\bibinfo {volume} {12}},\
  \bibinfo {pages} {2577--2585} (\bibinfo {year} {2013})}\BibitemShut {NoStop}%
\bibitem [{\citenamefont {Zhu}\ and\ \citenamefont {Ye}(2015)}]{Zhu2015}%
  \BibitemOpen
  \bibfield  {author} {\bibinfo {author} {\bibfnamefont {Meng-Zheng}\
  \bibnamefont {Zhu}}\ and\ \bibinfo {author} {\bibfnamefont {Liu}\
  \bibnamefont {Ye}},\ }\bibfield  {title} {\enquote {\bibinfo {title}
  {Efficient entanglement purification for greenberger--horne--zeilinger states
  via the distributed parity-check detector},}\ }\href {\doibase
  https://doi.org/10.1016/j.optcom.2014.07.090} {\bibfield  {journal} {\bibinfo
   {journal} {Optics Communications}\ }\textbf {\bibinfo {volume} {334}},\
  \bibinfo {pages} {51--57} (\bibinfo {year} {2015})}\BibitemShut {NoStop}%
\bibitem [{\citenamefont {Liu}\ \emph {et~al.}(2018)\citenamefont {Liu},
  \citenamefont {Zhou}, \citenamefont {Zhong},\ and\ \citenamefont
  {Sheng}}]{Liu2018}%
  \BibitemOpen
  \bibfield  {author} {\bibinfo {author} {\bibfnamefont {Jiu}\ \bibnamefont
  {Liu}}, \bibinfo {author} {\bibfnamefont {Lan}\ \bibnamefont {Zhou}},
  \bibinfo {author} {\bibfnamefont {Wei}\ \bibnamefont {Zhong}}, \ and\
  \bibinfo {author} {\bibfnamefont {Yu-Bo}\ \bibnamefont {Sheng}},\ }\bibfield
  {title} {\enquote {\bibinfo {title} {Logic bell state concentration with
  parity check measurement},}\ }\href {\doibase 10.1007/s11467-018-0866-z}
  {\bibfield  {journal} {\bibinfo  {journal} {Frontiers of Physics}\ }\textbf
  {\bibinfo {volume} {14}},\ \bibinfo {pages} {21601} (\bibinfo {year}
  {2018})}\BibitemShut {NoStop}%
\bibitem [{\citenamefont {Daiss}\ \emph {et~al.}(2019)\citenamefont {Daiss},
  \citenamefont {Welte}, \citenamefont {Hacker}, \citenamefont {Li},\ and\
  \citenamefont {Rempe}}]{Daiss2019}%
  \BibitemOpen
  \bibfield  {author} {\bibinfo {author} {\bibfnamefont {Severin}\ \bibnamefont
  {Daiss}}, \bibinfo {author} {\bibfnamefont {Stephan}\ \bibnamefont {Welte}},
  \bibinfo {author} {\bibfnamefont {Bastian}\ \bibnamefont {Hacker}}, \bibinfo
  {author} {\bibfnamefont {Lin}\ \bibnamefont {Li}}, \ and\ \bibinfo {author}
  {\bibfnamefont {Gerhard}\ \bibnamefont {Rempe}},\ }\bibfield  {title}
  {\enquote {\bibinfo {title} {Single-photon distillation via a photonic parity
  measurement using cavity qed},}\ }\href@noop {} {\bibfield  {journal}
  {\bibinfo  {journal} {Physical review letters}\ }\textbf {\bibinfo {volume}
  {122}},\ \bibinfo {pages} {133603} (\bibinfo {year} {2019})}\BibitemShut
  {NoStop}%
\bibitem [{\citenamefont {Pittman}\ \emph {et~al.}(2002)\citenamefont
  {Pittman}, \citenamefont {Jacobs},\ and\ \citenamefont
  {Franson}}]{Pittman2002}%
  \BibitemOpen
  \bibfield  {author} {\bibinfo {author} {\bibfnamefont {T.~B.}\ \bibnamefont
  {Pittman}}, \bibinfo {author} {\bibfnamefont {B.~C.}\ \bibnamefont {Jacobs}},
  \ and\ \bibinfo {author} {\bibfnamefont {J.~D.}\ \bibnamefont {Franson}},\
  }\bibfield  {title} {\enquote {\bibinfo {title} {Demonstration of
  nondeterministic quantum logic operations using linear optical elements},}\
  }\href {\doibase 10.1103/PhysRevLett.88.257902} {\bibfield  {journal}
  {\bibinfo  {journal} {Phys. Rev. Lett.}\ }\textbf {\bibinfo {volume} {88}},\
  \bibinfo {pages} {257902} (\bibinfo {year} {2002})}\BibitemShut {NoStop}%
\bibitem [{\citenamefont {Browne}\ and\ \citenamefont
  {Rudolph}(2005)}]{Browne2005}%
  \BibitemOpen
  \bibfield  {author} {\bibinfo {author} {\bibfnamefont {Daniel~E.}\
  \bibnamefont {Browne}}\ and\ \bibinfo {author} {\bibfnamefont {Terry}\
  \bibnamefont {Rudolph}},\ }\bibfield  {title} {\enquote {\bibinfo {title}
  {Resource-efficient linear optical quantum computation},}\ }\href {\doibase
  10.1103/PhysRevLett.95.010501} {\bibfield  {journal} {\bibinfo  {journal}
  {Phys. Rev. Lett.}\ }\textbf {\bibinfo {volume} {95}},\ \bibinfo {pages}
  {010501} (\bibinfo {year} {2005})}\BibitemShut {NoStop}%
\bibitem [{\citenamefont {Barrett}\ \emph {et~al.}(2005)\citenamefont
  {Barrett}, \citenamefont {Kok}, \citenamefont {Nemoto}, \citenamefont
  {Beausoleil}, \citenamefont {Munro},\ and\ \citenamefont
  {Spiller}}]{Bennett2005}%
  \BibitemOpen
  \bibfield  {author} {\bibinfo {author} {\bibfnamefont {S.~D.}\ \bibnamefont
  {Barrett}}, \bibinfo {author} {\bibfnamefont {Pieter}\ \bibnamefont {Kok}},
  \bibinfo {author} {\bibfnamefont {Kae}\ \bibnamefont {Nemoto}}, \bibinfo
  {author} {\bibfnamefont {R.~G.}\ \bibnamefont {Beausoleil}}, \bibinfo
  {author} {\bibfnamefont {W.~J.}\ \bibnamefont {Munro}}, \ and\ \bibinfo
  {author} {\bibfnamefont {T.~P.}\ \bibnamefont {Spiller}},\ }\bibfield
  {title} {\enquote {\bibinfo {title} {Symmetry analyzer for nondestructive
  bell-state detection using weak nonlinearities},}\ }\href {\doibase
  10.1103/PhysRevA.71.060302} {\bibfield  {journal} {\bibinfo  {journal} {Phys.
  Rev. A}\ }\textbf {\bibinfo {volume} {71}},\ \bibinfo {pages} {060302}
  (\bibinfo {year} {2005})}\BibitemShut {NoStop}%
\bibitem [{\citenamefont {Sheng}\ \emph {et~al.}(2008)\citenamefont {Sheng},
  \citenamefont {Deng},\ and\ \citenamefont {Zhou}}]{Sheng2008}%
  \BibitemOpen
  \bibfield  {author} {\bibinfo {author} {\bibfnamefont {Yu-Bo}\ \bibnamefont
  {Sheng}}, \bibinfo {author} {\bibfnamefont {Fu-Guo}\ \bibnamefont {Deng}}, \
  and\ \bibinfo {author} {\bibfnamefont {Hong-Yu}\ \bibnamefont {Zhou}},\
  }\bibfield  {title} {\enquote {\bibinfo {title} {Nonlocal entanglement
  concentration scheme for partially entangled multipartite systems with
  nonlinear optics},}\ }\href {\doibase 10.1103/PhysRevA.77.062325} {\bibfield
  {journal} {\bibinfo  {journal} {Phys. Rev. A}\ }\textbf {\bibinfo {volume}
  {77}},\ \bibinfo {pages} {062325} (\bibinfo {year} {2008})}\BibitemShut
  {NoStop}%
\bibitem [{\citenamefont {Lin}\ and\ \citenamefont {Li}(2009)}]{Qing2009}%
  \BibitemOpen
  \bibfield  {author} {\bibinfo {author} {\bibfnamefont {Qing}\ \bibnamefont
  {Lin}}\ and\ \bibinfo {author} {\bibfnamefont {Jian}\ \bibnamefont {Li}},\
  }\bibfield  {title} {\enquote {\bibinfo {title} {Quantum control gates with
  weak cross-kerr nonlinearity},}\ }\href {\doibase 10.1103/PhysRevA.79.022301}
  {\bibfield  {journal} {\bibinfo  {journal} {Phys. Rev. A}\ }\textbf {\bibinfo
  {volume} {79}},\ \bibinfo {pages} {022301} (\bibinfo {year}
  {2009})}\BibitemShut {NoStop}%
\bibitem [{\citenamefont {Guo}\ \emph {et~al.}(2011)\citenamefont {Guo},
  \citenamefont {Bai}, \citenamefont {Cheng}, \citenamefont {Shao},
  \citenamefont {Wang},\ and\ \citenamefont {Zhang}}]{Qi2011}%
  \BibitemOpen
  \bibfield  {author} {\bibinfo {author} {\bibfnamefont {Qi}~\bibnamefont
  {Guo}}, \bibinfo {author} {\bibfnamefont {Juan}\ \bibnamefont {Bai}},
  \bibinfo {author} {\bibfnamefont {Liu-Yong}\ \bibnamefont {Cheng}}, \bibinfo
  {author} {\bibfnamefont {Xiao-Qiang}\ \bibnamefont {Shao}}, \bibinfo {author}
  {\bibfnamefont {Hong-Fu}\ \bibnamefont {Wang}}, \ and\ \bibinfo {author}
  {\bibfnamefont {Shou}\ \bibnamefont {Zhang}},\ }\bibfield  {title} {\enquote
  {\bibinfo {title} {Simplified optical quantum-information processing via weak
  cross-kerr nonlinearities},}\ }\href {\doibase 10.1103/PhysRevA.83.054303}
  {\bibfield  {journal} {\bibinfo  {journal} {Phys. Rev. A}\ }\textbf {\bibinfo
  {volume} {83}},\ \bibinfo {pages} {054303} (\bibinfo {year}
  {2011})}\BibitemShut {NoStop}%
\bibitem [{\citenamefont {Kok}(2008)}]{Kok2008}%
  \BibitemOpen
  \bibfield  {author} {\bibinfo {author} {\bibfnamefont {Pieter}\ \bibnamefont
  {Kok}},\ }\bibfield  {title} {\enquote {\bibinfo {title} {Effects of
  self-phase-modulation on weak nonlinear optical quantum gates},}\ }\href
  {\doibase 10.1103/PhysRevA.77.013808} {\bibfield  {journal} {\bibinfo
  {journal} {Phys. Rev. A}\ }\textbf {\bibinfo {volume} {77}},\ \bibinfo
  {pages} {013808} (\bibinfo {year} {2008})}\BibitemShut {NoStop}%
\bibitem [{\citenamefont {Fushman}\ \emph {et~al.}(2008)\citenamefont
  {Fushman}, \citenamefont {Englund}, \citenamefont {Faraon}, \citenamefont
  {Stoltz}, \citenamefont {Petroff},\ and\ \citenamefont
  {Vučković}}]{Fushman2008}%
  \BibitemOpen
  \bibfield  {author} {\bibinfo {author} {\bibfnamefont {Ilya}\ \bibnamefont
  {Fushman}}, \bibinfo {author} {\bibfnamefont {Dirk}\ \bibnamefont {Englund}},
  \bibinfo {author} {\bibfnamefont {Andrei}\ \bibnamefont {Faraon}}, \bibinfo
  {author} {\bibfnamefont {Nick}\ \bibnamefont {Stoltz}}, \bibinfo {author}
  {\bibfnamefont {Pierre}\ \bibnamefont {Petroff}}, \ and\ \bibinfo {author}
  {\bibfnamefont {Jelena}\ \bibnamefont {Vučković}},\ }\bibfield  {title}
  {\enquote {\bibinfo {title} {{Controlled Phase Shifts with a Single Quantum
  Dot}},}\ }\href {\doibase 10.1126/science.1154643} {\bibfield  {journal}
  {\bibinfo  {journal} {Science}\ }\textbf {\bibinfo {volume} {320}},\ \bibinfo
  {pages} {769--772} (\bibinfo {year} {2008})}\BibitemShut {NoStop}%
\bibitem [{\citenamefont {Loo}\ \emph {et~al.}(2012)\citenamefont {Loo},
  \citenamefont {Arnold}, \citenamefont {Gazzano}, \citenamefont
  {Lema\^{\i}tre}, \citenamefont {Sagnes}, \citenamefont {Krebs}, \citenamefont
  {Voisin}, \citenamefont {Senellart},\ and\ \citenamefont {Lanco}}]{Loo2012}%
  \BibitemOpen
  \bibfield  {author} {\bibinfo {author} {\bibfnamefont {V.}~\bibnamefont
  {Loo}}, \bibinfo {author} {\bibfnamefont {C.}~\bibnamefont {Arnold}},
  \bibinfo {author} {\bibfnamefont {O.}~\bibnamefont {Gazzano}}, \bibinfo
  {author} {\bibfnamefont {A.}~\bibnamefont {Lema\^{\i}tre}}, \bibinfo {author}
  {\bibfnamefont {I.}~\bibnamefont {Sagnes}}, \bibinfo {author} {\bibfnamefont
  {O.}~\bibnamefont {Krebs}}, \bibinfo {author} {\bibfnamefont
  {P.}~\bibnamefont {Voisin}}, \bibinfo {author} {\bibfnamefont
  {P.}~\bibnamefont {Senellart}}, \ and\ \bibinfo {author} {\bibfnamefont
  {L.}~\bibnamefont {Lanco}},\ }\bibfield  {title} {\enquote {\bibinfo {title}
  {Optical nonlinearity for few-photon pulses on a quantum dot-pillar cavity
  device},}\ }\href {\doibase 10.1103/PhysRevLett.109.166806} {\bibfield
  {journal} {\bibinfo  {journal} {Phys. Rev. Lett.}\ }\textbf {\bibinfo
  {volume} {109}},\ \bibinfo {pages} {166806} (\bibinfo {year}
  {2012})}\BibitemShut {NoStop}%
\bibitem [{\citenamefont {Kim}\ \emph {et~al.}(2013)\citenamefont {Kim},
  \citenamefont {Bose}, \citenamefont {Shen}, \citenamefont {Solomon},\ and\
  \citenamefont {Waks}}]{kim2013quantum}%
  \BibitemOpen
  \bibfield  {author} {\bibinfo {author} {\bibfnamefont {Hyochul}\ \bibnamefont
  {Kim}}, \bibinfo {author} {\bibfnamefont {Ranojoy}\ \bibnamefont {Bose}},
  \bibinfo {author} {\bibfnamefont {Thomas~C}\ \bibnamefont {Shen}}, \bibinfo
  {author} {\bibfnamefont {Glenn~S}\ \bibnamefont {Solomon}}, \ and\ \bibinfo
  {author} {\bibfnamefont {Edo}\ \bibnamefont {Waks}},\ }\bibfield  {title}
  {\enquote {\bibinfo {title} {A quantum logic gate between a solid-state
  quantum bit and a photon},}\ }\href@noop {} {\bibfield  {journal} {\bibinfo
  {journal} {Nature Photonics}\ }\textbf {\bibinfo {volume} {7}},\ \bibinfo
  {pages} {373--377} (\bibinfo {year} {2013})}\BibitemShut {NoStop}%
\bibitem [{\citenamefont {Sun}\ \emph {et~al.}(2016)\citenamefont {Sun},
  \citenamefont {Kim}, \citenamefont {Solomon},\ and\ \citenamefont
  {Waks}}]{Sun2016}%
  \BibitemOpen
  \bibfield  {author} {\bibinfo {author} {\bibfnamefont {Shuo}\ \bibnamefont
  {Sun}}, \bibinfo {author} {\bibfnamefont {Hyochul}\ \bibnamefont {Kim}},
  \bibinfo {author} {\bibfnamefont {Glenn~S.}\ \bibnamefont {Solomon}}, \ and\
  \bibinfo {author} {\bibfnamefont {Edo}\ \bibnamefont {Waks}},\ }\bibfield
  {title} {\enquote {\bibinfo {title} {A quantum phase switch between a single
  solid-state spin and a photon},}\ }\href {\doibase 10.1038/nnano.2015.334}
  {\bibfield  {journal} {\bibinfo  {journal} {Nature Nanotechnology}\ }\textbf
  {\bibinfo {volume} {11}},\ \bibinfo {pages} {539--544} (\bibinfo {year}
  {2016})}\BibitemShut {NoStop}%
\bibitem [{\citenamefont {Tiarks}\ \emph {et~al.}(2016)\citenamefont {Tiarks},
  \citenamefont {Schmidt}, \citenamefont {Rempe},\ and\ \citenamefont
  {D{\"u}rr}}]{Tiarks2016}%
  \BibitemOpen
  \bibfield  {author} {\bibinfo {author} {\bibfnamefont {Daniel}\ \bibnamefont
  {Tiarks}}, \bibinfo {author} {\bibfnamefont {Steffen}\ \bibnamefont
  {Schmidt}}, \bibinfo {author} {\bibfnamefont {Gerhard}\ \bibnamefont
  {Rempe}}, \ and\ \bibinfo {author} {\bibfnamefont {Stephan}\ \bibnamefont
  {D{\"u}rr}},\ }\bibfield  {title} {\enquote {\bibinfo {title} {Optical $\pi$
  phase shift created with a single-photon pulse},}\ }\href {\doibase
  10.1126/sciadv.1600036} {\bibfield  {journal} {\bibinfo  {journal} {Science
  Advances}\ }\textbf {\bibinfo {volume} {2}},\ \bibinfo {pages} {e1600036}
  (\bibinfo {year} {2016})}\BibitemShut {NoStop}%
\bibitem [{\citenamefont {Sun}\ \emph {et~al.}(2018)\citenamefont {Sun},
  \citenamefont {Kim}, \citenamefont {Luo}, \citenamefont {Solomon},\ and\
  \citenamefont {Waks}}]{Sun2018}%
  \BibitemOpen
  \bibfield  {author} {\bibinfo {author} {\bibfnamefont {Shuo}\ \bibnamefont
  {Sun}}, \bibinfo {author} {\bibfnamefont {Hyochul}\ \bibnamefont {Kim}},
  \bibinfo {author} {\bibfnamefont {Zhouchen}\ \bibnamefont {Luo}}, \bibinfo
  {author} {\bibfnamefont {Glenn~S.}\ \bibnamefont {Solomon}}, \ and\ \bibinfo
  {author} {\bibfnamefont {Edo}\ \bibnamefont {Waks}},\ }\bibfield  {title}
  {\enquote {\bibinfo {title} {A single-photon switch and transistor enabled by
  a solid-state quantum memory},}\ }\href {\doibase 10.1126/science.aat3581}
  {\bibfield  {journal} {\bibinfo  {journal} {Science}\ }\textbf {\bibinfo
  {volume} {361}},\ \bibinfo {pages} {57--60} (\bibinfo {year}
  {2018})}\BibitemShut {NoStop}%
\bibitem [{\citenamefont {Wells}\ \emph {et~al.}(2019)\citenamefont {Wells},
  \citenamefont {Kalliakos}, \citenamefont {Villa}, \citenamefont {Ellis},
  \citenamefont {Stevenson}, \citenamefont {Bennett}, \citenamefont {Farrer},
  \citenamefont {Ritchie},\ and\ \citenamefont {Shields}}]{Wells2019}%
  \BibitemOpen
  \bibfield  {author} {\bibinfo {author} {\bibfnamefont {L.M.}\ \bibnamefont
  {Wells}}, \bibinfo {author} {\bibfnamefont {S.}~\bibnamefont {Kalliakos}},
  \bibinfo {author} {\bibfnamefont {B.}~\bibnamefont {Villa}}, \bibinfo
  {author} {\bibfnamefont {D.J.P.}\ \bibnamefont {Ellis}}, \bibinfo {author}
  {\bibfnamefont {R.M.}\ \bibnamefont {Stevenson}}, \bibinfo {author}
  {\bibfnamefont {A.J.}\ \bibnamefont {Bennett}}, \bibinfo {author}
  {\bibfnamefont {I.}~\bibnamefont {Farrer}}, \bibinfo {author} {\bibfnamefont
  {D.A.}\ \bibnamefont {Ritchie}}, \ and\ \bibinfo {author} {\bibfnamefont
  {A.J.}\ \bibnamefont {Shields}},\ }\bibfield  {title} {\enquote {\bibinfo
  {title} {Photon phase shift at the few-photon level and optical switching by
  a quantum dot in a microcavity},}\ }\href {\doibase
  10.1103/PhysRevApplied.11.061001} {\bibfield  {journal} {\bibinfo  {journal}
  {Phys. Rev. Applied}\ }\textbf {\bibinfo {volume} {11}},\ \bibinfo {pages}
  {061001} (\bibinfo {year} {2019})}\BibitemShut {NoStop}%
\bibitem [{\citenamefont {Nielsen}\ and\ \citenamefont
  {Chuang}(2010)}]{NielsenBook}%
  \BibitemOpen
  \bibfield  {author} {\bibinfo {author} {\bibfnamefont {Michael~A.}\
  \bibnamefont {Nielsen}}\ and\ \bibinfo {author} {\bibfnamefont {Isaac~L.}\
  \bibnamefont {Chuang}},\ }\href {\doibase 10.1017/CBO9780511976667} {\emph
  {\bibinfo {title} {Quantum Computation and Quantum Information: 10th
  Anniversary Edition}}}\ (\bibinfo  {publisher} {Cambridge University Press},\
  \bibinfo {year} {2010})\BibitemShut {NoStop}%
\bibitem [{\citenamefont {Wang}\ \emph {et~al.}(2013)\citenamefont {Wang},
  \citenamefont {Wen}, \citenamefont {Zhu}, \citenamefont {Zhang},\ and\
  \citenamefont {Yeon}}]{Wang2013}%
  \BibitemOpen
  \bibfield  {author} {\bibinfo {author} {\bibfnamefont {Hong-Fu}\ \bibnamefont
  {Wang}}, \bibinfo {author} {\bibfnamefont {Jing-Ji}\ \bibnamefont {Wen}},
  \bibinfo {author} {\bibfnamefont {Ai-Dong}\ \bibnamefont {Zhu}}, \bibinfo
  {author} {\bibfnamefont {Shou}\ \bibnamefont {Zhang}}, \ and\ \bibinfo
  {author} {\bibfnamefont {Kyu-Hwang}\ \bibnamefont {Yeon}},\ }\bibfield
  {title} {\enquote {\bibinfo {title} {Deterministic cnot gate and entanglement
  swapping for photonic qubits using a quantum-dot spin in a double-sided
  optical microcavity},}\ }\href {\doibase
  https://doi.org/10.1016/j.physleta.2013.09.005} {\bibfield  {journal}
  {\bibinfo  {journal} {Physics Letters A}\ }\textbf {\bibinfo {volume}
  {377}},\ \bibinfo {pages} {2870--2876} (\bibinfo {year} {2013})}\BibitemShut
  {NoStop}%
\bibitem [{\citenamefont {Ren}\ \emph {et~al.}(2013)\citenamefont {Ren},
  \citenamefont {Wei}, \citenamefont {Hua}, \citenamefont {Li},\ and\
  \citenamefont {Deng}}]{Ren2013}%
  \BibitemOpen
  \bibfield  {author} {\bibinfo {author} {\bibfnamefont {Bao-Cang}\
  \bibnamefont {Ren}}, \bibinfo {author} {\bibfnamefont {Hai-Rui}\ \bibnamefont
  {Wei}}, \bibinfo {author} {\bibfnamefont {Ming}\ \bibnamefont {Hua}},
  \bibinfo {author} {\bibfnamefont {Tao}\ \bibnamefont {Li}}, \ and\ \bibinfo
  {author} {\bibfnamefont {Fu-Guo}\ \bibnamefont {Deng}},\ }\bibfield  {title}
  {\enquote {\bibinfo {title} {Photonic spatial bell-state analysis for robust
  quantum secure direct communication using quantum dot-cavity systems},}\
  }\href {\doibase 10.1140/epjd/e2012-30626-x} {\bibfield  {journal} {\bibinfo
  {journal} {The European Physical Journal D}\ }\textbf {\bibinfo {volume}
  {67}},\ \bibinfo {pages} {30} (\bibinfo {year} {2013})}\BibitemShut {NoStop}%
\bibitem [{\citenamefont {Shapiro}(2006)}]{Shapiro2006}%
  \BibitemOpen
  \bibfield  {author} {\bibinfo {author} {\bibfnamefont {Jeffrey~H.}\
  \bibnamefont {Shapiro}},\ }\bibfield  {title} {\enquote {\bibinfo {title}
  {Single-photon kerr nonlinearities do not help quantum computation},}\ }\href
  {\doibase 10.1103/PhysRevA.73.062305} {\bibfield  {journal} {\bibinfo
  {journal} {Phys. Rev. A}\ }\textbf {\bibinfo {volume} {73}},\ \bibinfo
  {pages} {062305} (\bibinfo {year} {2006})}\BibitemShut {NoStop}%
\bibitem [{\citenamefont {Gea-Banacloche}(2010)}]{GBanacloche2010}%
  \BibitemOpen
  \bibfield  {author} {\bibinfo {author} {\bibfnamefont {Julio}\ \bibnamefont
  {Gea-Banacloche}},\ }\bibfield  {title} {\enquote {\bibinfo {title}
  {Impossibility of large phase shifts via the giant kerr effect with
  single-photon wave packets},}\ }\href {\doibase 10.1103/PhysRevA.81.043823}
  {\bibfield  {journal} {\bibinfo  {journal} {Phys. Rev. A}\ }\textbf {\bibinfo
  {volume} {81}},\ \bibinfo {pages} {043823} (\bibinfo {year}
  {2010})}\BibitemShut {NoStop}%
\bibitem [{\citenamefont {Campbell}\ \emph {et~al.}(2007)\citenamefont
  {Campbell}, \citenamefont {Fitzsimons}, \citenamefont {Benjamin},\ and\
  \citenamefont {Kok}}]{Campbell_PRA2007}%
  \BibitemOpen
  \bibfield  {author} {\bibinfo {author} {\bibfnamefont {Earl~T.}\ \bibnamefont
  {Campbell}}, \bibinfo {author} {\bibfnamefont {Joseph}\ \bibnamefont
  {Fitzsimons}}, \bibinfo {author} {\bibfnamefont {Simon~C.}\ \bibnamefont
  {Benjamin}}, \ and\ \bibinfo {author} {\bibfnamefont {Pieter}\ \bibnamefont
  {Kok}},\ }\bibfield  {title} {\enquote {\bibinfo {title} {Adaptive strategies
  for graph-state growth in the presence of monitored errors},}\ }\href
  {\doibase 10.1103/PhysRevA.75.042303} {\bibfield  {journal} {\bibinfo
  {journal} {Phys. Rev. A}\ }\textbf {\bibinfo {volume} {75}},\ \bibinfo
  {pages} {042303} (\bibinfo {year} {2007})}\BibitemShut {NoStop}%
\bibitem [{\citenamefont {Frantzeskakis}\ \emph {et~al.}(2023)\citenamefont
  {Frantzeskakis}, \citenamefont {Liu}, \citenamefont {Raissi}, \citenamefont
  {Barnes},\ and\ \citenamefont {Economou}}]{RafailWork}%
  \BibitemOpen
  \bibfield  {author} {\bibinfo {author} {\bibfnamefont {Rafail}\ \bibnamefont
  {Frantzeskakis}}, \bibinfo {author} {\bibfnamefont {Chenxu}\ \bibnamefont
  {Liu}}, \bibinfo {author} {\bibfnamefont {Zahra}\ \bibnamefont {Raissi}},
  \bibinfo {author} {\bibfnamefont {Edwin}\ \bibnamefont {Barnes}}, \ and\
  \bibinfo {author} {\bibfnamefont {Sophia~E.}\ \bibnamefont {Economou}},\
  }\bibfield  {title} {\enquote {\bibinfo {title} {Extracting perfect ghz
  states from imperfect weighted graph states via entanglement
  concentration},}\ }\href {\doibase 10.1103/PhysRevResearch.5.023124}
  {\bibfield  {journal} {\bibinfo  {journal} {Phys. Rev. Res.}\ }\textbf
  {\bibinfo {volume} {5}},\ \bibinfo {pages} {023124} (\bibinfo {year}
  {2023})}\BibitemShut {NoStop}%
\bibitem [{\citenamefont {Gilchrist}\ \emph {et~al.}(2005)\citenamefont
  {Gilchrist}, \citenamefont {Langford},\ and\ \citenamefont
  {Nielsen}}]{Gilchrist2005}%
  \BibitemOpen
  \bibfield  {author} {\bibinfo {author} {\bibfnamefont {Alexei}\ \bibnamefont
  {Gilchrist}}, \bibinfo {author} {\bibfnamefont {Nathan~K.}\ \bibnamefont
  {Langford}}, \ and\ \bibinfo {author} {\bibfnamefont {Michael~A.}\
  \bibnamefont {Nielsen}},\ }\bibfield  {title} {\enquote {\bibinfo {title}
  {Distance measures to compare real and ideal quantum processes},}\ }\href
  {\doibase 10.1103/PhysRevA.71.062310} {\bibfield  {journal} {\bibinfo
  {journal} {Phys. Rev. A}\ }\textbf {\bibinfo {volume} {71}},\ \bibinfo
  {pages} {062310} (\bibinfo {year} {2005})}\BibitemShut {NoStop}%
\bibitem [{\citenamefont {Androvitsaneas}\ \emph {et~al.}(2019)\citenamefont
  {Androvitsaneas}, \citenamefont {Young}, \citenamefont {Lennon},
  \citenamefont {Schneider}, \citenamefont {Maier}, \citenamefont {Hinchliff},
  \citenamefont {Atkinson}, \citenamefont {Harbord}, \citenamefont {Kamp},
  \citenamefont {H{\"o}fling}, \citenamefont {Rarity},\ and\ \citenamefont
  {Oulton}}]{Androvitsaneas2019}%
  \BibitemOpen
  \bibfield  {author} {\bibinfo {author} {\bibfnamefont {P.}~\bibnamefont
  {Androvitsaneas}}, \bibinfo {author} {\bibfnamefont {A.~B.}\ \bibnamefont
  {Young}}, \bibinfo {author} {\bibfnamefont {J.~M.}\ \bibnamefont {Lennon}},
  \bibinfo {author} {\bibfnamefont {C.}~\bibnamefont {Schneider}}, \bibinfo
  {author} {\bibfnamefont {S.}~\bibnamefont {Maier}}, \bibinfo {author}
  {\bibfnamefont {J.~J.}\ \bibnamefont {Hinchliff}}, \bibinfo {author}
  {\bibfnamefont {G.~S.}\ \bibnamefont {Atkinson}}, \bibinfo {author}
  {\bibfnamefont {E.}~\bibnamefont {Harbord}}, \bibinfo {author} {\bibfnamefont
  {M.}~\bibnamefont {Kamp}}, \bibinfo {author} {\bibfnamefont {S.}~\bibnamefont
  {H{\"o}fling}}, \bibinfo {author} {\bibfnamefont {J.~G.}\ \bibnamefont
  {Rarity}}, \ and\ \bibinfo {author} {\bibfnamefont {R.}~\bibnamefont
  {Oulton}},\ }\bibfield  {title} {\enquote {\bibinfo {title} {Efficient
  quantum photonic phase shift in a low q-factor regime},}\ }\href {\doibase
  10.1021/acsphotonics.8b01380} {\bibfield  {journal} {\bibinfo  {journal} {ACS
  Photonics}\ }\textbf {\bibinfo {volume} {6}},\ \bibinfo {pages} {429--435}
  (\bibinfo {year} {2019})}\BibitemShut {NoStop}%
\bibitem [{Note1()}]{Note1}%
  \BibitemOpen
  \bibinfo {note} {The reason we choose this value is that it is close to the
  CPhase gates demonstrated in Ref.~\cite {Wells2019}.}\BibitemShut {Stop}%
\bibitem [{\citenamefont {De~Lange}\ \emph {et~al.}(2010)\citenamefont
  {De~Lange}, \citenamefont {Wang}, \citenamefont {Riste}, \citenamefont
  {Dobrovitski},\ and\ \citenamefont {Hanson}}]{de2010universal}%
  \BibitemOpen
  \bibfield  {author} {\bibinfo {author} {\bibfnamefont {Gijs}\ \bibnamefont
  {De~Lange}}, \bibinfo {author} {\bibfnamefont {Zhi-Hui}\ \bibnamefont
  {Wang}}, \bibinfo {author} {\bibfnamefont {D}~\bibnamefont {Riste}}, \bibinfo
  {author} {\bibfnamefont {VV}~\bibnamefont {Dobrovitski}}, \ and\ \bibinfo
  {author} {\bibfnamefont {R}~\bibnamefont {Hanson}},\ }\bibfield  {title}
  {\enquote {\bibinfo {title} {Universal dynamical decoupling of a single
  solid-state spin from a spin bath},}\ }\href@noop {} {\bibfield  {journal}
  {\bibinfo  {journal} {Science}\ }\textbf {\bibinfo {volume} {330}},\ \bibinfo
  {pages} {60--63} (\bibinfo {year} {2010})}\BibitemShut {NoStop}%
\bibitem [{\citenamefont {Naydenov}\ \emph {et~al.}(2011)\citenamefont
  {Naydenov}, \citenamefont {Dolde}, \citenamefont {Hall}, \citenamefont
  {Shin}, \citenamefont {Fedder}, \citenamefont {Hollenberg}, \citenamefont
  {Jelezko},\ and\ \citenamefont {Wrachtrup}}]{naydenov2011dynamical}%
  \BibitemOpen
  \bibfield  {author} {\bibinfo {author} {\bibfnamefont {Boris}\ \bibnamefont
  {Naydenov}}, \bibinfo {author} {\bibfnamefont {Florian}\ \bibnamefont
  {Dolde}}, \bibinfo {author} {\bibfnamefont {Liam~T}\ \bibnamefont {Hall}},
  \bibinfo {author} {\bibfnamefont {Chang}\ \bibnamefont {Shin}}, \bibinfo
  {author} {\bibfnamefont {Helmut}\ \bibnamefont {Fedder}}, \bibinfo {author}
  {\bibfnamefont {Lloyd~CL}\ \bibnamefont {Hollenberg}}, \bibinfo {author}
  {\bibfnamefont {Fedor}\ \bibnamefont {Jelezko}}, \ and\ \bibinfo {author}
  {\bibfnamefont {J{\"o}rg}\ \bibnamefont {Wrachtrup}},\ }\bibfield  {title}
  {\enquote {\bibinfo {title} {Dynamical decoupling of a single-electron spin
  at room temperature},}\ }\href@noop {} {\bibfield  {journal} {\bibinfo
  {journal} {Physical Review B}\ }\textbf {\bibinfo {volume} {83}},\ \bibinfo
  {pages} {081201} (\bibinfo {year} {2011})}\BibitemShut {NoStop}%
\bibitem [{\citenamefont {Wang}\ \emph
  {et~al.}(2012{\natexlab{b}})\citenamefont {Wang}, \citenamefont {De~Lange},
  \citenamefont {Rist{\`e}}, \citenamefont {Hanson},\ and\ \citenamefont
  {Dobrovitski}}]{wang2012comparison}%
  \BibitemOpen
  \bibfield  {author} {\bibinfo {author} {\bibfnamefont {Zhi-Hui}\ \bibnamefont
  {Wang}}, \bibinfo {author} {\bibfnamefont {G}~\bibnamefont {De~Lange}},
  \bibinfo {author} {\bibfnamefont {D}~\bibnamefont {Rist{\`e}}}, \bibinfo
  {author} {\bibfnamefont {R}~\bibnamefont {Hanson}}, \ and\ \bibinfo {author}
  {\bibfnamefont {VV}~\bibnamefont {Dobrovitski}},\ }\bibfield  {title}
  {\enquote {\bibinfo {title} {Comparison of dynamical decoupling protocols for
  a nitrogen-vacancy center in diamond},}\ }\href@noop {} {\bibfield  {journal}
  {\bibinfo  {journal} {Physical Review B}\ }\textbf {\bibinfo {volume} {85}},\
  \bibinfo {pages} {155204} (\bibinfo {year} {2012}{\natexlab{b}})}\BibitemShut
  {NoStop}%
\bibitem [{\citenamefont {Shim}\ \emph {et~al.}(2012)\citenamefont {Shim},
  \citenamefont {Niemeyer}, \citenamefont {Zhang},\ and\ \citenamefont
  {Suter}}]{shim2012robust}%
  \BibitemOpen
  \bibfield  {author} {\bibinfo {author} {\bibfnamefont {JH}~\bibnamefont
  {Shim}}, \bibinfo {author} {\bibfnamefont {I}~\bibnamefont {Niemeyer}},
  \bibinfo {author} {\bibfnamefont {J}~\bibnamefont {Zhang}}, \ and\ \bibinfo
  {author} {\bibfnamefont {D}~\bibnamefont {Suter}},\ }\bibfield  {title}
  {\enquote {\bibinfo {title} {Robust dynamical decoupling for arbitrary
  quantum states of a single nv center in diamond},}\ }\href@noop {} {\bibfield
   {journal} {\bibinfo  {journal} {Europhysics Letters}\ }\textbf {\bibinfo
  {volume} {99}},\ \bibinfo {pages} {40004} (\bibinfo {year}
  {2012})}\BibitemShut {NoStop}%
\bibitem [{\citenamefont {Sukachev}\ \emph {et~al.}(2017)\citenamefont
  {Sukachev}, \citenamefont {Sipahigil}, \citenamefont {Nguyen}, \citenamefont
  {Bhaskar}, \citenamefont {Evans}, \citenamefont {Jelezko},\ and\
  \citenamefont {Lukin}}]{Sukachev2017}%
  \BibitemOpen
  \bibfield  {author} {\bibinfo {author} {\bibfnamefont {D.~D.}\ \bibnamefont
  {Sukachev}}, \bibinfo {author} {\bibfnamefont {A.}~\bibnamefont {Sipahigil}},
  \bibinfo {author} {\bibfnamefont {C.~T.}\ \bibnamefont {Nguyen}}, \bibinfo
  {author} {\bibfnamefont {M.~K.}\ \bibnamefont {Bhaskar}}, \bibinfo {author}
  {\bibfnamefont {R.~E.}\ \bibnamefont {Evans}}, \bibinfo {author}
  {\bibfnamefont {F.}~\bibnamefont {Jelezko}}, \ and\ \bibinfo {author}
  {\bibfnamefont {M.~D.}\ \bibnamefont {Lukin}},\ }\bibfield  {title} {\enquote
  {\bibinfo {title} {Silicon-vacancy spin qubit in diamond: A quantum memory
  exceeding 10 ms with single-shot state readout},}\ }\href {\doibase
  10.1103/PhysRevLett.119.223602} {\bibfield  {journal} {\bibinfo  {journal}
  {Phys. Rev. Lett.}\ }\textbf {\bibinfo {volume} {119}},\ \bibinfo {pages}
  {223602} (\bibinfo {year} {2017})}\BibitemShut {NoStop}%
\bibitem [{\citenamefont {Wood}\ \emph {et~al.}(2022)\citenamefont {Wood},
  \citenamefont {Stimpson}, \citenamefont {March}, \citenamefont {Lekhai},
  \citenamefont {Stephen}, \citenamefont {Green}, \citenamefont {Frangeskou},
  \citenamefont {Gin{\'e}s}, \citenamefont {Mandal}, \citenamefont {Williams}
  \emph {et~al.}}]{wood2022long}%
  \BibitemOpen
  \bibfield  {author} {\bibinfo {author} {\bibfnamefont {BD}~\bibnamefont
  {Wood}}, \bibinfo {author} {\bibfnamefont {GA}~\bibnamefont {Stimpson}},
  \bibinfo {author} {\bibfnamefont {JE}~\bibnamefont {March}}, \bibinfo
  {author} {\bibfnamefont {YND}\ \bibnamefont {Lekhai}}, \bibinfo {author}
  {\bibfnamefont {CJ}~\bibnamefont {Stephen}}, \bibinfo {author} {\bibfnamefont
  {BL}~\bibnamefont {Green}}, \bibinfo {author} {\bibfnamefont
  {AC}~\bibnamefont {Frangeskou}}, \bibinfo {author} {\bibfnamefont
  {L}~\bibnamefont {Gin{\'e}s}}, \bibinfo {author} {\bibfnamefont
  {S}~\bibnamefont {Mandal}}, \bibinfo {author} {\bibfnamefont
  {OA}~\bibnamefont {Williams}},  \emph {et~al.},\ }\bibfield  {title}
  {\enquote {\bibinfo {title} {Long spin coherence times of nitrogen vacancy
  centers in milled nanodiamonds},}\ }\href@noop {} {\bibfield  {journal}
  {\bibinfo  {journal} {Physical Review B}\ }\textbf {\bibinfo {volume}
  {105}},\ \bibinfo {pages} {205401} (\bibinfo {year} {2022})}\BibitemShut
  {NoStop}%
\bibitem [{\citenamefont {Varwig}\ \emph {et~al.}(2016)\citenamefont {Varwig},
  \citenamefont {Evers}, \citenamefont {Greilich}, \citenamefont {Yakovlev},
  \citenamefont {Reuter}, \citenamefont {Wieck}, \citenamefont {Meier},
  \citenamefont {Zrenner},\ and\ \citenamefont {Bayer}}]{varwig2016advanced}%
  \BibitemOpen
  \bibfield  {author} {\bibinfo {author} {\bibfnamefont {S}~\bibnamefont
  {Varwig}}, \bibinfo {author} {\bibfnamefont {E}~\bibnamefont {Evers}},
  \bibinfo {author} {\bibfnamefont {A}~\bibnamefont {Greilich}}, \bibinfo
  {author} {\bibfnamefont {DR}~\bibnamefont {Yakovlev}}, \bibinfo {author}
  {\bibfnamefont {Dirk}\ \bibnamefont {Reuter}}, \bibinfo {author}
  {\bibfnamefont {AD}~\bibnamefont {Wieck}}, \bibinfo {author} {\bibfnamefont
  {Torsten}\ \bibnamefont {Meier}}, \bibinfo {author} {\bibfnamefont {Artur}\
  \bibnamefont {Zrenner}}, \ and\ \bibinfo {author} {\bibfnamefont
  {M}~\bibnamefont {Bayer}},\ }\bibfield  {title} {\enquote {\bibinfo {title}
  {Advanced optical manipulation of carrier spins in ({In, Ga}) {As} quantum
  dots},}\ }\href@noop {} {\bibfield  {journal} {\bibinfo  {journal} {Applied
  Physics B}\ }\textbf {\bibinfo {volume} {122}},\ \bibinfo {pages} {1--11}
  (\bibinfo {year} {2016})}\BibitemShut {NoStop}%
\bibitem [{\citenamefont {Kosarev}\ \emph {et~al.}(2022)\citenamefont
  {Kosarev}, \citenamefont {Trifonov}, \citenamefont {Yugova}, \citenamefont
  {Yanibekov}, \citenamefont {Poltavtsev}, \citenamefont {Kamenskii},
  \citenamefont {Scholz}, \citenamefont {Sgroi}, \citenamefont {Ludwig},
  \citenamefont {Wieck} \emph {et~al.}}]{kosarev2022extending}%
  \BibitemOpen
  \bibfield  {author} {\bibinfo {author} {\bibfnamefont {Alexander~N}\
  \bibnamefont {Kosarev}}, \bibinfo {author} {\bibfnamefont {Artur~V}\
  \bibnamefont {Trifonov}}, \bibinfo {author} {\bibfnamefont {Irina~A}\
  \bibnamefont {Yugova}}, \bibinfo {author} {\bibfnamefont {Iskander~I}\
  \bibnamefont {Yanibekov}}, \bibinfo {author} {\bibfnamefont {Sergey~V}\
  \bibnamefont {Poltavtsev}}, \bibinfo {author} {\bibfnamefont {Alexander~N}\
  \bibnamefont {Kamenskii}}, \bibinfo {author} {\bibfnamefont {Sven~E}\
  \bibnamefont {Scholz}}, \bibinfo {author} {\bibfnamefont {Carlo~Alberto}\
  \bibnamefont {Sgroi}}, \bibinfo {author} {\bibfnamefont {Arne}\ \bibnamefont
  {Ludwig}}, \bibinfo {author} {\bibfnamefont {Andreas~D}\ \bibnamefont
  {Wieck}},  \emph {et~al.},\ }\bibfield  {title} {\enquote {\bibinfo {title}
  {Extending the time of coherent optical response in ensemble of
  singly-charged {InGaAs} quantum dots},}\ }\href@noop {} {\bibfield  {journal}
  {\bibinfo  {journal} {Communications Physics}\ }\textbf {\bibinfo {volume}
  {5}},\ \bibinfo {pages} {144} (\bibinfo {year} {2022})}\BibitemShut {NoStop}%
\bibitem [{\citenamefont {Zaporski}\ \emph {et~al.}(2023)\citenamefont
  {Zaporski}, \citenamefont {Shofer}, \citenamefont {Bodey}, \citenamefont
  {Manna}, \citenamefont {Gillard}, \citenamefont {Appel}, \citenamefont
  {Schimpf}, \citenamefont {Covre~da Silva}, \citenamefont {Jarman},
  \citenamefont {Delamare} \emph {et~al.}}]{zaporski2023ideal}%
  \BibitemOpen
  \bibfield  {author} {\bibinfo {author} {\bibfnamefont {Leon}\ \bibnamefont
  {Zaporski}}, \bibinfo {author} {\bibfnamefont {Noah}\ \bibnamefont {Shofer}},
  \bibinfo {author} {\bibfnamefont {Jonathan~H}\ \bibnamefont {Bodey}},
  \bibinfo {author} {\bibfnamefont {Santanu}\ \bibnamefont {Manna}}, \bibinfo
  {author} {\bibfnamefont {George}\ \bibnamefont {Gillard}}, \bibinfo {author}
  {\bibfnamefont {Martin~Hayhurst}\ \bibnamefont {Appel}}, \bibinfo {author}
  {\bibfnamefont {Christian}\ \bibnamefont {Schimpf}}, \bibinfo {author}
  {\bibfnamefont {Saimon~Filipe}\ \bibnamefont {Covre~da Silva}}, \bibinfo
  {author} {\bibfnamefont {John}\ \bibnamefont {Jarman}}, \bibinfo {author}
  {\bibfnamefont {Geoffroy}\ \bibnamefont {Delamare}},  \emph {et~al.},\
  }\bibfield  {title} {\enquote {\bibinfo {title} {Ideal refocusing of an
  optically active spin qubit under strong hyperfine interactions},}\
  }\href@noop {} {\bibfield  {journal} {\bibinfo  {journal} {Nature
  Nanotechnology}\ }\textbf {\bibinfo {volume} {18}},\ \bibinfo {pages}
  {257--263} (\bibinfo {year} {2023})}\BibitemShut {NoStop}%
\end{thebibliography}%

\end{document}